\documentstyle[11pt,aaspp4]{article}

\def\fe2{\ion{Fe}{2}}
\def\hb{H${\beta}$}
\def\hg{H${\gamma}$}
\def\oiii{$[$\ion{O}{3}$]$}
\def\ox3{$[$\ion{O}{3}$]\lambda5007$}
\def\odoublet{$[$\ion{O}{3}$]\lambda\lambda4959,5007$}

\def\civ{\ion{C}{4}}
\def\ciii{\ion{C}{3}$]$}
\def\zrange{$2.0 \le z \le 2.5$}

\def\kms{km~s$^{-1}$}
\def\logr{${\log(R^{\prime})}$}

\def\chisq{$\chi^2$}
\def\redchisq{$\chi_{\nu}^2$}

\def\peak{Peak$\lambda5007$}

\slugcomment{Revised Draft (15 Oct 98)}

\begin{document}
\title{A Statistical Study of Rest-frame Optical Emission Properties in
Luminous Quasars at $2.0 \le z \le 2.5$\footnote{Observations reported here were
obtained at the Multiple Mirror Telescope Observatory, a facility operated
jointly by the University of Arizona and the Smithsonian Institution.}}

\author{
D. H. McIntosh\altaffilmark{2,3},
M. J. Rieke\altaffilmark{2},
H.-W. Rix\altaffilmark{2}
}
\author{C. B. Foltz\altaffilmark{4}}
\author{R. J. Weymann\altaffilmark{5}}

\altaffiltext{2}{Steward Observatory, University of Arizona, Tucson, AZ 85721}
\altaffiltext{3}{E-mail: dmac@as.arizona.edu}
\altaffiltext{4}{Multiple Mirror Telescope Observatory, University of Arizona, Tuscon, AZ 85721}
\altaffiltext{5}{Observatories of the Carnegie Institute of Washington, 813 Santa Barbara Street, Pasadena, CA 91101}

\begin{abstract}
We have obtained $H$-band spectra of $32$ luminous quasars at \zrange\ with
the Multiple Mirror Telescope.  The sample contains
$15$ radio-loud quasars (RLQs)
and 17 radio-quiet quasars (RQQs).  We have measured emission line 
properties from the rest-frame wavelength range of approximately $\lambda\lambda4500-5500$,
by fitting the data with composite model spectra.  Our analysis
includes comparison of RLQs versus RQQs, as well as comparison between the
broad-absorption-line quasar (BALQSO) and non-broad-absorption-line quasar
(nonBALQSO) subsets of the RQQ sample.  In addition, we calculated the
complete correlation matrix of the measured properties.  We 
combined our high redshift sample with the sample of $87$ low redshift 
quasars from Boroson \& Green (1992) to determine
the luminosity and redshift 
dependences of the measured emission properties.  

Our main results are:
  (1) The RLQ sample has significantly (at $>97.2\%$ confidence) stronger
$[$\ion{O}{3}$]\lambda5007$ emission
than the RQQ sample, which favors scenarios including 
two populations of quasars that are
intrinsically different.  We are not aware of a unified model based upon
orientation that can explain enhanced $[$\ion{O}{3}$]$ emission with
increased radio power.
  (2) The RLQ sample has significantly narrower (in full-width at half-maximum)
H$\beta$ broad component line profiles than the RQQ sample.
  (3) At the sensitivity of our observations, there are no statistically significant ($>95\%$) differences between the rest-frame optical
emission line properties of the BALQSO and nonBALQSO subsamples.
This result is consistent with the view that all RQQs
have broad-absorption-line clouds with a small ($\sim10-20\%$)
covering factor and that differences between the two types are merely a 
function of viewing angle and covering factor.  
  (4) The significant \oiii\ -- \fe2\ anti-correlation found in lower redshift
quasars holds at this higher redshift range; however, it is the \oiii\ emission
in this relationship that appears to be related to the physical distinction
between the RLQ and RQQ classes instead of the \fe2\ emission that distinguishes
at low redshifts and luminosities.
We also find significant 
relationships between (i)
the \oiii\ emission strength and the radio power, the 
broad-emission-line
widths, and the X-ray continuum shape; (ii) positive correlations relating the
strength of optical \fe2\ emission to broad-emission-line
widths and the shape of the ionizing
continuum; and (iii) similar relations for the strength and width of the
\hb\ emission.  Many of these correlations have been found in lower redshift
and luminosity studies.
  (5) We report a previously unknown luminosity and/or redshift 
dependence of the narrow-line-region
velocity width over the range $0<z<2.5$, such that emission line widths
{\it increase} with increasing luminosity.  We confirm a 
similar dependence for the \hb\ broad line width.  These findings may be
evidence for a physical connection between the continuum and line-emitting
regions at similar energies.  Furthermore, we find a ``Baldwin Effect'' for
the $[$\ion{O}{3}$]\lambda5007$ line in the RQQ-only sample over this same 
range in redshifts.
\end{abstract}


\section {Introduction}
Quasi-stellar objects (QSOs $\equiv$ quasars) are among the most luminous objects
in the Universe, thus they give us the unique opportunity
to observe processes taking place at early epochs of the Universe.
A ``standard model''
for QSOs has emerged where the continuum luminosity 
(up to $L \sim 10^{48}$ ergs s$^{-1}$) is primarily produced in
an accretion disk
surrounding a super-massive black hole.  This model is supported by several
points: (i)~the mass of the central object is suggested by the Eddington
limit $M_{Edd} = 8 \times 10^5 L_{44} M_{\sun}$ (where $L_{44}$ is the
luminosity in $10^{44}$ ergs s$^{-1}$), thus one would expect $M \sim 10^{10}
M_{\sun}$ (\cite{peterson97});
(ii)~continuum variability studies have shown conclusively that
the continuum is produced in a very small region ($r~\sim$~few light~days 
$\sim~10^{16}$~cm,
\cite{peterson93}); and (iii)~the extreme energy output and tightly
collimated radio jets that emerge from the nuclei of radio-loud 
QSOs 
suggest a strong gravitational source with
large angular momentum, both defining qualities of an accretion fed giant black
hole (\cite{brotherton96}). 

One key to understanding the central engines of quasars lies in
examining their local environment.  As most of the power of QSOs is
emitted in the ultraviolet (UV) to soft X-ray regime (\cite{laor97}), this flux 
will photo-ionize the surrounding gas in physically distinct regions.
The broad emission lines in quasar spectra are 
thought to originate 
from the photo-ionization of dense ($n_e \approx 10^{11}$~cm$^{-3}$, \cite{ferland92}), high velocity
($v_{FWHM} \lesssim 10^4$~km~s$^{-1}$) clouds that populate the unresolved 
broad-emission-line-region (BELR)
within $r \sim 0.1-1.0$ pc of the nucleus
(\cite{kaspi96}).  Line diagnostics indicate that the BELR ionized
gas clouds have temperatures of a few $10^4$ K.  Since thermal and pressure
broadening are negligible for the inferred density and temperature 
(\cite{blandford82}), the lines must be broadened by bulk motions
of the BELR clouds ({\it e.g.}~\cite{woltjer59}; and \cite{burbidge67}).
Reverberation studies ({\it e.g.}~\cite{clavel91}; and \cite{peterson91}) 
have shown a time-delayed response
between emission line strengths and continuum variations, indicating a
radially stratified ionization structure and
confirming that the BELR is indeed
photo-ionized (\cite{baldwin97}).
A fairly successful model that reproduces the mean QSO broad emission spectrum
and ties many of these details together, proposes
that the BELR is an ensemble of locally optimally emitting clouds (LOCs) with a
modest covering factor (\cite{baldwin95}).

The narrow-line-region (NLR, $v_{FWHM} \sim 10^3$~km~s$^{-1}$), at 
$r \sim 1$ kpc, is believed to be
photo-ionized by continuum radiation extending from the Lyman
edge to the soft X-rays (\cite{wilson97}).  Forbidden
transitions, such as \odoublet\ , dominate the NLR emission and are radiating near their
critical densities (\cite{filippenko84}), indicating NLR densities of
$n_e \sim 10^{3}$~cm$^{-3}$.
Recent Hubble Space Telescope (HST) observations of a handful of local
active-galactic-nuclei (AGN) have resolved the NLR and shown it to have a
roughly bi-conical geometry
({\it e.g.}~\cite{wilson93}; \cite{bower95}; and 
\cite{simpson96}).

A significant issue in understanding the plethora of AGN ``types'' is the 
question of {\it intrinsic} vs. {\it orientation}
differences.  The radio-loud QSO (RLQ)/
radio-quiet QSO (RQQ) dichotomy is an important example of 
the intrinsic vs. orientation dilemma in 
QSO astronomy.  Kellermann~{\it et al.} (1989) originally defined the
radio-to-optical ratio $R_{r-o} = \frac{F_{\nu}(5GHz)}{F_{\nu}(4400\AA)}$ and called QSOs with
$R_{r-o} > 10$ radio-loud.  This ratio, regardless of specific radio and optical
flux apertures, appears bimodal with an absence of objects in the
$R_{r-o} =1$ to $10$ range (\cite{kellermann89}; \cite{visnovsky92};
\cite{stocke92}; and \cite{hooper95}).  The distribution of QSOs at higher
redshift ($1.8<z<2.5$) also appears bimodal in
terms of the $5$ GHz radio luminosity only (\cite{miller90}).  The fraction
of RLQs ($\sim 10\%$) in optically selected samples does not evolve
significantly from $z=0.2$ to redshifts approaching $5$ (\cite{hooper96}).
In terms of morphology, many RLQs have radio lobes and energetic jets of beamed radio emission
extending to hundreds of kpcs,
while similar radio features are absent in RQQs.  This characteristic
bright compact radio emission is thought to originate from non-thermal
synchrotron sources due to the nearly flat slopes and typically high
($\sim 10^{11-12}$ K) brightness temperatures (\cite{peterson97}).
Elvis~{\it et al.} (1994) produced the mean spectral energy distributions (SEDs)
for a set of RLQs and RQQs and showed that they have roughly the
same SEDs from $1000$\AA \ to $100$\micron, but that the RLQs are $\sim 1000$ 
times brighter at
radio wavelengths and $\sim 3$ times brighter in X-rays, compared to RQQs.  
RLQs also show a flatter, harder soft X-ray slope, and it is believed that
this extra X-ray luminosity is due to a hard, non-thermal
component produced by the inverse Compton
scattering of radio emission along the associated jet (\cite{elvis94}; and
\cite{green95}).
 
The physical process that is responsible for the difference in radio emission
between RLQs and RQQs remains unknown.  The spectral similarities 
at most wave-bands imply that
the black hole mass and the accretion rate should be quite similar in RLQs and
RQQs (\cite{hooper96}).  
The angular momentum of the super-massive black hole might provide a potential
difference such that large spin energy could produce 
powerful radio emission 
(\cite{blandford90}).
Rapidly rotating massive black holes may form in the merger of two black holes
of similar masses (\cite{wilson95}).
Alternatively, RLQs may constitute a short-lived phase during the
lifetimes of all quasars (\cite{wills96a}).

Other differences between RLQs and RQQs have been observed: (i)~in rest UV
emission line strengths (\cite{corbin91}; \cite{francis93}, hereafter FHI93; \cite{corbin94}; 
\cite{brotherton94a}); (ii)~in rest UV line widths (\cite{jackson91}; \cite{corbin91}; 
FHI93; \cite{wills93}; \cite{corbin94}; \cite{brotherton94a}; 
\cite{baker95}; and \cite{vestergaard97}); (iii)~
in optical rest-frame broad lines (Boroson \& Green 1992 -- hereafter BG92;
 and \cite{corbin96}); and (iv)~as
stronger \oiii\
emission in RLQs (BG92; and \cite{wills96b}).

Another important example of the intrinsic vs. orientation problem concerns 
the broad-absorption-line QSO (BALQSO)
phenomenon.  About $10\%$ of optically selected QSOs have UV rest spectra
showing deep, broad ($v_{FWHM} \lesssim 20,000$~km~s$^{-1}$) absorption lines 
extending blueward from 
the corresponding emission lines in the high ionization transitions of
\ion{C}{4}$\lambda1549$, \ion{Si}{4}$\lambda1400$, \ion{N}{5}$\lambda1240$, 
and \ion{O}{6}$\lambda1035$ (\cite{green95}). 
In particular,
the strength and proximity of the \civ\ absorption relative
to its corresponding \ion{C}{4}$\lambda1549$ emission line, provides the basis
for determining the strength (BALnicity) of the BAL phenomenon in individual
QSOs (\cite{weym91}; hereafter WMFH91).  In addition,
a small fraction ($<10\%$) of BALQSOs, known as loBALs or \ion{Mg}{2} BALQSOs, 
show additional broad-absorption-line (BAL) features in the
low ionization species at \ion{Mg}{2}$\lambda2798$, \ion{Al}{2}$\lambda1671$, and \ion{Al}{3}$\lambda\lambda1855,1863$ 
(WMFH91; and \cite{peterson97}).

It has been demonstrated that the rest-frame UV emission line properties of BALQSOs and
radio quiet non-broad-absorption-line QSOs (nonBALQSOs) 
are quite similar (WMFH91).  Furthermore, the total
lack of radio loud BALQSOs (\cite{stocke92}; except possibly 1556+3517
\cite{becker97}) 
establishes a 
strong anti-correlation between
luminous radio sources and the BALQSO phenomenon.  This
is consistent with current QSO unification theories
that propose that BALQSOs do not form an 
intrinsically different class of objects
from radio quiet nonBALQSOs.  The standard unification model places an ensemble
of BAL clouds in the neighborhood of the BELR of the ``standard model''.
Spectropolarimetry studies place the broad-absorption-line-region (BALR)
location just exterior to the
BELR at $r\sim1$ pc (\cite{cohen95}; and \cite{glenn94}).  Scattering constraints
between the \civ\ emission and absorption establish a low covering factor, 
roughly equal
to the BALQSO detection rate ($\sim 10\%$), for the 
BAL clouds (\cite{hamann93}).
BALQSOs are known to be weak in X-ray flux (\cite{turnshek84}; \cite{green95}
\& 1996);
therefore, a RQQ is classified a BALQSO when viewed through the small angle
containing the BAL clouds which may be sloughed off the surface of the
accretion disk, thus any minor emission-line differences between
BALQSOs and nonBALQSOs is due to obscuration associated with the BAL clouds.
In this picture, any emission produced on larger spatial scales such as 
forbidden \oiii\
emission lines from the NLR should be isotropic in BALQSOs and radio quiet 
nonBALQSOs.  

Assuming that \oiii\
emission is indeed an isotropic property, Boroson \& Meyers (1992; see also
\cite{turnshek94}) argued
that loBALs may represent
a rare population of QSOs physically distinct from RQQs and hiBALs --- the
differences due to much larger (up to $100\%$, hence roughly spherically
symmetric) BAL covering factors that would inhibit ionization of the narrow
\oiii\ emitting gas.  The strong \ion{N}{5}$\lambda1240$ 
absorption associated with the absence of Ly$\alpha$ emission in loBALs also
suggests large covering factors (\cite{peterson97}).
Others have argued for unification of both BALQSO types ---
suggesting the differences are the result of lines of sight through hot dust 
(\cite{sprayberry92}; \cite{hines95}; \cite{goodrich95}; and \cite{wills96b}),
possibly due to ongoing nuclear starbursts (\cite{lipari94}).

In this paper, we have selected a sample of $32$ very luminous QSOs 
with redshifts
spanning the range $2.0 \le z \le 2.5$.  This sample was magnitude limited 
($V \lesssim 18.0$) so that sufficient signal-to-noise (S/N) near-infrared 
(NIR) $H$-band spectra could be
obtained with a $4$-meter class telescope.  These observational parameters
allowed us an unprecedented view of the rest-frame optical emission
lines of high redshift and high luminosity quasars.  The spectra
were centered near
the interesting H$\beta$, \odoublet\ and blended \fe2\ features.
Consequently, we present the first detailed study of the
important \oiii\ NLR emission at these redshifts, when the Universe was
$\sim 75\%$ younger and QSOs were in their ``heyday'' --- they existed in
the largest numbers and the brightest QSOs
were $10-100$ times more luminous 
than their corresponding present-day counterparts.
The recent advances of NIR spectrographs 
with adequate sensitivity and array size have just made these types of 
moderate resolution studies feasible.  Until now, 
only a handful of 
very low resolution NIR spectra have been obtained for
$z\sim2$ QSOs (\cite{carswell91}; \cite{hill93}; and \cite{baker94}).

From the rest-frame optical emission line properties, combined with continuum
parameters from the radio, optical and soft X-ray band-passes, plus rest-frame
UV emission line data, we can
perform a complete high redshift QSO property correlation analysis similar to
the important low redshift study carried out by BG92.  Our selection of
approximately equal numbers of RLQs ($15$) and RQQs ($17$) allows us to
study the statistical differences between these types in terms of observed 
characteristics that have been previously poorly examined at high redshift.
The same justification applies to our subsamples of BALQSOs ($7$) and
radio quiet nonBALQSOs ($10$).

The organization of this paper is as follows:  In $\S2$ we describe the
selection and observation of this sample.  We also calculate and tabulate
the continuum parameters: (i)~the ratio of radio-to-optical flux; (ii)~the
rest-frame $V$ band luminosity density; and (iii)~the monochromatic soft
X-ray luminosity density plus optical-to-X-ray spectral index.  We
compile the rest-frame UV line widths and equivalent widths (EWs).
In $\S3$ we describe the raw data
reduction procedures, the systemic redshift determination, the construction
of a composite model spectrum and the fit of this model to the data, and the
measurement of the emission line properties from the data.  In $\S4$ we 
calculate the Spearman Rank, and the Kendall $\tau$ (to check for consistency),
correlation matrices for the measured and compiled rest-frame emission
properties of our combined RLQ + RQQ sample.  The significant results are 
compared to
findings from the literature.  In $\S5$ we fit our model
spectrum to BG92's low redshift sample, then combine the results with the
similar optical rest-frame measurements from our high redshift sample.  We
use this combined sample spanning the redshift range $0<z<2.5$ to
determine the
dependencies on rest-frame $V$ band luminosity and/or redshift for each of the
measured properties.
In $\S6$ we draw the RLQ and RQQ subsamples
from our data and perform statistical tests to determine
the differences in emission parameters between these.  The significant results
we find for the rest-frame optical wavelengths are added with
rest-frame UV
differences from the literature to test the validity of the proposed QSO
model that RLQs and RQQs are separate classes of QSO.  
In $\S7$ we perform the same analysis as $\S6$ on the BALQSO and nonBALQSO
subsets drawn from the RQQ sample.  We use our findings to test the validity
that BALQSOs are a subset of all RQQs and that observed differences are due to
orientation coupled to coverage of the BAL material.  We also note specific
object-to-object spectral differences in light of the loBAL phenomenon.
In $\S8$ we summarize our results, and in $\S9$ we discuss our results in the
context of two simple models, one RLQ and the other RQQ.

\section {Observations}

A total of $32$ QSOs brighter than $V=18.0$ mag., with redshifts
between {$2.0 \le z \le 2.5$, were observed.
The sample consists of $7$ BALQSOs ($5$
from WMFH91), $10$ nonBALQSOs, and $15$ RLQs.
One object, Q1148-001, has redshift
$z=1.980$ slightly out of our specified range and thus no \hb\ 
measurements were obtained.

All observations were made in the {\it H}-band
with the long-slit near-IR spectrometer, FSPEC (\cite{fspec93}), at
the $4.5$-meter Multiple Mirror Telescope (MMT).  FSPEC
uses a
$256\times256$ NICMOS3 HgCdTe array and a $75$ grooves per millimeter
low resolution grating 
providing a two-pixel resolution
of about $700$.  
Observations were taken through an {\it H}(${1.6\micron}$) filter at 
four positions along a
$1.2\arcsec$ by $30\arcsec$ slit which produced an instrumental resolution 
of about $550$~\kms\ and
spectral coverage of about $0.30\micron$.  The limiting $V$ magnitude 
corresponds to $H \sim 16$ mag. which allowed a S/N $\approx 10$ to
be achieved in
exposure times of a few hours.
The slit was positioned along a line of constant azimuth for all 
observations and we attempted to
observe at air masses less than $1.60$.
Spectra of atmospheric transmission standard
stars, used to remove telluric atmosphere absorption features, 
were obtained at least
twice a night for each object observed.  Exposure times were
typically between $960$ and $7200$ seconds.  Observations were made
over many nights between November 1993 and April 1996.  The log of observations
is presented in Table~1.  Columns (3) and (4) contain the published 
apparent $V$-band magnitude and redshift, respectively.  Column (5) lists the
systemic redshift measured from the forbidden narrow \ox3\ line in this
study (see $\S3.1$).
Additional photometric properties 
compiled from the literature are presented in Table~2.
Column (2) classifies each QSO as radio-quiet (RQ) or BAL (BALQSO) such that
\logr\ $<1$, or radio-loud RL (\logr\ $>1$).  Columns (3) gives the logarithm of
$R^{\prime}$, the ratio of the {\it observed} radio ($5$ GHz) 
flux density to the
{\it observed} optical {\it V}-band flux density assuming zero magnitude
in {\it V} is $3880$ Jy (\cite{johnson66}).  Though this
$R^{\prime}$ parameter differs slightly 
from other radio-to-optical ({\it R}) parameters normally used to
classify AGN (see for example \cite{kellermann89}, 
\cite{hooper95}, \cite{stocke92}),
the resultant bimodal distribution and classifications
are consistent with those from the literature.  Column (4) lists
the $5$ GHz flux reference(s), in some cases $F_{\rm 5GHz}$ was extrapolated from
$F_{\rm 1.4GHz}$, $F_{\rm 4.85GHz}$, or $F_{\rm 8.4GHz}$ assuming 
$F_{\nu} \propto \nu^{-\case{1}{2}}$ over radio frequencies (FHI93).
Column (5) gives the logarithm of the luminosity density
(in ergs ${\rm s}^{-1} {\rm Hz}^{-1}$) emitted at rest-frame {\it V} given by
\begin{equation}
L_{\nu}(V) = 4\pi D_L^2(z) F_{\nu}(H) (1+z)^{-1} .
\end{equation}
\begin{equation}
D_L(z) = \frac{2cz}{{\rm H_0}g}(1+\frac{z}{g})
\end{equation}
is the luminosity distance from Hall~{\it et al.} (1997), with
\begin{equation}
g = 1 + \sqrt{1+2zq_0} ,
\end{equation}
$c$ is the velocity of light, and
$z = z_{\rm sys}$ is the systemic redshift measured in this study.
$F_{\nu}(H)$ is the observed {\it H}-band flux density 
(in ergs~cm$^{-2}~{\rm s}^{-1}~{\rm Hz}^{-1}$) derived from 
{\it H}-band magnitudes
and assuming zero magnitude in {\it H} is $1075$ Jy (\cite{campins85}).
Note that 
$\frac{\lambda_{H}}{\lambda_{V}} \approx 1 + \langle z \rangle$,
and hence no K-corrections were necessary.
Following BG92
we adopted an H$_{0} = 50$~km~s$^{-1}$~Mpc$^{-1}$
and $q_0 = 0.1$ cosmology 
throughout this paper.
Columns (6) and (7) list the {\it H}-band magnitude and its reference.  
Column (8) gives
the logarithm of the monochromatic soft X-ray luminosity density 
(in ergs ${\rm s}^{-1} {\rm Hz}^{-1}$) emitted at rest-frame $2$ keV given by
\begin{equation}
L_{\nu}({\rm 2keV}) = \sqrt{2}\pi D_L^2(z) F_{\nu}({\rm 1keV}) \sqrt{1+z} ,
\end{equation}
where $F_{\nu}$(1keV) is the monochromatic ($1$ keV) flux density 
(in ergs~${\rm cm}^{-2}~{\rm s}^{-1}~{\rm Hz}^{-1}$) derived from the {\it observed} flux
density and assuming $F_{\nu} \propto \nu^{-\case{3}{2}}$ over 
soft X-ray frequencies (\cite{wilkes94}).  The factor of root $(1+z)$ is
related to the difference between the frequency corresponding to $1$~keV
and the {\it observed} frequency.  Column (9) contains the mean spectral slope
$\alpha_{ox}$ between the rest-frame optical ({\it V}-band) and rest-frame
$2$ keV soft X-ray emission.  This spectral index is defined such that 
$L_{\nu} \propto \nu^{-\alpha_{ox}}$, or
\begin{equation}
\alpha_{ox} = - \frac{1}{a} \log{\frac{L_{\nu}({\rm 2keV})}{L_{\nu}(V)}} ,
\end{equation}
where the constant $a=2.948$ is the logarithm of the ratio of
$2$ keV to {\it V}-band frequencies.  The luminosity densities, 
$L_{\nu}(V)$ and $L_{\nu}$(2keV), are
from equations (1) and (4).
Column (10) lists the soft X-ray references.

\begin{scriptsize}
\begin{deluxetable}{lcccccccc}
\tablewidth{0pt}
\tablenum{1}
\tablecaption{Log~of~Observations}
\tablehead{\multicolumn{1}{l}{Object~Name} &
\colhead{Other~Name} & \colhead{$V$\tablenotemark{a}} & \colhead{$z_{\rm lit}$\tablenotemark{b}} & \colhead{$z_{\rm sys}$} & \colhead{RA(1950.0)\tablenotemark{c}} & \colhead{Dec(1950.0)\tablenotemark{c}} & \colhead{Date(s)} & \colhead{$t_{\rm exp}$
}\\
\colhead{} & \colhead{} & \colhead{(mag.)} & \colhead{} & \colhead{} & \colhead{} & \colhead{} & \colhead{} & \colhead{(sec.)
}}
\startdata
Q0043+008 & UM275 & 17. & 2.143 & 2.146 & 00h43m39.59s & +00\arcdeg48\arcmin02\farcs40 & 1993~Nov~29 & 3360 \nl
\nodata & \nodata & \nodata & \nodata & \nodata & \nodata & \nodata & 1993~Nov~30 & 3360 \nl
Q0049+007 & UM287 & 17.8 & 2.268 & 2.279 & 00h49m28.43s & +00\arcdeg45\arcmin11\farcs30 & 1994~Sept~22 & 4800 \nl
Q0049+014 & UM288 & 17. & 2.310 & 2.307 & 00h49m59.56s & +01\arcdeg24\arcmin23\farcs30 & 1993~Nov~30 & 4800 \nl
Q0109+022 & UM87 & 17.8 & 2.350 & 2.351 & 01h09m42.31s & +02\arcdeg13\arcmin53\farcs1 & 1994~Sept~22 & 4080 \nl
Q0123+257 & 4c25.05 & 17.5 & 2.358 & 2.370 & 01h23m57.26s & +25\arcdeg43\arcmin27\farcs88 & 1995~Nov~10 & 5040 \nl
Q0153+744 & \nodata & 16.0 & 2.338 & 2.341 & 01h53m04.33s & +74\arcdeg28\arcmin05\farcs58 & 1994~Nov~17 & 2880 \nl
\nodata & \nodata & \nodata & \nodata & \nodata & \nodata & \nodata & 1995~Nov~2 & 1920 \nl
\nodata & \nodata & \nodata & \nodata & \nodata & \nodata & \nodata & 1995~Nov~6 & 3600 \nl
Q0226-104 & \nodata & 17.0 & 2.256 & 2.268 & 02h26m00s & -10\arcdeg24\farcm0 & 1993~Nov~29 & 3360 \nl
Q0226-038 & PHL1305 & 16.96 & 2.066 & 2.073 & 02h26m22.10s & -03\arcdeg50\arcmin58\farcs98 & 1994~Jan~31 & 4320 \nl
Q0421+019 & \nodata & 17.04 & 2.055 & 2.056 & 04h21m32.67s & +01\arcdeg57\arcmin32\farcs70s & 1993~Nov~30 & 5760 \nl
Q0424-131 & \nodata & 17.5 & 2.165 & 2.168 & 04h24m47.85s & -13\arcdeg09\arcmin33\farcs40 & 1994~Jan~29 & 4080 \nl
Q0552+398 & \nodata & 18. & 2.365 & 2.363 & 05h52m01.40s & +39\arcdeg48\arcmin21\farcs94 & 1995~Nov~10 & 4680 \nl
Q0836+710 & 4c71.07 & 16.5 & 2.170 & 2.218 & 08h36m21.54s & +71\arcdeg04\arcmin22\farcs54 & 1993~Nov~29 & 4320 \nl
\nodata & \nodata & \nodata & \nodata & \nodata & \nodata & \nodata & 1993~Nov~30 & 4680 \nl
Q0842+345 & CSO203 & 17. & 2.126 & 2.163\tablenotemark{d} & 08h42m30.37s & +34\arcdeg31\arcmin41\farcs0 & 1994~Jan~29 & 6240 \nl
\nodata & \nodata & \nodata & \nodata & \nodata & \nodata & \nodata & 1995~Mar~19 & 5760 \nl
Q1011+091 & \nodata & 17.8 & 2.268 & 2.305\tablenotemark{d} & 10h11m03.35s & +09\arcdeg06\arcmin19\farcs90 & 1993~Nov~30 & 4320 \nl
\nodata & \nodata & \nodata & \nodata & \nodata & \nodata & \nodata & 1994~Apr~30 & 960 \nl
\nodata & \nodata & \nodata & \nodata & \nodata & \nodata & \nodata & 1995~Mar~19 & 4320 \nl
Q1104-181 & HE1104-1805A & 16.2 & 2.319 & 2.318 & 11h04m04.95s & -18\arcdeg05\arcmin10\farcs07 & 1996~Apr~3 & 4200 \nl
Q1148-001 & UM458 & 17.14 & 1.980 & 1.980 & 11h48m10.13s & -00\arcdeg07\arcmin13\farcs01 & 1996~Apr~7 & 7200 \nl
Q1158-187 & Pox42 & 16.93 & 2.453 & 2.462\tablenotemark{d} & 11h58m11.27s & -18\arcdeg43\arcmin02\farcs70 & 1995~Apr~10 & 5280 \nl
Q1222+228 & Ton1530 & 15.49 & 2.048 & 2.058 & 12h22m56.58s & +22\arcdeg51\arcmin49\farcs00 & 1994~Apr~29 & 3840 \nl
Q1225+317 & Ton0618 & 15.84 & 2.219 & 2.226\tablenotemark{d} & 12h25m55.94s & +31\arcdeg45\arcmin12\farcs60 & 1994~May~20 & 1920 \nl
Q1228+077 & \nodata & 17.59 & 2.391 & 2.389 & 12h28m48.02s & +07\arcdeg42\arcmin26\farcs40 & 1994~May~22 & 3120 \nl
Q1246-057 & \nodata & 16.73 & 2.236 & 2.243\tablenotemark{d} & 12h46m38.69s & -05\arcdeg42\arcmin58\farcs9 & 1994~May~26 & 2880 \nl
\nodata & \nodata & \nodata & \nodata & \nodata & \nodata & \nodata & 1995~Mar~19 & 3840 \nl
Q1247+267 & PG1247+267 & 15.8 & 2.043 & 2.042 & 12h47m39.09s & +26\arcdeg47\arcmin27\farcs10 & 1994~Apr~29 & 2400 \nl
Q1309-056 & \nodata & 17.44 & 2.188 & 2.220\tablenotemark{d} & 13h09m00.75s & -05\arcdeg36\arcmin43\farcs40 & 1994~Apr~30 & 6240 \nl
\nodata & \nodata & \nodata & \nodata & \nodata & \nodata & \nodata & 1995~Mar~19 & 4320 \nl
Q1331+170 & \nodata & 16.71 & 2.084 & 2.097 & 13h31m10.10s & +17\arcdeg04\arcmin25\farcs0 & 1996~Apr~1 & 2880 \nl
Q1346-036 & \nodata & 17.27 & 2.349 & 2.362 & 13h46m08.32s & -03\arcdeg38\arcmin30\farcs80 & 1994~May~21 & 1440 \nl
\nodata & \nodata & \nodata & \nodata & \nodata & \nodata & \nodata & 1994~May~22 & 2760 \nl
\nodata & \nodata & \nodata & \nodata & \nodata & \nodata & \nodata & 1994~May~26 & 3840 \nl
Q1416+091 & \nodata & 17.0 & 2.015 & 2.017 & 14h16m23.30s & +09\arcdeg06\arcmin14\farcs0 & 1994~Apr~29 & 4800 \nl
\nodata & \nodata & \nodata & \nodata & \nodata & \nodata & \nodata & 1995~Mar~19 & 3840 \nl
Q1435+638 & \nodata & 15. & 2.068 & 2.066 & 14h35m37.25s & +63\arcdeg49\arcmin35\farcs97 & 1996~Apr~6 & 6240 \nl
Q1448-232 & \nodata & 16.96 & 2.215 & 2.220 & 14h48m09.31s & -23\arcdeg17\arcmin11\farcs30 & 1996~Apr~7 & 5760 \nl
Q1704+710 & \nodata & 17.5 & 2.015 & 2.010 & 17h05m00.60s & +71\arcdeg01\arcmin34\farcs0 & 1994~May~26 & 3360 \nl
\nodata & \nodata & \nodata & \nodata & \nodata & \nodata & \nodata & 1994~Sept~22 & 2160 \nl
Q2212-179 & \nodata & 18.3 & 2.280 & 2.228 & 22h12m48.30s & -17\arcdeg59\arcmin03\farcs1 & 1993~Nov & 30 \nl
\nodata & \nodata & \nodata & \nodata & \nodata & \nodata & \nodata & 1994~Nov~15 & 3120 \nl
\nodata & \nodata & \nodata & \nodata & \nodata & \nodata & \nodata & 1994~Nov~17 & 4800 \nl
Q2251+244 & 4c24.61 & 17.8 & 2.327 & 2.359 & 22h51m44.57s & +24\arcdeg29\arcmin23\farcs80 & 1994~Sept~22 & 4320 \nl
\nodata & \nodata & \nodata & \nodata & \nodata & \nodata & \nodata & 1995~Nov~3 & 7080 \nl
Q2310+385 & \nodata & 17.5 & 2.170 & 2.181 & 23h10m36.18s & +38\arcdeg31\arcmin22\farcs69 & 1993~Nov~29 & 4440 \nl
\enddata
\tablenotetext{a}{$V$-band magnitudes from Hewitt \& Burbidge 1993; except for Q0226-104 and HE1104-1805A from NED; and Q0421+019, Q1148-001, Q1158-187, Q1228+077, Q1331+170, and Q1346-036 from Adam 1985.}
\tablenotetext{b}{Published redshifts from Hewitt \& Burbidge 1993; except for Q0226-104 from WMFH91; and HE1104-1805A from Wisotzki {\it et al.} 1993.}
\tablenotetext{c}{Epoch 1950.0 coordinates from NED.}
\tablenotetext{d}{Uncertain measurement due to low $[$\ion{O}{3}$]$ EW and/or poor S/N spectrum.}
\end{deluxetable}
\end{scriptsize}

\begin{scriptsize}
\begin{deluxetable}{lccccccccc}
\tablewidth{0pt}
\tablenum{2}
\tablecaption{Photometric~Properties}
\tablehead{
\multicolumn{1}{l}{QSO} &
\colhead{Type} & \colhead{$\log(R^{\prime})$} & \colhead{Refs.\tablenotemark{a,d}} & \colhead{$\log(L_{\nu}(V))$} & \colhead{$H$} & \colhead{Refs.\tablenotemark{b}} & \colhead{$\log(L_{\nu}$(2keV))} & \colhead{$\alpha_{ox}$} & \colhead{Refs.\tablenotemark{c,d}
}\\
\colhead{} & \colhead{} & \colhead{} & \colhead{} & \colhead{(ergs ${\rm s}^{-1} {\rm Hz}^{-1})$} & \colhead{(mag.)} & \colhead{} & \colhead{(ergs ${\rm s}^{-1} {\rm Hz}^{-1})$} & \colhead{} & \colhead{
}}
\startdata
Q0043+008 & BAL & 0.824 & 1,2 & 32.24 & 15.26 & $12$\tablenotemark{h} & $<27.30$ & $<1.675$ & $17$\tablenotemark{i} \nl
Q0049+007 & RQ & $<0.230$ & $3$\tablenotemark{e} & 32.08 & 15.79 & 12 & 26.84 & $1.777$ & 18 \nl
Q0049+014 & RQ & $<0.325$ & $4$\tablenotemark{f} & 32.30 & 15.26 & $12$\tablenotemark{h} & 28.16 & $1.405$ & 18 \nl
Q0109+022 & RQ & 0.133 & $3$\tablenotemark{e} & 32.24 & 15.46 & 12 & $<28.26$ & $<1.350$ & 18 \nl
Q0123+257 & RL & 3.393 & 5,6 & 32.13 & 15.76 & $12$\tablenotemark{h} & 28.84 & $1.115$ & $19$\tablenotemark{j} \nl
Q0153+744 & RL & 2.990 & 5,7 & 32.72 & 14.26 & $12$\tablenotemark{h} & 28.40 & $1.464$ & $19$\tablenotemark{j} \nl
Q0226-104 & BAL & $<-0.229$ & 2 & 32.57 & 14.55 & 13 & $<27.55$ & $<1.703$ & $17$\tablenotemark{i} \nl
Q0226-038 & RL & 3.046 & 6 & 32.22 & 15.22 & $12$\tablenotemark{h} & 28.45 & $1.280$ & 18 \nl
Q0421+019 & RL & 3.090 & 6 & 32.18 & 15.30 & $12$\tablenotemark{h} & \nodata & \nodata & \nodata \nl
Q0424-131 & RL & 2.864 & 6,8 & 32.05 & 15.76 & $12$\tablenotemark{h} & 27.98 & $1.379$ & 18 \nl
Q0552+398 & RL & 4.344 & 5 & 31.92 & 16.26 & $12$\tablenotemark{h} & 29.12 & $0.951$ & $19$\tablenotemark{j} \nl
Q0836+710 & RL & 3.425 & 5,7 & 32.47 & 14.76 & $12$\tablenotemark{h} & 29.57 & $0.983$ & $19$\tablenotemark{j},$20$\tablenotemark{j} \nl
Q0842+345 & BAL & $<0.325$ & $4$\tablenotemark{f} & 32.24 & 15.26 & $12$\tablenotemark{h} & \nodata & \nodata & \nodata \nl
Q1011+091 & BAL & $<0.008$ & 2 & 32.28 & 15.32 & 13 & \nodata & \nodata & \nodata \nl
Q1104-181 & RQ & $<0.005$ & $4$\tablenotemark{f} & 32.63 & 14.46 & $12$\tablenotemark{h} & 29.05 & $1.213$ & $21$\tablenotemark{j} \nl
Q1148-001 & RL & 3.546 & 5,6,7 & 32.06 & 15.53 & 12 & 28.60 & $1.173$ & $22$\tablenotemark{i} \nl
Q1158-187 & RQ & $<0.297$ & $4$\tablenotemark{f} & 32.39 & 15.19 & $12$\tablenotemark{h} & \nodata & \nodata & \nodata \nl
Q1222+228 & RQ & 0.668 & 1,9 & 32.58 & 14.3 & 14 & 28.29 & $1.457$ & 18 \nl
Q1225+317 & RL & 2.276 & $10$\tablenotemark{g},$11$\tablenotemark{g} & 32.69 & 14.20 & 15 & 28.53 & $1.413$ & $19$\tablenotemark{j} \nl
Q1228+077 & RL & 2.003 & $11$\tablenotemark{g} & 32.10 & 15.85 & $12$\tablenotemark{h} & 28.33 & $1.278$ & 18 \nl
Q1246-057 & BAL & 0.057 & 2 & 32.11 & 15.67 & 13 & 26.93 & $1.758$ & $17$\tablenotemark{i} \nl
Q1247+267 & RQ & -0.269 & 9 & 32.66 & 14.1 & 14 & 28.42 & $1.437$ & $19$\tablenotemark{j} \nl
Q1309-056 & BAL & $<-0.534$ & 2 & 32.11 & 15.65 & 13 & 27.90 & $1.428$ & 18 \nl
Q1331+170 & RL & 2.909 & $10$\tablenotemark{g},$11$\tablenotemark{g} & 32.53 & 14.49 & 12 & 28.17 & $1.477$ & 18 \nl
Q1346-036 & RQ & $<0.433$ & $4$\tablenotemark{f} & 32.22 & 15.53 & $12$\tablenotemark{h} & $<28.13$ & $<1.386$ & 18 \nl
Q1416+091 & RQ & $<0.325$ & $4$\tablenotemark{f} & 32.18 & 15.26 & $12$\tablenotemark{h} & \nodata & \nodata & \nodata \nl
Q1435+638 & RL & 2.498 & 5,7 & 33.00 & 13.26 & $12$\tablenotemark{h} & $<28.31$ & $<1.592$ & $19$\tablenotemark{j} \nl
Q1448-232 & RL & 2.687 & 6 & 32.42 & 14.87 & 16 & $<28.47$ & $<1.342$ & 18 \nl
Q1704+710 & RQ & $<0.525$ & $4$\tablenotemark{f} & 31.98 & 15.76 & $12$\tablenotemark{h} & \nodata & \nodata & \nodata \nl
Q2212-179 & BAL & $<0.250$ & 2 & 32.32 & 15.14 & 12 & 28.59 & $1.265$ & $22$\tablenotemark{i} \nl
Q2251+244 & RL & 3.446 & 6,8 & 32.00 & 16.06 & $12$\tablenotemark{h} & 28.75 & $1.104$ & $19$\tablenotemark{j} \nl
Q2310+385 & RL & 3.110 & $10$\tablenotemark{g},$11$\tablenotemark{g} & 32.05 & 15.76 & $12$\tablenotemark{h} & \nodata & \nodata & \nodata \nl
\enddata
\tablenotetext{a}{The $5$ GHz flux references.}
\tablenotetext{b}{The $H$-band magnitude references.}
\tablenotetext{c}{The soft X-ray monochromatic (1 keV) flux density references.}
\tablenotetext{d}{Multiple references indicate that the value is an average.}
\tablenotetext{e}{Reference contained $8.4$ GHz fluxes.}
\tablenotetext{f}{Reference contained $1.4$ GHz fluxes.}
\tablenotetext{g}{Reference contained $4.85$ GHz fluxes.}
\tablenotetext{h}{Calculated $H$ magnitudes from an average color $(V-H)=1.74$, derived from a sample of $53$ QSOs within our redshift range, supplied by Hewett \& Foltz (1997).}
\tablenotetext{i}{Derived result from luminosity density for a $q_0=0.5$ cosmology.}
\tablenotetext{j}{Derived $F_{\nu}$(1keV) from ROSAT flux integrated between $0.1$ and $2.4$ keV, assuming $f_{\nu} \propto \nu^{-1.5}$.}
\tablerefs{$(1)$ Barvainis {\it et~al.} 1996; $(2)$ Stocke {\it et~al.} 1992; $(3)$ Hooper {\it et~al.} 1995; $(4)$ Condon {\it et~al.} 1996 (NVSS); $(5)$ Johnston {\it et~al.} 1995; $(6)$ Wright \& Otrupcek 1990 (Parkes Survey); $(7)$ Stickel {\it et~al.} 1994; $(8)$ Lonsdale {\it et~al.} 1993; $(9)$ Kellermann {\it et~al.} 1989; $(10)$ Gregory \& Condon 1991 (GB87 Survey); $(11)$ Gregory {\it et~al.} 1996 (GB6 Survey); $(12)$ Hewett \& Foltz 1997; $(13)$ Rieke \& Weymann 1995; $(14)$ Neugebauer {\it et~al.} 1987; $(15)$ Sitko {\it et~al.} 1982; $(16)$ Hyland \& Allen 1982; $(17)$ Green \& Mathur 1996; $(18)$ Wilkes {\it et~al.} 1994; $(19)$ Brinkmann {\it et~al.} 1997; $(20)$ Schartel {\it et~al.} 1996; $(21)$ Reimers {\it et~al.} 1995; $(22)$ Green {\it et~al.} 1995.}
\end{deluxetable}
\end{scriptsize}

\section {Data Reduction}

Basic reduction of the science frames, using standard IRAF\footnote{IRAF is
distributed by the National Optical Astronomical Observatories, which are
operated by AURA, Inc. under contract to the NSF.} routines,
included subtraction of the dark current
and offset level, and division by normalized flatfield frames obtained
with the same grating and wavelength settings as the object frames.  Fewer than
$1\%$ of the array's pixels were bad and these were removed.
In the near-IR, OH airglow is particularly unstable.  Thus sky subtraction
was performed by interactively scaling the background images before subtracting.
The background is merely the adjacent frames with the QSO at another position
along the spectrometer slit.
The individual frames were aligned and their relative weights were calculated
interactively.  With this information the frames were combined into a single,
averaged frame.  Once combined, the residual OH emission, usually from the
three strong Q-branch features, was removed by subtracting a background residual
frame produced by median combining the $100$ columns centered on the slit in the
object frame.  One dimensional sky-subtracted spectra were extracted from the
final averaged frames using a spatial width defined by $\sim 90\%$
FWFM (full-width at full-maximum)
of the flux
centered on the nucleus.  Each spectrum was divided by the reduced spectrum of
its corresponding standard to remove atmospheric absorption features.  We
chose spectral type A0V -- G3V standards, with close proximity to the sample
QSOs, from the Bright Star Catalogue (\cite{hoffleit82}).  The
spectra were then wavelength calibrated using atmospheric OH airglow lines and
rebinned to a linear dispersion of $\sim 11.5$~\AA \ per pixel.  
For objects that were
observed on multiple nights, a single combined spectrum was produced from a
S/N weighted average of the reduced spectra from separate
observations.  The completely reduced spectra of the $32$
QSOs are shown, along with their best fit model (see $\S3.2$), 
in each top panel of Figure 1.  The horizontal axis is the rest-frame
wavelength (in \AA), and the verticle axis is an arbitrary flux level that is
normalized to
the square of the S/N.  We did not attempt to calibrate the
spectra to an absolute flux scale since the 
narrowness of the slit did not provide a sufficient enough
aperture for collecting the entire nuclear flux from each QSO.

\subsection {Systemic Redshift Determination}
To measure the systemic redshift $z_{\rm sys}$
of each QSO, the spectra were boxcar smoothed by $3$ pixels and 
the upper
half of the \ox3\ line was fit by a Gaussian, except in the case of
six objects (Q0842+345, Q1011+091, Q1158-187, Q1225+317, Q1246-057, 
and Q1309-056).  Due to the low $[$\ion{O}{3}$]$ EW and
the poor S/N of these spectra, the peak or centroid of the
\ox3\ line was used, except for Q1246-057 which had no \oiii\
emission and thus the peak of the \hb\ line was used.  Both Q1225+317 and
Q1246-057 have strong, blended \fe2\ emission and the location of minima of this
emission blueward of \hb\ ($\lambda_{\rm min} \sim 4750$~\AA) and redward of 
\oiii\ ($\lambda_{\rm min} \sim~5090$~\AA)
gave an approximate confirmation of the redshifts for these objects.
Finally, the spectra were moved to the rest-frame by dividing their linear dispersion by
$(1+z_{\rm sys})$ and then transformed to a linear logarithmic wavelength
scale with a dispersion of $3.5$~\AA \ per pixel
necessary for compatibility with the fitting routine.  The measured systemic
redshifts are presented in column (5) of Table~1.

\subsection {Fitting the Spectra with a Multi-component Model}

The spectral features around the \hb\ and \odoublet\
complex, as well as the continuum level, are complicated by the presence of many
blended \fe2\ emission lines.  In particular, there are two broad features at
$\lambda\lambda4450-4700$ and $\lambda\lambda5150-5350$, and a
strong multiplet (42)
of three lines at $\lambda\lambda4924, 5018, 5169$.  As BG92 showed, 
it is
necessary to subtract this \fe2\ emission from the spectra before accurate
measurements of the \hb\ and \ox3\ lines can be obtained, and before the
continuum can be defined.
Instead of subtracting a variety of different width
and strength \fe2\ templates, looking for the best fit ``by eye'', and then
removing the continuum  by fitting a low order polynomial to the regions where
there was \fe2\ contamination (BG92), we developed a least squares
$\chi^2$ minimization routine that simultaneously fit for line strength,
width and continuum slope, thereby producing the 
best fit model spectrum for each QSO.  This application was particularly
favorable given the low S/N of our data set.

Our spectral model was the composite of several templates.  First was an 
object-specific
\fe2\ emission template (originally from I Zw 1, courtesy BG92), broadened 
by a Gaussian of the
same full-width at half-maximum (FWHM) as observed for the \ion{C}{4}$\lambda1549$
broad emission line taken from the 
literature.  Due to the high redshift of our sample of QSOs, the 
\ion{C}{4}$\lambda1549$
emission line was the observed permitted broad line with the 
least amount of contamination by other species.  Thus, we tacitly assumed that the
broad component of both \civ\ and \fe2\ were emitted from roughly the same
region and therefore have the same line
width.  We assumed, as did BG92, that the relative strengths of the \fe2\ lines,
within
each multiplet and between multiplets, were the same for all QSOs.
For the two cases, Q1416+091 and Q1704+710, where the \civ\ line
width was not available, we used the published 
\ion{C}{3}$]\lambda1909$ line width with the 
same basic assumptions instead.  For one of the seven BALQSOs (Q0226-104), we
used a \civ\ FWHM defined by
the product of doubling the red HWHM (half-width at half-maximum)
to account for possible biasing by the broad absorption trough.  The other six
BALQSOs were all found to have absorption troughs {\it detached} from the 
emission line.

The \fe2\
template was combined with sets of templates representing hydrogen Balmer 
and
\oiii\ emission, as well as a set of $3$ orthonormal vectors (flat,
positively sloped, and negatively sloped) that 
reproduced the
sloped continuum approximated by a first order power-law.  There 
were $5$ templates for hydrogen
Balmer emission, each one a spectrum of two Gaussian line profiles, 
H$\gamma\lambda4340$ and
H$\beta\lambda4861$ with the ratio of H$\beta =1.70\times $H$\gamma$
from the composite QSO
spectrum of Francis~{\it et al.} (1991). 
The \oiii\ emission was represented as $5$
templates as well, each one a spectrum of two Gaussian lines, $\lambda5007$ and 
$\lambda4959$, with a relative strength ratio of $3$ to $1$ given by the ratio
of the two transition probabilities from the $^{1}D$ level (\cite{osterb89}).  
In AGN, narrow
forbidden lines (such as \oiii\ ) are generally found to have widths of several
hundred to a few thousand~\kms\ FWHM, and broad permitted
lines (such as H${\beta}$) are seen with widths up to $10^4$~\kms\
FWHM (see for instance \cite{peterson97}).
The minimum width used for the hydrogen Balmer and the \oiii\ 
template sets was
equal to FSPEC's instrumental line-width of $550$~\kms\ FWHM.  From initial test
fitting of our sample we found that a maximum template width for 
$[$\ion{O}{3}$]$ of 
$2580$~\kms\ FWHM was necessary to fit the broadest forbidden lines (those of
Q1158-187).
The three remaining $[$\ion{O}{3}$]$ templates have widths intermediate between 
the 
minimum and maximum values by factors of $\case{1}{\sqrt{2}}$ of the
maximum ($912$, $1290$ and $1820$~\kms\ FWHM).  
For the hydrogen Balmer broad lines, we chose
a maximum template width of $10320$~\kms\ FWHM, which is four times the maximum
narrow line width, as well as nearly equal to the broadest \civ\ line width from
the literature ($10350$~\kms\ for Q1225+317).  The remaining Balmer 
templates have
intermediate widths that are factors of $\case{1}{2}$ below the maximum
($1290$, $2580$ and $5160$~\kms\ FWHM).
We did not include the $[$\ion{O}{3}$]\lambda4363$ line, nor 
the \ion{He}{2}$\lambda4686$ line,
in our templates.  These lines fall within the
spectral range of our sample but were too weak to detect given the low
S/N of the observations.

A further complication in constructing Balmer template sets for each QSO in our
sample was the nonzero difference between $z_{\rm sys}$ determined by the \ox3\ 
line and the redshift from the \hb\ line.  This effect
is of obvious importance --- for example, determining the 
center-of-mass rest-frame of QSOs and of their host galaxies is 
crucial in estimating the
inter-galactic radiation field via the Proximity Effect.  The redshift
difference between the NLR and the BELR in this sample will be addressed
in a separate paper (\cite{mcintosh99}).  To correct
for this effect in our fitting routine,
we measured the individual \hb\ line
centers by direct examination of each boxcar
smoothed, rest-frame spectrum, then we used this value to shift each object's
set of $5$ Balmer templates to match its measured line center.

A total of $14$ component templates ($1$ \ion{Fe}{2}, $5$ $[$\ion{O}{3}$]$, 
$5$ Balmer 
and $3$ power-law)
make up our model spectrum.
To have complete spectral coverage of our entire sample, 
each template spanned the range from $4196$ -- $5855$~\AA, except for the
\fe2\ template which began at $4250$~\AA.  Furthermore, each template was
linearly rebinned to a logarithmic wavelength scale
($3.5$~\AA \ per pixel dispersion) over a total of
$475$ pixels.

Construction of each best fit model spectrum was done by simultaneously
fitting, in pixel space, a linear combination of the minimum number of template
spectra to the object spectrum.  To do this we adapted a non-negative
least-squares routine (see \cite{rix95}) that minimizes the
goodness-of-fit parameter 
\begin{equation}
\chi^2 = \sum_{i=1}^{N} \left[ \frac{1}{\sigma(x_i)} \left(y(x_i) - \sum_{k=1}^{m} a_k T_k(x_i) \right) \right]^2 .
\end{equation}
First the set of $m=14$ templates $T_k$ were trimmed to match the wavelength
range (same initial pixel $x_1$ and number $N$ of pixels) for each individual
object spectrum $y$.  The object and templates were
weighted by the object spectrum noise $\sigma(x_i) \simeq \sqrt{y(x_i)}$
(Poisson statistics).  Then an iterative process multiplied the
templates by scalar coefficients $a_k$, calculated the linear combination
(composite model) and compared it to the input QSO spectrum until the
optimum fit was obtained.  The optimum fit was such that the reduced
chi-squared
\begin{equation}
\chi_{\nu}^2 = \frac{\chi^2}{N_{\rm d.o.f.}} \approx 1 ,
\end{equation}
where the number of degrees of freedom is $N_{\rm d.o.f.} = N - m^{\prime}$, the
total
number of pixels minus the number of
templates used (those with nonzero coefficients $a_k$).

The final composite model spectrum was comprised of
four components - broad \hb\ emission, narrow \oiii\
emission, broad \fe2\ emission and a sloped continuum.
For the \hb\ and \oiii\ components, a
linear combination of Gaussians at equivalent line centers, over a range
of widths, resulted in a
symmetric, pseudo-Lorentzian line profile for \hb\ and a blended, double
line profile for $[$\ion{O}{3}$]\lambda\lambda4959,5007$.  
The summation of many blended broad lines
gave the \fe2\ component the look of a high order polynomial continuum.  The
sloped continuum was a superposition of the flat continuum template and one of
the two sloped, either negative or positive, templates.
The best fit model and its individual components for each QSO 
are plotted in the bottom panels of Figure 1.  The parameters used
to determine each best fit model are presented in Table~3.  Column (2) lists
the \fe2\ template FWHM in \kms\ .  Column (3) gives the Balmer
template \hb\ line center in \AA.  The number $m^{\prime}$ of templates 
used to construct the model spectrum is given in column (4).  The final
best fit \redchisq\ and the number of degrees of freedom $N_{\rm d.o.f.}$ 
are tabulated in column (5).  Column (6) cites the reference(s)
of the \ion{C}{4}$\lambda1549$ line width used to broaden each \fe2\ template.
 
\clearpage
\begin{scriptsize}
\begin{deluxetable}{lccccc}
\tablewidth{0pt}
\tablenum{3}
\tablecaption{Best~Fit~Template~Specifications}
\tablehead{\multicolumn{1}{l}{QSO} &
\colhead{\ion{Fe}{2}~FWHM} & \colhead{Balmer~Center} & \colhead{$m^{\prime}$\tablenotemark{a}} & \colhead{$\chi_{\nu}^2$~($N_{\rm d.o.f.}$)} & \colhead{Refs.\tablenotemark{b,c}
}\\
\colhead{$$} & \colhead{(km~s$^{-1}$)} & \colhead{(\AA)} & \colhead{$$} & \colhead{$$} & \colhead{$
$}}
\startdata
Q0043+008 & $3000$ & $4870$ & $10$ & $1.005(232)$ & $1$ \nl
Q0049+007 & $8300$ & $4864$ & $8$ & $1.008(247)$ & $2$\tablenotemark{d} \nl
Q0049+014 & $8200$ & $4865$ & $9$ & $0.999(226)$ & $2$\tablenotemark{d} \nl
Q0109+022 & $5260$ & $4859$ & $7$ & $1.001(217)$ & $3$ \nl
Q0123+257 & $4700$ & $4865$ & $11$ & $1.011(238)$ & $1$ \nl
Q0153+744 & $5250$ & $4872$ & $9$ & $1.001(251)$ & $4$ \nl
Q0226-104 & $6772$ & $4869$ & $9$ & $1.000(226)$ & $5$\tablenotemark{e} \nl
Q0226-038 & $4500$ & $4864$ & $10$ & $0.992(259)$ & $1,6,7,8$ \nl
Q0421+019 & $5100$ & $4868$ & $10$ & $1.003(254)$ & $9$ \nl
Q0424-131 & $2820$ & $4865$ & $11$ & $1.000(251)$ & $7,8$ \nl
Q0552+398 & $3000$ & $4870$ & $7$ & $1.003(222)$ & $2$\tablenotemark{d} \nl
Q0836+710 & $6740$ & $4858$ & $8$ & $1.001(250)$ & $4$ \nl
Q0842+345 & $6800$ & $4862$ & $9$ & $0.998(266)$ & $10$\tablenotemark{d} \nl
Q1011+091 & $10000$ & $4851$ & $8$ & $1.007(261)$ & $1$ \nl
Q1104-181 & $6400$ & $4870$ & $9$ & $0.996(197)$ & $11$\tablenotemark{d} \nl
Q1148-001 & $2740$ & $4861$ & $9$ & $1.000(248)$ & $1,7,8$ \nl
Q1158-187 & $4045$ & $4847$ & $8$ & $0.999(213)$ & $12$ \nl
Q1222+228 & $4840$ & $4857$ & $10$ & $0.998(252)$ & $1,8,13$\tablenotemark{d} \nl
Q1225+317 & $10350$ & $4864$ & $7$ & $1.000(258)$ & $1,8$ \nl
Q1228+077 & $4000$ & $4861$ & $8$ & $1.003(235)$ & $1$ \nl
Q1246-057 & $5000$ & $4861$ & $9$ & $1.000(253)$ & $1$ \nl
Q1247+267 & $3610$ & $4875$ & $10$ & $1.001(250)$ & $1,8,13$\tablenotemark{d} \nl
Q1309-056 & $5000$ & $4871$ & $9$ & $1.004(250)$ & $1$ \nl
Q1331+170 & $5210$ & $4861$ & $7$ & $0.999(262)$ & $1,8,14$ \nl
Q1346-036 & $10000$ & $4874$ & $9$ & $1.000(230)$ & $1$ \nl
Q1416+091 & $7500$\tablenotemark{f} & $4867$ & $8$ & $0.997(250)$ & $7$ \nl
Q1435+638 & $4200$ & $4872$ & $9$ & $1.001(267)$ & $7,8$ \nl
Q1448-232 & $3610$ & $4874$ & $10$ & $0.997(267)$ & $12$ \nl
Q1704+710 & $6000$\tablenotemark{f} & $4871$ & $10$ & $1.013(254)$ & $7$ \nl
Q2212-179 & $3000$ & $4863$ & $8$ & $0.998(256)$ & $1$ \nl
Q2251+244 & $6600$ & $4855$ & $9$ & $1.000(197)$ & $15$ \nl
Q2310+385 & $4300$ & $4875$ & $9$ & $0.990(251)$ & $16$\tablenotemark{d} \nl
\enddata
\tablenotetext{a}{The number of non-zero weighted templates, out of a total possible 14, used to achieve the best fit model spectrum.}
\tablenotetext{b}{The \ion{C}{4}$\lambda1549$ line width references.}
\tablenotetext{c}{Multiple references indicate the use of an average line width.}
\tablenotetext{d}{The \ion{C}{4}$\lambda1549$ line width was measured directly off the published spectrum.}
\tablenotetext{e}{FWHM of \ion{C}{4} line derived from red HWHM.}
\tablenotetext{f}{Used the \ion{C}{3}$]\lambda1909$ line width since no \ion{C}{4}$\lambda1549$ data was available.}
\tablerefs{$(1)$ Turnshek 1984; $(2)$ Wolfe {\it et al.} 1986; $(3)$ Schneider {\it et al.} 1994; $(4)$ Lawrence {\it et al.} 1996; $(5)$ WMFH91; $(6)$ Tytler \& Fan 1992; $(7)$ Brotherton {\it et al.} 1994a; $(8)$ Wills {\it et al.} 1993; $(9)$ Baldwin {\it et al.} 1989; $(10)$ Thompson {\it et al.} 1989; $(11)$ Wisotzki {\it et al.} 1993; $(12)$ Ulrich 1989; $(13)$ Sargent {\it et al.} 1988; $(14)$ Corbin 1992; $(15)$ Corbin 1991; $(16)$ Wills 1997.}
\end{deluxetable}
\end{scriptsize}
\clearpage

\subsection {Deriving Emission Line Properties from the Model}

We derived emission line properties directly from the individual components
of each composite model spectrum. 
Since the model represented the best fit to the
rest-frame spectrum, all calculated emission line properties refer to
the rest-frame.
The EWs in this paper were measured relative to the fitted continuum component 
represented by the sloped line in the bottom panels of Figure 1.
From the \hb\ and \oiii\ components, the
strength (EW in \AA) and width (FWHM in \kms\ ) of
the \hb\ and \ox3\ lines were calculated.
In the case of the blended \odoublet\
emission, the EW of the \ox3\ line was equal to $\case{3}{4}$
the total EW of \oiii\ based on the known line ratio.  In a few cases a small
negative correction, arising from Brackett absorption lines in the telluric
standard, was applied to the $[$\ion{O}{3}$]\lambda5007$ EW.  In no case
was this correction $>3.5$ \AA.
For both \hb\ and \ox3\, the symmetry of the line profiles allowed the 
extrapolation of both the strength and the width of a line by calculating 
either the redward or
blueward half and doubling it.  This method was used for those QSOs with an
\hb\ line too close to the blue edge of the spectrum, or conversely for those
with \ox3\ too
close to the red edge.  The FWHM of \hb\ measured from the composite model  
represented the {\it total}, narrow plus broad components,
line width (H$\beta_{{\rm tot}}$ FWHM).  For five objects (Q0049+007, Q1011+091,
Q1158-187, Q1704+710 and Q2310+385) this
quantity may have been underestimated.  In these cases the line profile of
the \hb\ component had a narrow spike on top of a broad hump due to the
fitting routine possibly fitting noise with the narrowest Balmer template.
A {\it relative} strength of the
\fe2\ emission was determined by measuring the EW 
of the blended line complex between $4810$ and $5090$~\AA.  This spectral 
range was chosen since it encompasses the \hb\ and \odoublet\ emission features.
The width of the {\it broad} component of the \hb\ line 
(H$\beta_{{\rm broad}}$ FWHM) 
was derived by fitting a single Gaussian to the component of the line profile
defined by the 
relative flux $\leq \case{3}{4}$ of the maximum.  
We separated the
actual \hb\ emission by subtracting the other three
best fit model components ($[$\ion{O}{3}$]\lambda5007$, \fe2\ and continuum) 
from the original
spectrum.  Then we used the Levenberg-Marquardt nonlinear \chisq\
minimization method (see \cite{press}) to produce the fit and calculate 
the broad component FWHM as well as its line center.
As an example, the reduced rest-frame spectra with extracted
\hb\ lines, for a RLQ (Q0424-131), a RQQ (Q1346-036) and a 
BALQSO (Q1246-057), are shown in Figure 2.

The measured emission line
properties of each object are presented in Table~4.  Columns (2)~thru~(4) list
the EWs of the $[$\ion{O}{3}$]\lambda5007$, \hb\ and {\it relative} \fe2\ emission.  
Based on studies of broad line asymmetries (see for instance \cite{corbin95}),
we may  have systematically overestimated the strength of the \ox3\ line by 
using a {\it symmetric} \hb\ line profile component; however, the low
S/N and resolution of our spectra did not justify the use of more
complicated model line profiles.
The FWHM of the forbidden \ox3\ narrow line is given in column (5).  This
line width was not tabulated for the three objects 
(Q1225+317, Q1246-057 and Q1309-056) with no detectable 
(above noise) \ox3\ emission.  Column
(6) lists the total FWHM of H${\beta}$, while column (7) gives the FWHM of the
broad component of the \hb\ line.  All three line widths are the intrinsic
widths with the instrumental resolution ($550$~km~s$^{-1}$) removed.
Columns (8) and (9) give the ratios
of the EWs of \ox3\ and \fe2\ to that of H${\beta}$.
Column (10) lists the ratio of the
peak flux of the \ox3\ line to that of the \hb\ line, the quantity
``Peak$\lambda5007$'' from BG92.  \peak\ is not as
quantitative as EWs or line ratios, but we include it
to compare with the low redshift data.  QSO 1148-001 has only
\ox3\ EW and FWHM measurements since the \hb\ line fell
outside of the spectral coverage of the detector.

To establish the confidence of the emission line property measurements, a
Monte Carlo analysis was performed to calculate the errors associated with
all quantities tabulated in Table~4 except for those in column (7).  The error  
affiliated with H$\beta_{{\rm broad}}$ FWHM was
derived from the estimated covariance matrix 
in the Levenberg-Marquardt fitting routine and was about $8\%$ on
average.  For the 
remaining eight properties, the best fit model spectrum of each object was used
to create $n=250$ synthetic spectra.  We added random Poisson noise, at a gain
of unity, 
to each normalized composite model in order to produce an artificial 
spectrum with similar
S/N as found in the original data.
Then each synthetic spectrum 
was fitted with
our fitting routine, the optimal fit reduced chi-squared \redchisq\ was
found, and the emission line properties $p_i(n)$ (such that $i=$ 
emission line property type) were
calculated.  For each synthetic spectrum, the eight parameters plus the
\redchisq\ were
tabulated in separate Monte Carlo distributions.
Each distribution
was sorted and the central $68\%$ of its values ($p_i(40)$ to $p_i(210)$) 
were determined.  We
assumed that the most likely value of each distribution roughly corresponded 
to the measured result ($p_i(result)$) tabulated in Table~4.  Therefore, the
left ($-$) and right ($+$)
sigmas for each emission line property were found by:
\begin{equation}
\sigma_i(-) = p_i({\rm result}) - p_i(40)
\end{equation}
and
\begin{equation}
\sigma_i(+) = p_i(210) - p_i({\rm result}) .
\end{equation}
Figure 3 gives a representative example of the nine Monte Carlo 
distributions, for PG1247+267.  A solid
dot representing the parameter's measured value is plotted above each
distribution, as are its left and right sigmas.
The uncertaintities associated with each of the eight emission 
line properties are
tabulated along with the measured result in Table~4.  
In general, the Monte Carlo
distributions were approximately Gaussian, and on average the mean $1\sigma$ 
for each property was roughly $23\%$ of its measured value.

\clearpage
\begin{scriptsize}
\begin{deluxetable}{lccccccccc}
\tablewidth{0pt}
\tablenum{4}
\tablecaption{Measured~Emission~Line~Properties}
\tablehead{
\multicolumn{1}{l}{QSO} &
\multicolumn{3}{c}{Equivalent Widths} &
\multicolumn{3}{c}{FWHM} &
\colhead{} &
\colhead{} &
\colhead{} \nl
\colhead{} &
\colhead{$\lambda5007$} &
\colhead{H$\beta$} &
\colhead{\ion{Fe}{2}\tablenotemark{a}} &
\colhead{$\lambda5007$} &
\colhead{H$\beta_{\rm total}$\tablenotemark{b}} &
\colhead{H$\beta_{\rm broad}$} &
\colhead{$[$\ion{O}{3}$]$/H$\beta$} &
\colhead{\ion{Fe}{2}/H$\beta$} &
\colhead{Peak $\lambda5007$} \nl
\colhead{} &
\colhead{(\AA)} &
\colhead{(\AA)} &
\colhead{(\AA)} &
\colhead{(km ${\rm s}^{-1}$)} &
\colhead{(km ${\rm s}^{-1}$)} &
\colhead{(km ${\rm s}^{-1}$)} &
\colhead{} &
\colhead{} &
\colhead{}}
\startdata
Q0043+008 & $16.0^{+2.4}_{-1.9}$ & $108.9^{+4.8}_{-11.5}$ & $28.9^{+4.3}_{-7.3}$ & $910^{+190}_{-300}$ & $4330^{+450}_{-660}$ & $9950^{+560}_{-560}$ & $0.15^{+0.03}_{-0.02}$ & $0.27^{+0.04}_{-0.05}$ & $0.66^{+0.11}_{-0.07}$ \nl
\tablevspace{.6ex}
Q0049+007 & $4.0^{+3.4}_{-1.1}$ & $94.6^{+5.6}_{-13.3}$ & $64.4^{+5.1}_{-12.2}$ & $480^{+470}_{-210}$ & $4160^{+1040}_{-1720}$ & $13440^{+1040}_{-1040}$ & $0.04^{+0.04}_{-0.01}$ & $0.68^{+0.06}_{-0.08}$ & $0.48^{+0.14}_{-0.13}$ \nl
\tablevspace{.6ex}
Q0049+014 & $5.0^{+1.3}_{-0.7}$ & $55.4^{+4.8}_{-5.6}$ & $18.9^{+5.0}_{-5.3}$ & $570^{+270}_{-280}$ & $5890^{+960}_{-1410}$ & $13660^{+950}_{-950}$ & $0.09^{+0.03}_{-0.01}$ & $0.34^{+0.07}_{-0.08}$ & $0.69^{+0.17}_{-0.15}$ \nl
\tablevspace{.6ex}
Q0109+022 & $26.2^{+3.0}_{-1.5}$ & $32.5^{+6.4}_{-4.6}$ & $16.7^{+5.5}_{-5.2}$ & $1390^{+60}_{-200}$ & $7020^{+940}_{-2120}$ & $9560^{+1180}_{-1180}$ & $0.81^{+0.17}_{-0.14}$ & $0.51^{+0.11}_{-0.15}$ & $4.52^{+0.36}_{-0.96}$ \nl
\tablevspace{.6ex}
Q0123+257 & $28.1^{+4.4}_{-2.3}$ & $55.5^{+8.3}_{-9.0}$ & $21.5^{+7.2}_{-8.0}$ & $440^{+160}_{-110}$ & $1160^{+380}_{-390}$ & $5020^{+580}_{-580}$ & $0.51^{+0.12}_{-0.08}$ & $0.39^{+0.10}_{-0.13}$ & $1.34^{+0.21}_{-0.18}$ \nl
\tablevspace{.6ex}
Q0153+744 & $19.5^{+1.5}_{-1.5}$ & $74.4^{+5.7}_{-5.1}$ & $17.8^{+4.0}_{-3.8}$ & $1190^{+190}_{-220}$ & $5650^{+410}_{-690}$ & $11470^{+570}_{-570}$ & $0.26^{+0.03}_{-0.03}$ & $0.24^{+0.04}_{-0.04}$ & $1.14^{+0.09}_{-0.12}$ \nl
\tablevspace{.6ex}
Q0226-104 & $2.3^{+1.4}_{-0.8}$ & $77.9^{+6.3}_{-4.9}$ & $40.8^{+6.5}_{-5.8}$ & $1670^{+410}_{-780}$ & $5490^{+450}_{-950}$ & $12850^{+660}_{-660}$ & $0.03^{+0.02}_{-0.01}$ & $0.52^{+0.06}_{-0.06}$ & $0.25^{+0.08}_{-0.05}$ \nl
\tablevspace{.6ex}
Q0226-038 & $31.9^{+4.4}_{-2.3}$ & $58.3^{+19.9}_{-8.2}$ & $20.1^{+9.2}_{-6.9}$ & $1120^{+220}_{-290}$ & $2780^{+840}_{-1130}$ & $9130^{+1180}_{-1180}$ & $0.55^{+0.10}_{-0.12}$ & $0.34^{+0.11}_{-0.12}$ & $1.66^{+0.21}_{-0.36}$ \nl
\tablevspace{.6ex}
Q0421+019 & $51.1^{+2.3}_{-2.0}$ & $130.3^{+7.9}_{-11.3}$ & $31.6^{+4.3}_{-5.4}$ & $1370^{+190}_{-190}$ & $4660^{+710}_{-800}$ & $14310^{+650}_{-650}$ & $0.39^{+0.03}_{-0.02}$ & $0.24^{+0.03}_{-0.03}$ & $1.49^{+0.14}_{-0.12}$ \nl
\tablevspace{.6ex}
Q0424-131 & $29.8^{+2.3}_{-1.6}$ & $80.3^{+11.3}_{-3.2}$ & $9.9^{+4.4}_{-4.0}$ & $1190^{+90}_{-190}$ & $4380^{+510}_{-600}$ & $9560^{+570}_{-570}$ & $0.37^{+0.03}_{-0.06}$ & $0.12^{+0.04}_{-0.05}$ & $1.44^{+0.16}_{-0.14}$ \nl
\tablevspace{.6ex}
Q0552+398 & $17.2^{+2.7}_{-2.1}$ & $35.5^{+6.0}_{-4.7}$ & $2.7^{+4.1}_{-2.7}$ & $1470^{+230}_{-380}$ & $2730^{+150}_{-540}$ & $4700^{+570}_{-570}$ & $0.48^{+0.09}_{-0.08}$ & $0.07^{+0.10}_{-0.07}$ & $0.94^{+0.15}_{-0.16}$ \nl
\tablevspace{.6ex}
Q0836+710 & $1.1^{+1.1}_{-0.6}$ & $40.2^{+3.3}_{-4.4}$ & $28.2^{+3.4}_{-3.9}$ & $1270^{+190}_{-620}$ & $3410^{+1000}_{-930}$ & $13070^{+1000}_{-1000}$ & $0.03^{+0.04}_{-0.01}$ & $0.70^{+0.07}_{-0.07}$ & $0.42^{+0.14}_{-0.09}$ \nl
\tablevspace{.6ex}
Q0842+345 & $6.4^{+1.5}_{-1.3}$ & $56.6^{+3.7}_{-5.7}$ & $48.6^{+5.3}_{-5.8}$ & $1270^{+880}_{-520}$ & $8540^{+620}_{-2590}$ & $15350^{+1440}_{-1440}$ & $0.11^{+0.03}_{-0.02}$ & $0.86^{+0.10}_{-0.07}$ & $0.55^{+0.14}_{-0.14}$ \nl
\tablevspace{.6ex}
Q1011+091 & $6.2^{+3.0}_{-1.2}$ & $90.3^{+13.4}_{-13.6}$ & $77.1^{+16.7}_{-15.4}$ & $930^{+400}_{-520}$ & $7510^{+1460}_{-2800}$ & $13500^{+1190}_{-1190}$ & $0.07^{+0.04}_{-0.01}$ & $0.85^{+0.09}_{-0.10}$ & $0.47^{+0.24}_{-0.14}$ \nl
\tablevspace{.6ex}
Q1104-181 & $14.9^{+3.1}_{-1.6}$ & $89.6^{+10.0}_{-8.5}$ & $25.7^{+7.3}_{-9.3}$ & $1420^{+280}_{-410}$ & $3950^{+460}_{-660}$ & $10490^{+720}_{-720}$ & $0.17^{+0.05}_{-0.02}$ & $0.29^{+0.07}_{-0.09}$ & $0.55^{+0.12}_{-0.09}$ \nl
\tablevspace{.6ex}
Q1148-001 & $22.9^{+2.9}_{-0.9}$ & \nodata & \nodata & $760^{+140}_{-110}$ & \nodata & \nodata & \nodata & \nodata & \nodata \nl
\tablevspace{.6ex}
Q1158-187 & $16.3^{+5.4}_{-1.5}$ & $82.7^{+25.1}_{-11.6}$ & $16.9^{+13.8}_{-8.8}$ & $2370^{+90}_{-280}$ & $3710^{+1020}_{-1160}$ & $15280^{+1080}_{-1080}$ & $0.20^{+0.09}_{-0.06}$ & $0.20^{+0.13}_{-0.14}$ & $0.59^{+0.27}_{-0.20}$ \nl
\tablevspace{.6ex}
Q1222+228 & $10.8^{+2.8}_{-3.1}$ & $83.1^{+8.3}_{-10.5}$ & $31.1^{+4.6}_{-6.1}$ & $1040^{+150}_{-540}$ & $7060^{+1740}_{-1470}$ & $13650^{+750}_{-750}$ & $0.13^{+0.04}_{-0.04}$ & $0.37^{+0.04}_{-0.07}$ & $0.65^{+0.15}_{-0.08}$ \nl
\tablevspace{.6ex}
Q1225+317 & $0.6^{+0.5}_{-0.2}$ & $42.3^{+2.9}_{-2.8}$ & $46.8^{+3.6}_{-3.3}$ & \nodata & $8300^{+420}_{-1060}$ & $13420^{+660}_{-660}$ & $0.01^{+0.01}_{-0.01}$ & $1.11^{+0.05}_{-0.05}$ & $0.21^{+0.09}_{-0.09}$ \nl
\tablevspace{.6ex}
Q1228+077 & $9.7^{+1.5}_{-1.5}$ & $46.4^{+8.0}_{-8.6}$ & $4.1^{+5.3}_{-4.5}$ & $960^{+410}_{-320}$ & $4030^{+1210}_{-970}$ & $5780^{+730}_{-730}$ & $0.21^{+0.02}_{-0.01}$ & $0.09^{+0.04}_{-0.04}$ & $0.78^{+0.13}_{-0.08}$ \nl
\tablevspace{.6ex}
Q1246-057 & $3.0^{+0.7}_{-0.7}$ & $59.6^{+2.4}_{-3.1}$ & $47.1^{+2.4}_{-2.5}$ & \nodata & $5870^{+1130}_{-680}$ & $14820^{+540}_{-540}$ & $0.05^{+0.01}_{-0.01}$ & $0.79^{+0.03}_{-0.03}$ & $0.14^{+0.07}_{-0.02}$ \nl
\tablevspace{.6ex}
Q1247+267 & $10.4^{+3.1}_{-1.3}$ & $77.6^{+11.7}_{-0.9}$ & $31.4^{+7.5}_{-4.1}$ & $940^{+240}_{-430}$ & $4210^{+450}_{-1010}$ & $7460^{+220}_{-220}$ & $0.13^{+0.05}_{-0.05}$ & $0.40^{+0.12}_{-0.09}$ & $0.56^{+0.17}_{-0.20}$ \nl
\tablevspace{.6ex}
Q1309-056 & $2.1^{+1.3}_{-0.2}$ & $55.6^{+4.2}_{-4.7}$ & $51.2^{+3.8}_{-4.3}$ & \nodata & $3220^{+540}_{-600}$ & $9620^{+570}_{-570}$ & $0.04^{+0.03}_{-0.00}$ & $0.92^{+0.07}_{-0.06}$ & $0.16^{+0.09}_{-0.05}$ \nl
\tablevspace{.6ex}
Q1331+170 & $18.6^{+1.2}_{-0.9}$ & $67.8^{+6.9}_{-5.7}$ & $19.1^{+3.3}_{-3.3}$ & $1770^{+290}_{-260}$ & $7480^{+240}_{-380}$ & $14550^{+840}_{-840}$ & $0.28^{+0.02}_{-0.02}$ & $0.28^{+0.03}_{-0.03}$ & $1.14^{+0.06}_{-0.06}$ \nl
\tablevspace{.6ex}
Q1346-036 & $2.8^{+1.5}_{-0.9}$ & $72.7^{+5.6}_{-6.2}$ & $51.3^{+3.2}_{-4.1}$ & $370^{+410}_{-370}$ & $3470^{+760}_{-1650}$ & $10250^{+380}_{-380}$ & $0.04^{+0.03}_{-0.02}$ & $0.70^{+0.05}_{-0.06}$ & $0.37^{+0.10}_{-0.20}$ \nl
\tablevspace{.6ex}
Q1416+091 & $6.1^{+1.1}_{-0.9}$ & $60.8^{+3.3}_{-4.3}$ & $11.3^{+2.4}_{-5.4}$ & $1740^{+310}_{-60}$ & $4610^{+460}_{-520}$ & $10900^{+1580}_{-1580}$ & $0.10^{+0.02}_{-0.01}$ & $0.19^{+0.02}_{-0.05}$ & $0.31^{+0.05}_{-0.07}$ \nl
\tablevspace{.6ex}
Q1435+638 & $15.2^{+2.1}_{-1.5}$ & $77.4^{+10.7}_{-9.4}$ & $26.8^{+6.9}_{-6.1}$ & $1020^{+290}_{-790}$ & $6280^{+550}_{-970}$ & $11650^{+900}_{-900}$ & $0.20^{+0.04}_{-0.03}$ & $0.35^{+0.09}_{-0.09}$ & $0.96^{+0.11}_{-0.07}$ \nl
\tablevspace{.6ex}
Q1448-232 & $19.9^{+2.1}_{-2.1}$ & $78.5^{+14.1}_{-9.2}$ & $20.8^{+7.5}_{-6.2}$ & $1670^{+300}_{-350}$ & $3230^{+550}_{-1280}$ & $7770^{+380}_{-380}$ & $0.25^{+0.04}_{-0.04}$ & $0.26^{+0.07}_{-0.06}$ & $0.59^{+0.10}_{-0.18}$ \nl
\tablevspace{.6ex}
Q1704+710 & $19.7^{+1.9}_{-1.0}$ & $95.7^{+5.1}_{-6.2}$ & $10.6^{+3.6}_{-4.0}$ & $940^{+90}_{-260}$ & $1560^{+220}_{-370}$ & $10200^{+1100}_{-1100}$ & $0.21^{+0.04}_{-0.02}$ & $0.11^{+0.03}_{-0.04}$ & $0.68^{+0.08}_{-0.06}$ \nl
\tablevspace{.6ex}
Q2212-179 & $20.5^{+5.5}_{-2.2}$ & $51.5^{+21.5}_{-17.7}$ & $13.3^{+15.2}_{-10.5}$ & $1200^{+370}_{-290}$ & $6150^{+490}_{-600}$ & $11160^{+1020}_{-1020}$ & $0.40^{+0.07}_{-0.03}$ & $0.26^{+0.12}_{-0.11}$ & $1.83^{+0.14}_{-0.12}$ \nl
\tablevspace{.6ex}
Q2251+244 & $14.7^{+1.8}_{-1.4}$ & $28.0^{+3.7}_{-6.0}$ & $34.3^{+2.2}_{-4.2}$ & $840^{+240}_{-300}$ & $4910^{+290}_{-1590}$ & $9190^{+970}_{-970}$ & $0.53^{+0.07}_{-0.04}$ & $1.23^{+0.05}_{-0.08}$ & $2.65^{+0.28}_{-0.30}$ \nl
\tablevspace{.6ex}
Q2310+385 & $28.7^{+1.3}_{-1.7}$ & $92.2^{+4.9}_{-3.3}$ & $75.2^{+5.3}_{-3.3}$ & $1130^{+140}_{-170}$ & $3050^{+850}_{-1460}$ & $8520^{+810}_{-810}$ & $0.31^{+0.08}_{-0.12}$ & $0.82^{+0.14}_{-0.14}$ & $1.17^{+0.41}_{-0.45}$ \nl
\enddata
\tablenotetext{a}{Relative EW measured over $4810-5090$\AA.}
\tablenotetext{b}{Total (narrow + broad) FWHM.}
\end{deluxetable}
\end{scriptsize}
\clearpage

\section {Correlation Analysis}
Having measured and tabulated a host of continuum and emission line properties
for this QSO sample, we proceeded to explore whether the various
parameters correlate with one another.  To this end, we calculated the Spearman 
rank-order correlation matrix, along with its significance matrix,
for the following set of properties: (i)~the
apparent $V$ magnitude and systemic redshift from Table~1; (ii)~the 
five continuum parameters from Table~2; (iii)~the
nine rest-frame optical measurements from Table~4; and (iv)~the
rest-frame UV emission FWHMs and EWs
compiled from the literature and tabulated in Tables~5 and 6, respectively.
The complete correlation
coefficient matrix is shown in Table~7.  The correlation coefficients 
$r_s$ were computed using only those objects that had both involved
values tabulated.
In addition to the
Spearman statistic, we calculated the Kendall
nonparametric ($\tau$) correlation coefficient and confidence matrices.  
 
For our total sample size of $N=32$ objects, a significant correlation at the
$\gtrsim 95\%$ confidence level corresponds to a coefficient of
$|r_s| \gtrsim 0.35$, with a negative coefficient indicating an 
anti-corrlation.  This $95\%$ significance was derived from the two-sided
probability ($=0.05$) of getting the {\it same} coefficient from an
uncorrelated
sample.  Measures of more significant correlations for the same sample size
are $|r_s| \gtrsim 0.47$ (for $\gtrsim 99\%$) and $|r_s| \gtrsim 0.57$ 
(for $\gtrsim 99.9\%$).  A set of 22 different properties results in $22 \times 
\frac{21}{2} = 231$ correlation coefficients; therefore, we would expect
$\lesssim3$ spurious events at $99\%$ confidence.  For the Spearman analysis, 
we find $34$ correlations (both positive and negative)
significant at the $\geq 99\%$ level: $9$ are
due to correlations between {\it dependent} parameters; $10$ are 
{\it degenerate}
correlations; and therefore, $15$ are {\it independent} correlations at 
the $\geq 99\%$ 
level.  The highly correlated ($r_s = +0.82$) $V$ and
$H$ apparent magnitudes is an example of a dependent correlation since $H$
was calculated directly from $V$ for most of the sample.  A correlation is
degenerate when multiple parameters have been used to quantify a single property
({\it e.g.}~\ox3\ EW, $[$\ion{O}{3}$]$/H$\beta$ ratio and \peak\ are all
measures of
\ox3\ emission strength), and these parameters {\it all} correlate with another
property, such as ${\log(R^{\prime})}$.  The $15$ independent correlations,
as well as other less significant independent relations confirmed in lower
redshift/luminosity studies, are tabulated in Table~8.  The number of
correlated pairs, plus the computed
coefficients and corresponding confidence levels for both the Spearman and
Kendall statistics, as well as the number of degenerate (at $\geq 95\%$ confidence) 
correlations are all presented for each result.  The relevance and
interpretation of the majority of these findings are discussed in detail below.

\clearpage
\begin{scriptsize}
\begin{deluxetable}{lcccc}
\tablewidth{0pt}
\tablenum{5}
\tablecaption{Rest-frame~Ultraviolet~Emission~Line~Widths}
\tablehead{
\multicolumn{1}{l}{QSO} & \colhead{\ion{C}{4}$\lambda1549$} & \colhead{Refs.\tablenotemark{a}} & \colhead{\ion{C}{3}$]\lambda1909$} & \colhead{Refs.\tablenotemark{a}}\\
\colhead{} & \colhead{(km ${\rm s}^{-1}$)} & \colhead{} & \colhead{(km ${\rm s}^{-1}$)} & \colhead{}\\}
\startdata
Q0043+008 & 3000 & 1 & 4450 & 1,2 \nl
Q0049+007 & 8300 & 3\tablenotemark{b} & 7400 & 4 \nl
Q0049+014 & 8200 & 3\tablenotemark{b} & 10810 & 5 \nl
Q0109+022 & 5260 & 5 & 8120 & 5 \nl
Q0123+257 & 4700 & 1 & 3600 & 1 \nl
Q0153+744 & 5250 & 6 & 5630 & 6 \nl
Q0226-104 & 6772 & 7\tablenotemark{c} & 9014 & 7 \nl
Q0226-038 & 4500 & 1,4,8,9 & 6750 & 1,4 \nl
Q0421+019 & 5100 & 10 & 5380 & 4,8 \nl
Q0424-131 & 2820 & 4,9 & 4650 & 1,4 \nl
Q0552+398 & 3000 & 3\tablenotemark{b} & \nodata & \nodata \nl
Q0836+710 & 6740 & 6 & 11200 & 6,11 \nl
Q0842+345 & 6800 & 12\tablenotemark{b} & \nodata & \nodata \nl
Q1011+091 & 10000 & 1 & 11000 & 1,2 \nl
Q1104-181 & 6400 & 13\tablenotemark{b} & 9900 & 13\tablenotemark{b} \nl
Q1148-001 & 2740 & 1,4,9 & 4100 & 4 \nl
Q1158-187 & 4045 & 14 & \nodata & \nodata \nl
Q1222+228 & 4840 & 1,9,15\tablenotemark{b} & 6000 & 1 \nl
Q1225+317 & 10350 & 1,16 & 12300 & 1,16 \nl
Q1228+077 & 4000 & 1 & \nodata & \nodata \nl
Q1246-057 & 5000 & 1 & 9200 & 1,2,4 \nl
Q1247+267 & 3610 & 1,9,15\tablenotemark{b} & 5500 & 1,10 \nl
Q1309-056 & 5000 & 1 & 11300 & 1,2,7 \nl
Q1331+170 & 5210 & 1,9,16 & 6500 & 1,16,17 \nl
Q1346-036 & 10000 & 1 & 11680 & 14 \nl
Q1416+091 & \nodata & \nodata & 7500 & 4 \nl
Q1435+638 & 4200 & 4,9 & 4242 & 4,18 \nl
Q1448-232 & 3610 & 14 & 4105 & 14 \nl
Q1704+710 & \nodata & \nodata & 6000 & 4 \nl
Q2212-179 & 3000 & 1 & \nodata & \nodata \nl
Q2251+244 & 6600 & 19 & 8700 & 19 \nl
Q2310+385 & 4300 & 20\tablenotemark{b} & \nodata & \nodata \nl
\enddata
\tablenotetext{a}{Multiple references indicate that the value is an average.}
\tablenotetext{b}{The line width was measured directly off the published spectrum.}
\tablenotetext{c}{FWHM derived from red HWHM.}
\tablerefs{$(1)$ Turnshek 1984; $(2)$ Hartig \& Baldwin 1986; $(3)$ Wolfe {\it et al.} 1986; $(4)$ Brotherton {\it et al.} 1994a; $(5)$ Schneider {\it et al.} 1994; $(6)$ Lawrence {\it et al.} 1996; $(7)$ WMFH91; $(8)$ Tytler \& Fan 1992; $(9)$ Wills {\it et al.} 1993; $(10)$ Baldwin {\it et al.} 1989; $(11)$ Stickel \& K\"{u}hr 1993; $(12)$ Thompson {\it et al.} 1989; $(13)$ Wisotzki {\it et al.} 1993; $(14)$ Ulrich 1989; $(15)$ Sargent {\it et al.} 1988; $(16)$ Corbin 1992; $(17)$ Carswell {\it et al.} 1991; $(18)$ Laor {\it et al.} 1995; $(19)$ Corbin 1991; $(20)$ Wills 1997.}
\end{deluxetable}
\end{scriptsize}
 
\begin{scriptsize}
\begin{deluxetable}{lcccccccc}
\tablewidth{0pt}
\tablenum{6}
\tablecaption{Rest-frame~Ultraviolet~Emission~Line~Equivalent~Widths}
\tablehead{\multicolumn{1}{l}{QSO} & \colhead{Ly$\alpha\lambda1216$} & \colhead{Refs.\tablenotemark{a}} & \colhead{\ion{C}{4}$\lambda1549$} & \colhead{Refs.\tablenotemark{a}} & \colhead{\ion{C}{3}$]\lambda1909$} & \colhead{Refs.\tablenotemark{a}} & \colhead{\ion{Mg}{2}$\lambda2798$} & \colhead{Refs.\tablenotemark{a}
}\\
\colhead{} & \colhead{(\AA)} & \colhead{} & \colhead{(\AA)} & \colhead{} & \colhead{(\AA)} & \colhead{} & \colhead{(\AA)} & \colhead{
}}
\startdata
Q0043+008 & 98.7 & 1 & 7.0 & 1,2\tablenotemark{b} & 15.0 & 2,3 & 52.0 & 3 \nl
Q0049+007 & 100.0 & 4\tablenotemark{c} & 29.0 & 4\tablenotemark{c} & 20.7 & 5 & \nodata & \nodata \nl
Q0049+014 & 92.0 & 4\tablenotemark{c} & 20.0 & 6,4\tablenotemark{c} & 17.0 & 6 & \nodata & \nodata \nl
Q0109+022 & \nodata & \nodata & 24.0 & 6 & 18.0 & 6 & \nodata & \nodata \nl
Q0123+257 & 83.0 & 4\tablenotemark{c} & \nodata & \nodata & \nodata & \nodata & \nodata & \nodata \nl
Q0153+744 & 48.7 & 7 & 38.0 & 7 & 26.3 & 7 & 17.2 & 7 \nl
Q0226-104 & \nodata & \nodata & 14.8 & 2\tablenotemark{b} & 24.6 & 2 & \nodata & \nodata \nl
Q0226-038 & 56.2 & 1,8 & 26.7 & 1,5,8,9 & 13.5 & 5,8 & 30.4 & 5 \nl
Q0421+019 & 42.2 & 1,8 & 16.6 & 1,8 & 15.2 & 5 & 27.5 & 5 \nl
Q0424-131 & 84.8 & 1 & 39.4 & 1,5,9 & 19.5 & 5 & 18.2 & 5 \nl
Q0552+398 & \nodata & \nodata & 6.0 & 4\tablenotemark{c} & \nodata & \nodata & \nodata & \nodata \nl
Q0836+710 & \nodata & \nodata & 10.5 & 10 & 20.1 & 10 & \nodata & \nodata \nl
Q0842+345 & \nodata & \nodata & 15.0 & 11\tablenotemark{c} & \nodata & \nodata & \nodata & \nodata \nl
Q1011+091 & \nodata & \nodata & 15.8 & 2\tablenotemark{b} & 12.0 & 2,3 & 18.0 & 3 \nl
Q1104-181 & \nodata & \nodata & 10.4 & 12 & 15.7 & 12 & \nodata & \nodata \nl
Q1148-001 & 112.0 & 8 & 27.4 & 5,8,9 & 10.4 & 5,8 & 31.1 & 5 \nl
Q1158-187 & \nodata & \nodata & 31.0 & 13 & \nodata & \nodata & \nodata & \nodata \nl
Q1222+228 & 78.0 & 8 & 25.7 & 8,9 & \nodata & \nodata & \nodata & \nodata \nl
Q1225+317 & \nodata & \nodata & 6.9 & 14 & \nodata & \nodata & \nodata & \nodata \nl
Q1228+077 & \nodata & \nodata & \nodata & \nodata & \nodata & \nodata & \nodata & \nodata \nl
Q1246-057 & \nodata & \nodata & 9.6 & 2\tablenotemark{b} & 17.9 & 2,3,5 & 29.0 & 3 \nl
Q1247+267 & 86.5 & 8 & 29.8 & 8,9 & 19.7 & 8 & \nodata & \nodata \nl
Q1309-056 & \nodata & \nodata & 16.2 & 2\tablenotemark{b} & 18.4 & 2,3 & 20.0 & 3 \nl
Q1331+170 & 63.0 & 14 & 23.9 & 15,14,9 & 25.0 & 15,14 & 16.0 & 15 \nl
Q1346-036 & 140.0 & 4\tablenotemark{c} & 32.3 & 4\tablenotemark{c},13 & 16.7 & 13 & \nodata & \nodata \nl
Q1416+091 & \nodata & \nodata & \nodata & \nodata & 18.2 & 5 & 28.2 & 5 \nl
Q1435+638 & 83.3 & 14 & 34.0 & 5,14,9 & 10.1 & 5,14,16 & 21.6 & 5,16 \nl
Q1448-232 & 105.0 & 14 & 21.2 & 14,13 & 16.7 & 14,13 & \nodata & \nodata \nl
Q1704+710 & \nodata & \nodata & \nodata & \nodata & 18.7 & 5 & 27.3 & 5 \nl
Q2212-179 & \nodata & \nodata & \nodata & \nodata & \nodata & \nodata & \nodata & \nodata \nl
Q2251+244 & 6.0 & 17 & 5.0 & 17 & 22.0 & 17 & \nodata & \nodata \nl
Q2310+385 & 110.0 & 18\tablenotemark{c} & 38.0 & 18\tablenotemark{c} & \nodata & \nodata & \nodata & \nodata \nl
\enddata
\tablenotetext{a}{Multiple references indicate that the value is an average.}
\tablenotetext{b}{Full EW derived from red half of emission line.}
\tablenotetext{c}{The line width was measured directly off the published spectrum.}
\tablerefs{$(1)$ Osmer {\it et al.} 1994; $(2)$ WMFH91; $(3)$ Hartig \& Baldwin 1986; $(4)$ Wolfe {\it et al.} 1986; $(5)$ Steidel \& Sargent 1991; $(6)$ Schneider {\it et al.} 1994; $(7)$ Lawrence {\it et al.} 1996; $(8)$ Baldwin {\it et al.} 1989; $(9)$ Wills {\it et al.} 1993; $(10)$ Stickel \& K\"{u}hr 1993; $(11)$ Thompson {\it al.} 1989; $(12)$ Wisotzki {\it et al.} 1993; $(13)$ Ulrich 1989;$(14)$ Corbin 1992; $(15)$ Carswell {\it et al.} 1991; $(16)$ Laor {\it et al.} 1995; $(17)$ Corbin 1991; $(18)$ Wills 1997.}
\end{deluxetable}
\end{scriptsize}

\begin{figure}[p]
\centering
\plotone{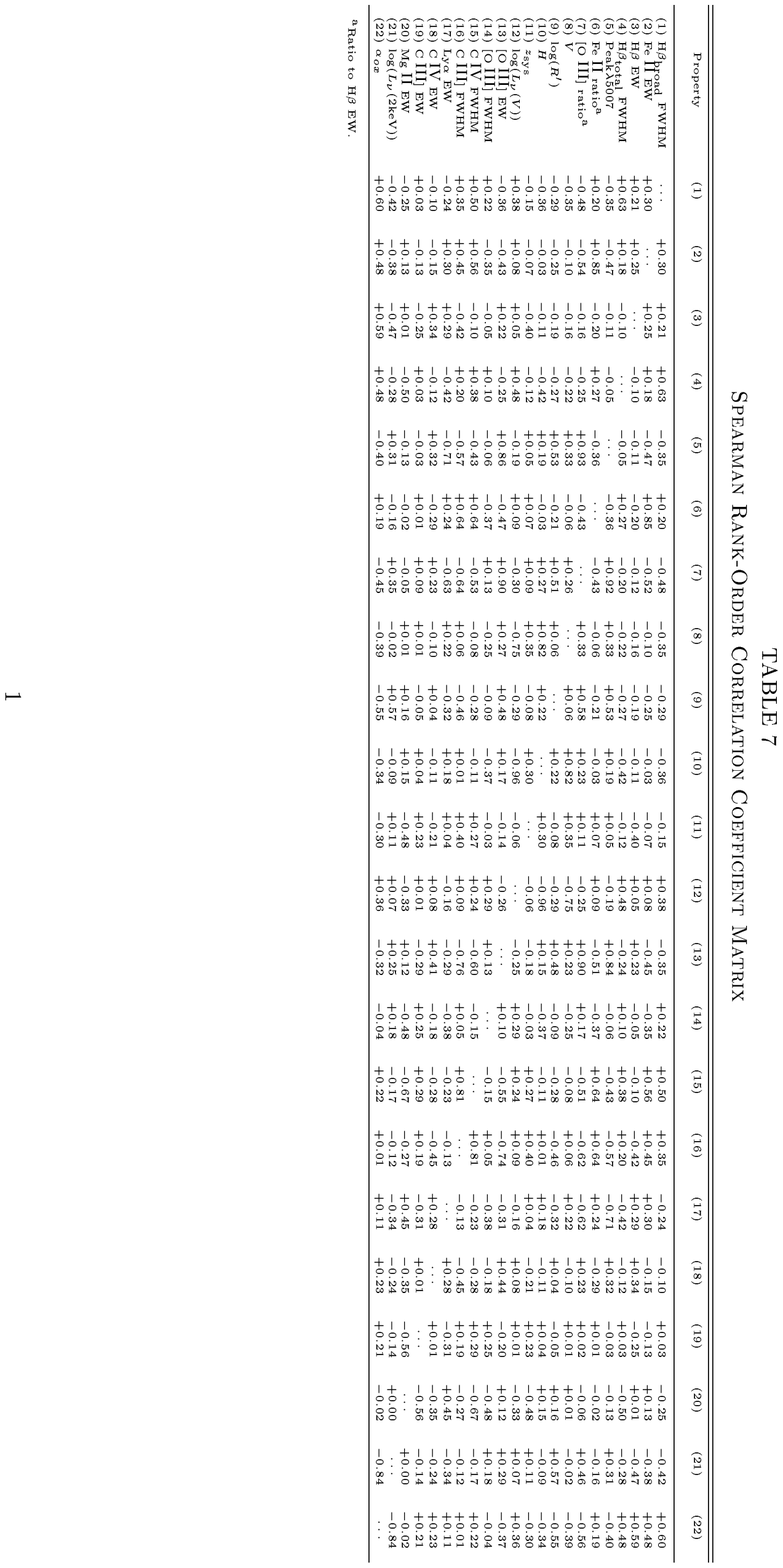}
\end{figure}

\begin{scriptsize}
\begin{deluxetable}{lccccccc}
\tablewidth{0pt}
\tablenum{8}
\tablecaption{Significant~Correlations}
\tablehead{\colhead{$$} & \colhead{$$} & \multicolumn{3}{c}{Spearman} & \multicolumn{3}{c}{Kendall}\\
\colhead{Correlated~Properties} & \colhead{$N_{\rm pairs}$} & \colhead{$r_s$} & \colhead{CL\tablenotemark{a}} & \colhead{$N_{\rm deg}$\tablenotemark{b}} & \colhead{$\tau$} & \colhead{CL\tablenotemark{a}} & \colhead{$N_{\rm deg}$\tablenotemark{b}
}}
\startdata
$[$\ion{O}{3}$]$~strength~-~\ion{Fe}{2}~strength & $31$ & $-0.524$ & $99.8$ & $5$ & $-0.366$ & $99.6$ & $5$ \nl
$[$\ion{O}{3}$]$~strength~-~$\log(R^{\prime})$ & $31$ & $+0.508$ & $99.6$ & $2$ & $+0.421$ & $99.9$ & $2$ \nl
$[$\ion{O}{3}$]$~strength~-~\ion{C}{3}$]$~FWHM & $26$ & $-0.762$ & $>99.9$ & $2$ & $-0.561$ & $>99.9$ & $2$ \nl
$[$\ion{O}{3}$]$~strength~-~\ion{C}{4}~FWHM & $30$ & $-0.599$ & $>99.9$ & $2$ & $-0.433$ & $99.9$ & $2$ \nl
$[$\ion{O}{3}$]$~strength~-~H$\beta_{\rm~broad}$~FWHM & $31$ & $-0.478$ & $99.4$ & $1$ & $-0.297$ & $98.1$ & $1$ \nl
$[$\ion{O}{3}$]$~strength~-~$\alpha_{ox}$ & $24$ & $-0.455$ & $97.4$ & $1$ & $-0.312$ & $96.7$ & $2$ \nl
$[$\ion{O}{3}$]$~EW~-~\ion{C}{4}~EW\tablenotemark{c} & $27$ & $+0.415$ & $96.9$ & $0$ & $+0.280$ & $95.9$ & $0$ \nl
$[$\ion{O}{3}$]$~strength~-~Ly$\alpha$~EW & $16$ & $-0.629$ & $99.1$ & $1$ & $-0.467$ & $98.8$ & $1$ \nl
H$\beta$~FWHM~-~Luminosity & $31$ & $+0.482$ & $99.4$ & $4$ & $+0.326$ & $99.0$ & $4$ \nl
H$\beta$~FWHM~-~$\alpha_{ox}$ & $24$ & $+0.598$ & $99.8$ & $2$ & $+0.449$ & $99.8$ & $2$ \nl
H$\beta$~FWHM~-~\ion{C}{4}~FWHM & $29$ & $+0.499$ & $99.4$ & $1$ & $+0.328$ & $98.7$ & $0$ \nl
H$\beta$~EW~-~$\alpha_{ox}$ & $24$ & $+0.590$ & $99.8$ & $1$ & $+0.428$ & $99.7$ & $1$ \nl
H$\beta$~EW~-~Redshift\tablenotemark{d} & $31$ & $-0.401$ & $97.5$ & $0$ & $-0.284$ & $97.5$ & $0$ \nl
\ion{Fe}{2}~EW~-~$\alpha_{ox}$ & $24$ & $+0.481$ & $98.3$ & $0$ & $+0.319$ & $97.1$ & $0$ \nl
\ion{Fe}{2}~strength~-~\ion{C}{4}~FWHM & $29$ & $+0.645$ & $>99.9$ & $1$ & $+0.476$ & $>99.9$ & $1$ \nl
\ion{Fe}{2}~strength~-~\ion{C}{3}$]$~FWHM & $25$ & $+0.645$ & $99.9$ & $1$ & $+0.457$ & $99.9$ & $1$ \nl
\ion{C}{4}~FWHM~-~\ion{C}{3}$]$~FWHM & $24$ & $+0.808$ & $>99.9$ & $0$ & $+0.630$ & $>99.9$ & $0$ \nl
$\log(R^{\prime})$~-~$\alpha_{ox}$ & $25$ & $-0.547$ & $99.5$ & $1$ & $-0.380$ & $99.2$ & $1$ \nl
\enddata
\tablenotetext{a}{Confidence Level}
\tablenotetext{b}{Number of additional ``like'' (degenerate) correlations, significant at $\geq 95\%$ confidence.}
\tablenotetext{c}{Confirmed at lower redshift by Corbin \& Boroson (1996).}
\tablenotetext{d}{Confirmed in $z\lesssim1$, RLQ only sample (Brotherton 1996).}
\end{deluxetable}
\end{scriptsize}
\clearpage

\subsection {Correlations with the Strength of $[$\ion{O}{3}$]\lambda5007$ Emission}
We find that all measures of \oiii\ emission strength are anti-correlated,
at $>95\%$ confidence, with the optical \ion{Fe}{2} 
emission strength.  The most 
significant example is 
$[$\ion{O}{3}$]$/H$\beta$ to \fe2\ EW
($99.8\%$ confidence), which is plotted in Figure 4a.  This anti-correlation
was the most significant one
found by BG92, and at lower
redshift by Corbin \& Boroson (1996).  It was also seen by
Brotherton (1996)
in an entirely RLQ sample at $z \leq 0.95$.  As an explanation, BG92 
and Wills \& Brotherton (1996) have proposed
that a large covering fraction by dense, high-speed \ion{Fe}{2}-rich clouds
will result in high \ion{Fe}{2} emission and at the same time will prevent much of the ionizing
radiation from reaching the more distant, low density NLR gas where the
forbidden \oiii\ emission originates.  BG92 suggested that this BELR covering
factor is dependent on the accretion rate and the mass of the black 
hole, while Wills (1996) added that greater nuclear obscuration by
dust associated with a torus could result in stronger \fe2\ emission.
Recently, dust has been detected in a handful of high 
redshift QSOs (\cite{cimatti98}).

We also find $>99\%$ confidence level correlations between all measures
of \oiii\ 
emission and the radio-to-optical flux ratio $R^{\prime}$,
measuring the degree of
radio loudness.  The most significant ($99.6\%$) result is the 
$[$\ion{O}{3}$]$/H$\beta$ to \logr\ correlation plotted in Figure 4b.
BG92 found similar, but weaker, correlations.  It is problematic to explain this
relationship in terms of orientation (thus, unification of RLQs and RQQs),
since Jackson \& Browne (1991) found 
that the EW of \oiii\ is anti-correlated with the ratio of core-to-lobe luminosity, which is
a strong indicator of the alignment between the radio axis and the line-of-sight
in low redshift RLQs.  They
proposed that near face-on views show a beamed continuum that swamps the NLR
emission, thus significantly reducing the \oiii\ EW.   Instead, we propose
that the intensity and direction of the radio power produced by the central
engine is closely tied to the intensity and direction of the NLR ionizing
radiation.  Possible physical mechanisms for
this intimate association between the NLR ionization and the radio component
are radio jet driven radiative shocks (\cite{wilson93}; though this is a very
inefficient mechanism \cite{laor98}), or anisotropic ionization
of the NLR by the nuclear continuum photons emitted preferentially along
the radio axis (\cite{bower95}, and \cite{simpson96}).  An extension of the latter idea
is that the nuclear ionizing continuum is collimated by the opening of
a dusty torus which is also roughly aligned with the radio axis 
(\cite{wilson97}).  Supporting this hypothesis are a host
of HST observations of local Seyfert galaxies in the light of 
$[$\ion{O}{3}$]\lambda5007$ ({\it e.g.}~\cite{wilson93};
\cite{bower95}; and \cite{simpson96}), showing
sharp edged, V-shaped profiles believed to be the projection of a biconical
NLR, and these cones are generally well aligned with the radio axis.
In addition, \oiii\ emission has been found to be partially obscured
in low redshift radio loud AGN
via polarization (\cite{disergo97}) and by the 
$[$\ion{O}{3}$]\lambda5007$/$[$\ion{O}{2}$]\lambda3727$ ratio 
(\cite{baker97};
and \cite{crawford97}).

Further, we find that the FWHM of the two rest-frame UV broad lines
(\ion{C}{4}$\lambda1549$ and \ion{C}{3}$]\lambda1909$) are anti-correlated
with several measures of \oiii\ emission strength.
Especially significant, at $>99.9\%$ confidence, are the
relations between \oiii\ EW and the FWHM of each carbon line: (i)~\ciii\ 
shown in Figure~4c; and (ii)~\civ\
presented in Figure~4d.  The former relation was predicted
and then found
in a similar $z \sim 2$ sample by Brotherton~({\it et al.} 1994b, 1997). 
A weak $[$\ion{O}{3}$]$~EW to \ion{C}{4}~FWHM 
anti-correlation was found 
by \cite{corbin96}.
Brotherton and Wills (\cite{wills93}; Brotherton~{\it et al.} 1994a,b; 
\cite{wills96b}; etc.) have argued that these UV broad emission line profiles 
(\ion{C}{4} and \ion{C}{3}$]$) have two components, a narrow
($v_{FWHM} \sim 2000$~km~s$^{-1}$) core arising from an intermediate-line-region
(ILR), and a very-broad-line-region (VBLR, $v_{FWHM} \gtrsim 7000$~km~s$^{-1}$) 
base.  They have proposed that the ILR is an inner extension of the NLR and that
this physical connection between the two regions would produce the observed 
anti-correlation, such that narrow
\ion{C}{3}$]\lambda1909$ and \ion{C}{4}$\lambda1549$ lines are the signature 
of strong ILR emission, combined with large \oiii\ EW indicative of strong
NLR emission.  

We also find an 
anti-correlation between $[$\ion{O}{3}$]$/H$\beta$ and broad \hb\ FWHM 
(at $99.4\%$) plotted in Figure~4e.  However, this inverse relation
cannot be explained with the VBLR+ILR
model since the strength of NLR emission anti-correlates only
with the {\it broad}
component of \hb\ and not with the total line width.  If there were an ILR
component of the \hb\ emission, one would expect the total line FWHM to
decrease while the total line EW increases with larger NLR strength.
An \oiii\ EW -- \hb\ FWHM relation was not found
in low redshift/luminosity studies (\cite{corbin96}, and \cite{brotherton96});
and, BG92 suggested they
observed only a slight negative trend between their main eigenvector
(the \oiii\ -- \ion{Fe}{2} anti-correlation) and the width of broad H${\beta}$.

The impact of the shape and strength of the ionizing continuum on the line
emission manifests itself in an anti-correlation ($97.4 \%$ significance)
between the ratio $[$\ion{O}{3}$]$/H$\beta$
and the optical-to-X-ray spectral slope $\alpha_{ox}$ 
(see Figure 4f), also found by BG92 in their
low redshift/luminosity sample.  This spectral index
is a rough
measure of the strength of the ionizing continuum emitted from the optical
to the soft X-ray, such that $\alpha_{ox} \gtrsim~
1.4$ corresponds to a {\it steeper} slope and thus a {\it softer} ionizing
continuum, whereas, $\alpha_{ox} \lesssim 1.4$ indicates a {\it flatter} and
{\it harder} ionizing continuum.  A flatter slope is thought to contain two
components: (i)~a harder non-thermal component of emission
produced by the inverse Comptonization of the beamed radio jet photons;
and (ii)~a softer thermal component emitted from the hot inner edges of the
accretion disk.  Therefore, this anti-correlation supports the hypothesis that
the NLR is ionized by the nuclear continuum.

It should be noted that the one obvious outlier seen in Figures 4a, 4b, 4e and
4f is
Q0109+022 (UM87).  This is a RQQ with anomalously strong $[$\ion{O}{3}$]$ compared to
H$\beta$, as well as weak \fe2\ emission, which are both indicative
of the RLQ class.  The radio flux measurement, though an upper limit
based on detection sensitivity, has been confirmed by the recent NRAO VLA Sky
Survey (NVSS -- \cite{condon96}).  Therefore, the placement of this data
point at the extreme end of $[$\ion{O}{3}$]$/H$\beta$ values is
due to the very weak H$\beta$ emission apparent in the observed spectrum (see
Figure 1d), which may be partially due to the low S/N of the data.

\subsection {Correlations with the Width of H$\beta$ Emission}
We find a significant ($99.4\%$ confidence) correlation between
the total fitted
\hb\ FWHM and the intrinsic rest-frame $V$-band luminosity density, $L_{\nu}(V)$
(see Figure 5a).
At lower redshifts ($z<1$) and luminosities, similar weak relations have been
found between \hb\ FWHM and the luminosity of the rest-frame $B$-band
(\cite{wang96a}), as well as between broad
\hb\ FWHM and the absolute $V$-band magnitude (BG92).  However,
Corbin \& Boroson (1996) did not find this correlation.  This observed trend
is consistent with BELR clouds in semi-flattened, gravitationally bound orbits
about the nucleus as favored by recent reverberation studies (\cite{wang96b}).
On average, an increased luminosity will correspond to a larger central black hole
mass, thus correlating with a larger orbital velocity.

We find that
the broad \hb\ line width and the mean spectral slope $\alpha_{ox}$ are
correlated at the $99.8\%$ level (see Figure 5b).  This correlation is
consistent with a viewing angle effect, if we assume that the BELR clouds orbit
and that $\alpha_{ox}$ is a good inclination indicator (\cite{wang96a}).
In a nearly face-on view one would observe a narrower BELR
emission line and a flatter (small $\alpha_{ox}$) ionizing continuum slope,
suggesting a direct view of the harder component of X-rays.
Wills \& Brotherton (1995; as well as 
\cite{baker95}) showed that \hb\ FWHM
is inversely correlated with another property measuring the beaming angle in
low redshift quasars.  Jackson \& Browne (1991) confirmed that the Balmer
line profile narrows systematically with increasing face-on orientation in a 
RLQ only study.  In contrast to the above results, 
both Wang, Brinkmann \& Bergeron (1996) and Laor
{\it et al.} (1997) have found anti-correlations between \hb\
line width and $\alpha_{ox}$ at low redshift/luminosity.
In addition, BG92 and Brotherton (1996) found no
correlation in their samples.  Note
that we do not find any correlations between $\alpha_{ox}$
and the other two broad line widths (\ciii\ and \ion{C}{4}), but 
recall that the 
width of these rest UV lines may be related to the fractional contribution from
the proposed ILR (see Wills and Brotherton references herein).

Lastly, we find that broad \hb\ FWHM and \ion{C}{4}$\lambda1549$ FWHM are
correlated at the $99.4\%$ level (see Figure 5c); however, we find the broad
component of the \hb\ line to be, on average wider, than the total \civ\ line.  We suggest
that the {\it broad} components of each line are emitted from the same VBLR,
but that the \civ\ line has an additional, narrower component produced in a separate
volume of gas, possibly the proposed ILR.  Presumably, the ILR component
superimposed on the VBLR component would result in a narrower total line
width measurement.  Similar correlations have been
found at low redshift in a RLQ/RQQ sample (\cite{corbin96}) and a RLQ only 
sample (\cite{marziani96}).  

\subsection {Correlations with the Strength of H$\beta$ Emission}
In Figure 6 we present the significant ($99.8\%$ confidence) 
positive correlation between the EW of the 
total fitted \hb\ emission and the slope of the ionizing continuum,
estimated by $\alpha_{ox}$.  
As with the \hb\ line width to ionizing continuum slope relation,
a significant but {\it opposite} result was found at low redshift/luminosity
(BG92 and \cite{wang96a}), while Brotherton (1996) and Wilkes~({\it et al.} 
1997) found no correlation with $\alpha_{ox}$.  

\subsection {Correlations with the Strength of Optical \ion{Fe}{2} Emission}
In addition to our confirmation of BG92's strongest result (see $\S4.1$), we
find the optical \fe2\ EW and the optical-to-X-ray spectral slope $\alpha_{ox}$ 
to be positively correlated at the $98.3\%$ level (see Figure 7a).
It has been shown that
strong \fe2\ emission is associated with softer X-ray spectra in PG QSOs
(\cite{laor94}) and in narrow line AGN (\cite{boller96}), yet
this correlation was not found in several low
redshift/luminosity QSO samples (BG92; \cite{brotherton96};
\cite{wang96a}; and \cite{wilkes97}).  Furthermore, our finding is
not in agreement with
photo-ionization models ({\it e.g.}~\cite{kwan81}; and \cite{netzer83}) 
which predict that hard, flat X-ray spectra are expected to produce strong
\fe2\ emission.

We also find that both parameters of optical \fe2\ emission strength 
(\fe2\ EW and \ion{Fe}{2}/H$\beta$) are correlated with the rest-frame
UV line widths of \ion{C}{4}$\lambda1549$ and \ion{C}{3}$]\lambda1909$
at $>97.7\%$ confidence.  In fact, the \ion{Fe}{2}/H$\beta$ ratio gives
correlations with both carbon line widths at $\geq99.9\%$ significance (see
Figures 7b and 7c).
A similarly strong \fe2\ strength -- \ciii\ width correlation was found in 
another $z \sim 2$ sample (\cite{brotherton97}).  Brotherton suggested that
this result was another consequence of the \oiii\ -- \fe2\ anti-correlation
which relates QSO properties from the radio to the X-ray such that variations
along this relation (BG92's first eigenvector) may be a ``fundamental plane''
for QSOs.  These strong positive correlations between
\fe2\ emission strength from the VBLR and the widths of the carbon lines 
from an ILR, combined with
the inverse correlations between NLR emission strength and ILR line widths
(see $\S4.1$), give credence to the VBLR+ILR
model proposed by Brotherton and Wills.  A word of
caution --- the FWHM of \ion{C}{4} (and of \ion{C}{3}$]$ when no \ion{C}{4}
measurement was available) was used to broaden the \ion{Fe}{2} template
used to fit this sample of spectra.  In addition, the \civ\ and \ciii\ line widths
are extremely well correlated ($r_s = +0.808$; see below).  Thus, the
\ion{Fe}{2} strength, determined by the best-fit model,
could have been biased in such a way as to produce, or at least reinforce, the 
aforementioned positive correlations.  Increased broadening of the iron
template may have resulted in confusion between the continuum and the very
blended \ion{Fe}{2} emission, and thus an overestimate of the \ion{Fe}{2} 
strength.  
However, we do find a strong correlation between the line widths of 
\ion{C}{4} and H$\beta$ (see $\S4.2$), which is
consistent with BG92's finding that the widths of
\ion{Fe}{2} and H$\beta$ are similar, and justifies our use of the
\civ\ line width for broadening the \fe2\ template.  
In addition, our Monte Carlo analysis did {\it not} show any particularly
large variations in \ion{Fe}{2} EW uncertaintities, which would have been
evidence for confusion between the continuum and \fe2\ templates.

\subsection {Other Significant Correlations}
As stated above, we find
a very significant ($>99.99\%$), correlation
between the FWHMs of the two UV, high ionization broad carbon lines taken from the
literature (\civ\ and \ion{C}{3}$]$).  The same 
equally strong
result has been found at $z\sim2$ (\cite{corbin91}) and at 
low redshift/luminosity (\cite{corbin94}; and
\cite{brotherton94a}).

And finally, we find that the degree of radio loudness ${\log(R^{\prime})}$
is anti-correlated with the slope of the ionizing continuum $\alpha_{ox}$, at
$99.5\%$ significance.  This is consistent with a physical
connection between the strength of the ionizing continuum and the strength of
the radio power, both believed to be produced by the nuclear source.

\section {Luminosity or Redshift Dependencies}
An important question that can be asked about
our correlation analysis of the previous section
is whether any of the emission line properties
are dependent on luminosity and/or
redshift.  Though the luminosity of our sample is fairly constant
(see Figure 8), and the redshift range \zrange\ only
spans a very small portion of the total age of the Universe $\tau_0$, 
any relationships
between QSO emission parameters and redshift or luminosity could be evidence for
evolutionary effects.  This might especially be true if we compared QSOs
at $z\sim2.5$, corresponding 
to a look back time of nearly $80\%$ (for $q_0 = 0.1$) of $\tau_0$, to QSOs
existing in the local Universe.  The QSOs of our
high redshift sample do not have similarly luminous counterparts in the
nearby Universe.  Thus, dependencies on redshift and luminosity are 
entangled.
With these points in mind, we attempted
to combine the measurements of the rest-frame optical spectral 
features of our high redshift, high luminosity data with similar 
measurements from the low redshift ($z\lesssim 0.5$) and lower luminosity
sample (see again Figure 8) of BG92.  This combined data set enabled us
to search for such dependencies
using the most luminous QSOs observed at the two ends of the redshift
range spanning $0<z<2.5$.

\subsection {The Low Redshift Sample}
To achieve our goal of creating a consistent $0<z<2.5$ QSO sample,
it was necessary to combine our high redshift
measurements with matching parameters from the low redshift sample.
Foremost in importance was the need for equivalent luminosity measurements,
therefore, we calculated the rest-frame $V$-band luminosity density
(in ergs~${\rm s}^{-1}~{\rm Hz}^{-1}$) of each low redshift QSO:
\begin{equation}
L_{\nu}(V) = 4\pi (10 {\rm pc})^2 F_{\nu}^i(V) ,
\end{equation}
where the {\it intrinsic} {\it V}-band flux density 
(in ergs~cm$^{-2}~{\rm s}^{-1}~{\rm Hz}^{-1}$) is
\begin{equation}
F_{\nu}^i(V) = 10^{-0.4M_V -22.41} ,
\end{equation}
derived from BG92's tabulated absolute magnitude ($M_V$), assuming that
zero magnitude in {\it V} is $3880$ Jy (\cite{johnson66}).   BG92 calculated 
$M_V$ from the apparent $V$ of Neugebauer~{\it et al.} (1987), assuming an
H$_{0} = 50$~km~s$^{-1}$~Mpc$^{-1}$ and $q_0 = 0.1$ cosmology.

Second, we obtained consistent emission line property measurements by running
our \chisq\ minimization composite model fitting routine 
on the entire sample of $87$
reduced, rest-frame spectra from BG92.  We used the same template
set as for the fitting of our high redshift sample, except the narrowest
\oiii\ and \hb\ templates were set to the instrumental
resolution of the BG92 spectra ($450$~km~s$^{-1}$).  As with
our sample, we inspected each spectrum, fitting the upper half of the 
\hb\ line with a Gaussian to determine the line center and then shifting the
Balmer template set accordingly for each object.  We also rebinned each
spectrum to a logarithmic dispersion of $3.5$~\AA \ per pixel 
for compatibility with the fitting routine.   The wavelength coverage of
the BG92 sample was greater than in ours, resulting in the detection of the
H$\gamma\lambda4340$ line in most of the objects.  From initial fitting tests
we found that our set of Balmer templates, with a fixed ratio between \hb\ 
and H$\gamma$, usually did not match the actual data, either producing an
overestimate or underestimate of the \hb\ line flux.  Thus it was necessary 
to trim the blue half of the \hg\ line from each spectrum prior to fitting.

Since these quasars are at
low redshifts, the probability of finding a \civ\ line width in the literature
is very small; therefore, we decided to use the \hb\ line FWHM of each
object to broaden its \fe2\ template.  This decision is consistent with BG92's
finding that the optical \fe2\ and Balmer hydrogen emission originate from the
same clouds in the BELR.  However,
we decided not to use their published values of the {\it broad} \hb\ FWHM 
because these did not include the narrow line component of the permitted
emission that the {\it total} line FWHM of our model does, plus there is
no evidence that the permitted \fe2\ emission is solely comprised of a
broad line component.  Instead we used a first iteration fit of each
spectrum with a larger set of templates to obtain the {\it total} \hb\
line width.  The
larger set of templates included the $3$ power-law, $5$ \oiii\ and $5$ Balmer
templates, plus a set of $8$ \fe2\ templates broadened by Gaussians of
$450$ (instrumental width),
$1370$, $1940$, $2740$, $3870$, $5480$, $7320$ and $10350$~\kms\ . 
These line widths were factors of $\case{1}{\sqrt{2}}$
of the minimum ($2740$~\kms\ ) and maximum ($10350$~\kms\ )
\civ\ line widths from our high redshift sample.  From the first iteration
fit of $21$ templates we obtained the {\it total}
\hb\ line width, which was used
to broaden the single, object-specific, \fe2\ template for the final ($14$
template) fitting iteration of each
quasar spectrum.  The final best fit composite models for
the BG92 sample were quite remarkable in their reproduction of most of the
spectral features.  The fit to the data and the
individual components for a representative object (PG1404+226)
are plotted in the top and bottom panels, respectively, of Figure 9.

As with our high redshift sample, the set of emission
line properties for each QSO were derived directly from the 
individual components
of its composite model spectrum.  In Figure 10 we plot our best fit
derived values versus the published BG92 results for the EWs
of \ox3\ and H$\beta$, the {\it relative} EW of optical \ion{Fe}{2}, and the
\peak\ parameter.  In each case the agreement is
excellent
for the two properties associated with the \ox3\ strength.  For \fe2\ 
EW there is a trend such that the BG92 value is greater than our model fit
value.  This trend is due to a difference in 
{\it relative} EW definitions: BG92 selected the strong \fe2\ emission
complex between $4434$ and $4684$~\AA; whereas we chose the spectral range
encompassing the \hb\ and \ox3\ lines where there is much less \fe2\ emission.
There is also a slight trend where the BG92 \hb\ EW is, on average, stronger 
than our fitted value.  We believe this trend is due to BG92's systematic
overestimation of the
\fe2\ strength, thereby producing a lower 
continuum level and thus a higher \hb\ EW measurement.  
A systematic underestimation of the continuum level would result in a more
pronounced effect for the EW of {\it broad} \hb\ than it would
for the EW of {\it narrow} $[$\ion{O}{3}$]\lambda5007$.

\subsection {Correlations with Luminosity and Redshift}
To determine whether any of the rest-frame optical emission properties
correlated with luminosity or redshift, we computed the Spearman rank-order
coefficients for the redshift and rest-frame $V$-band luminosity against
the set of nine parameters common between both the low redshift and high
redshift samples.  The broad \hb\ FWHM and $\alpha_{ox}$ of the $z\lesssim 0.5$
QSOs were tabulated in BG92, while the remaining parameters were measured
by the fitting routine.
As before, we also calculated the
Kendall $\tau$ and found the results to be consistent.  For the total combined
$z\lesssim 0.5$ and $z\sim 2$ sample of $N=119$ objects, a 
significant correlation at the $\gtrsim 99.7\%$
confidence level corresponds to a coefficient of
$|r_s| \gtrsim 0.285$.  The number of correlated pairs, their coefficients and
confidence levels,
for all nine parameters with respect to 
luminosity and redshift, are presented in Table~9.
 
\begin{scriptsize}
\begin{deluxetable}{lccccc}
\tablewidth{0pt}
\tablenum{9}
\tablecaption{Correlations~with~Luminosity~and~Redshift}
\tablehead{\colhead{$$} & \colhead{$$} & \multicolumn{2}{c}{$\log(L_{\nu}(V))$} & \multicolumn{2}{c}{Redshift}\\
\colhead{Emission~Property} & \colhead{$N_{\rm pairs}$\tablenotemark{a}} & \colhead{$r_s$} & \colhead{CL\tablenotemark{b}} & \colhead{$r_s$} & \colhead{CL\tablenotemark{b}
}}
\startdata
H$\beta_{\rm~broad}$~FWHM & $111$ & $+0.710$ & $>99.99$ & $+0.671$ & $>99.99$ \nl
$[$\ion{O}{3}$]$~FWHM & $110$ & $+0.703$ & $>99.99$ & $+0.662$ & $>99.99$ \nl
\ion{Fe}{2}~EW & $109$ & $+0.098$ & $<90.00$ & $+0.073$ & $<90.00$ \nl
\ion{Fe}{2}/H$\beta$ & $109$ & $+0.084$ & $<90.00$ & $+0.074$ & $<90.00$ \nl
H$\beta$~EW & $118$ & $-0.103$ & $<90.00$ & $-0.136$ & $<90.00$ \nl
$[$\ion{O}{3}$]$~EW & $116$ & $-0.203$ & $97.15$ & $-0.208$ & $97.47$ \nl
$[$\ion{O}{3}$]$/H$\beta$ & $115$ & $-0.214$ & $97.81$ & $-0.194$ & $96.20$ \nl
Peak$\lambda5007$ & $115$ & $-0.130$ & $<90.00$ & $-0.113$ & $<90.00$ \nl
$\alpha_{ox}$ & $80$ & $-0.071$ & $<90.00$ & $-0.083$ & $<90.00$ \nl
\enddata
\tablenotetext{a}{From a total sample of 87 ($z<0.5$) QSOs from BG92,
added to our sample of 32 ($2.0<z<2.5$) QSOs.}
\tablenotetext{b}{Confidence Level}
\end{deluxetable}
\end{scriptsize}
\clearpage

We find that the rest-frame FWHM of \ox3\ is positively correlated with luminosity, as
well as redshift, at the $>99.99\%$ confidence level.
Figure 11a shows a clear trend where
the forbidden narrow line width increases with the continuum 
luminosity emitted from the same spectral range.
However, the lack of high
redshift, low luminosity ({\it e.g.}~$30\leq \log(L_{\nu}(V)) \leq 32$) QSOs
in our sample prevent us from distinguishing which property, redshift or 
luminosity, \oiii\ width physically depends upon.

Furthermore, we find a similar luminosity and redshift dependency for the
rest-frame broad 
\hb\ FWHM at $>99.99\%$ significance.
Again, the
line width increases with the luminosity emitted at similar wavelengths 
(see Figure 11b), but
in this case it is an allowed transition from the BELR which is firmly
established to be physically distinct and distant from the NLR.  As with our
narrow line result, this relation could be biased by our selection of only
high luminosity QSOs at $z>2$.
In a sample of $41$ RLQs,
Brotherton (1996) found a \oiii\ FWHM correlation with increasing
rest-frame $V$ luminosity; however, he did not find a like relation for
\hb\ FWHM and suggested that this was evidence for
different line-broadening mechanisms in the BELR and NLR.  Similar
tests for same wavelength range luminosity dependencies of emission line
widths have been done for the rest-frame UV with mixed results.  One study
showed a positive correlation between the FWHM of three broad emission
lines (Ly$\alpha$, \ion{N}{5}, and \ion{C}{4}) and the luminosity emitted
at $1450$~\AA, in a $z>3$ QSO sample, but not in a $z>2$ sample 
(\cite{osmer94}), while another study showed
only Ly$\alpha$ FWHM dependent on UV luminosity at $z<1.5$ 
(\cite{green96}).  At $z<0.8$ the
Ly$\alpha$ full-width at zero-intensity (FWZI) correlated with the luminosity
emitted at $1549$~\AA \ (\cite{corbin96}), yet the FWHMs of 
Ly$\alpha$, \ion{C}{4}, and
\hb\ were all independent of this luminosity.  Two other samples at $z\leq2$
showed no correlations between emission line widths and luminosity emitted at
UV wavelengths
(\cite{corbin92}; and \cite{brotherton94a}).

\subsection {Baldwin Effects}
An important emission line luminosity dependence observed in QSOs, 
is the systematic
decrease in line EW with increasing continuum luminosity.  
This effect was first seen in the
\ion{C}{4}$\lambda1549$ line by Baldwin (1977), and has been found
to extend over a wide range of luminosities in a large sample (\cite{kinney90}).
The ``Baldwin Effect'' has also been found in other rest-frame UV 
lines such as Ly$\alpha$, \ion{N}{5}, \ion{O}{6}, \ion{He}{2}, \ion{C}{3}$]$, 
and \ion{Mg}{2} ({\it e.g.} \cite{tytler92}; \cite{green96}).  Furthermore,
this trend has been shown to be continuous over all redshifts up to $z=3.8$
for the \ion{C}{4} and Ly$\alpha$ lines (\cite{osmer94}).  

For the three rest-frame optical emission line EWs that we
correlated with optical continuum luminosity (see Table 9), we noticed a weak
($\sim 97\%$ confidence) ``Baldwin-like'' trend for the 
$[$\ion{O}{3}$]\lambda5007$ line and no trend for H$\beta$ or \ion{Fe}{2}.
However, since we found the EW
of the $[$\ion{O}{3}$]$ line to correlate positively with radio strength
(see $\S4.1$), we split the total QSO sample into RQQ and RLQ subsamples and
re-computed the Spearman rank-order coefficients against the rest-frame
$V$-band luminosity.  For the RQQ-only subsample of $87$ objects, we found
a more significant ($\gtrsim99\%$, $r_s=-0.291$) $[$\ion{O}{3}$]$ 
Baldwin Effect, with no effect present in the RLQ-only subsample.
In particular, the absence of luminous, high redshift RQQs with EWs 
larger than $30$~\AA \ is apparent in Figure 11c.  Again, we must caution the
reader that the EW of $[$\ion{O}{3}$]$ is equally well anti-correlated with
redshift in the RQQ-only subsample, and that the lack of low luminosity, high
redshift QSOs in this combined sample prevent the dis-entanglement of these
two properties.  In addition, we did not observe a
Baldwin Effect for the H$\beta$ or \ion{Fe}{2} lines, in either of the
subsamples.

\section {RLQ vs. RQQ Comparison}
One of the main objectives of this study was to determine if there were
significant differences between RLQs and RQQs in their rest-frame
optical spectra.  With the set of derived emission line parameters
in hand, combined with the data from the literature, 
we constructed $22$ distributions of the individual property's values for the
RLQ and RQQ subsamples.  We calculated the mean of each property set, plus
the right ($+$) and left ($-$) $1\sigma$ uncertainties in the mean via
bootstrap re-sampling (see \cite{press}).  
Next, we calculated the two-sided Kolmogorov-Smirnov
(K-S) probability that we could reject the null hypothesis that 
the RLQ and RQQ distributions of a given parameter were
drawn from the same parent sample.
The results of our comparison, including the number of objects, mean with
associated $1\sigma$ uncertainties, and K-S test probabilities for each 
pair of RLQ and RQQ property distributions are tabulated in Table~10.  Notice
that the two subsamples are well matched in mean intrinsic luminosity 
($L_{\nu}(V)$); therefore, parameter differences dependent on luminosity
should be ruled out.  We tested
the validity of the K-S test under these conditions of limited numbers
(typically $15$ vs. $17$) and determined that a difference probability of
$97.8\%$ is repeatable to within $\pm 0.3\%$.  We find four independent properties
different at $>97.8\%$ confidence, and these results are plotted in Figures
12a -- 12d.  The top panel of each figure shows the RLQ distribution,
while the RQQ distribution with its over-plotted BALQSO subsample is given
in the bottom panel.  The bin sizes approximate twice the mean of the individual
$1\sigma$ values from the Monte Carlo error analysis.  The large solid
dots represent the mean of each distribution, and the uncertainties 
($1,~2$~and~$3~\sigma$'s) of each mean are given by the diminishing error bars.
Note that one property, the measure of radio loudness (${\log(R^{\prime})}$)
plotted in Figure 12a, shows a clear bimodal RLQ/RQQ distribution due to
the choice of samples.  The other three differences
are discussed in detail below.

\clearpage
\begin{scriptsize}
\begin{deluxetable}{lccccc}
\tablewidth{0pt}
\tablenum{10}
\tablecaption{RLQ~vs.~RQQ~Emission~Property~Comparison}
\tablehead{\colhead{$$} & \multicolumn{2}{c}{RLQ~Distribution} & \multicolumn{2}{c}{RQQ~Distribution} & \colhead{$
$}\\
\colhead{Emission~Property} & \colhead{$N$} & \colhead{Mean} & \colhead{$N$} & \colhead{Mean} & \colhead{$(\%)_{\rm~diff}$\tablenotemark{a}
}}
\startdata
$[$\ion{O}{3}$]$~EW~(\AA) & $15$ & $20.6^{+2.6}_{-3.5}$ & $17$ & $10.2^{+1.9}_{-1.7}$ & $97.212$ \nl
\tablevspace{.6ex}
H$\beta$~EW~(\AA) & $14$ & $64.8^{+6.2}_{-6.9}$ & $17$ & $73.2^{+5.1}_{-3.7}$ & $57.072$ \nl
\tablevspace{.6ex}
\ion{Fe}{2}~EW~(\AA) & $14$ & $25.6^{+4.8}_{-4.9}$ & $17$ & $34.4^{+3.8}_{-4.2}$ & $58.842$ \nl
\tablevspace{.6ex}
$[$\ion{O}{3}$]$~FWHM~(km~${\rm~s}^{-1}$) & $14$ & $1160^{+90}_{-80}$ & $15$ & $1150^{+120}_{-130}$ & $32.185$ \nl
\tablevspace{.6ex}
H$\beta_{\rm~total}$~FWHM~(km~${\rm~s}^{-1}$) & $14$ & $4430^{+590}_{-520}$ & $17$ & $5100^{+350}_{-470}$ & $62.250$ \nl
\tablevspace{.6ex}
H$\beta_{\rm~broad}$~FWHM~(km~${\rm~s}^{-1}$) & $14$ & $9870^{+930}_{-970}$ & $17$ & $11890^{+450}_{-600}$ & $97.807$ \nl
\tablevspace{.6ex}
$[$\ion{O}{3}$]$/H$\beta$ & $14$ & $0.31^{+0.04}_{-0.05}$ & $17$ & $0.16^{+0.03}_{-0.05}$ & $99.906$ \nl
\tablevspace{.6ex}
\ion{Fe}{2}/H$\beta$ & $14$ & $0.45^{+0.10}_{-0.09}$ & $17$ & $0.49^{+0.06}_{-0.06}$ & $31.245$ \nl
\tablevspace{.6ex}
Peak$\lambda5007$ & $14$ & $1.14^{+0.12}_{-0.14}$ & $17$ & $0.79^{+0.21}_{-0.26}$ & $99.906$ \nl
\tablevspace{.6ex}
$z_{\rm~sys}$ & $15$ & $2.207^{+0.032}_{-0.038}$ & $17$ & $2.222^{+0.029}_{-0.033}$ & $38.978$ \nl
\tablevspace{.6ex}
$\log(R^{\prime})$ & $15$ & $3.042^{+0.144}_{-0.163}$ & $17$ & $0.198^{+0.076}_{-0.065}$ & $100.000$ \nl
\tablevspace{.6ex}
$\log(L_{\nu}(V))\;$~(ergs~${\rm~s}^{-1}~{\rm~Hz}^{-1}$) & $15$ & $32.30^{+0.07}_{-0.08}$ & $17$ & $32.30^{+0.05}_{-0.05}$ & $52.235$ \nl
\tablevspace{.6ex}
$V$~magn & $15$ & $16.94^{+0.20}_{-0.20}$ & $17$ & $17.06^{+0.19}_{-0.16}$ & $4.611$ \nl
\tablevspace{.6ex}
$H$~magn & $15$ & $15.16^{+0.18}_{-0.25}$ & $17$ & $15.17^{+0.13}_{-0.12}$ & $52.235$ \nl
\tablevspace{.6ex}
$\log(L_{\nu}$(2keV))$\;$~(ergs~${\rm~s}^{-1}~{\rm~Hz}^{-1}$) & $13$ & $28.58^{+0.10}_{-0.11}$ & $12$ & $27.95^{+0.22}_{-0.15}$ & $98.701$ \nl
\tablevspace{.6ex}
$\alpha_{ox}$ & $13$ & $1.27^{+0.06}_{-0.05}$ & $12$ & $1.49^{+0.06}_{-0.04}$ & $88.476$ \nl
\tablevspace{.6ex}
\ion{C}{4}~FWHM~(km~${\rm~s}^{-1}$) & $15$ & $4880^{+390}_{-460}$ & $15$ & $6020^{+670}_{-500}$ & $69.206$ \nl
\tablevspace{.6ex}
\ion{C}{3}$]$~FWHM~(km~${\rm~s}^{-1}$) & $12$ & $6430^{+720}_{-730}$ & $14$ & $8420^{+620}_{-620}$ & $91.474$ \nl
\tablevspace{.6ex}
Ly$\alpha$~EW~(\AA) & $11$ & $72.2^{+8.9}_{-8.3}$ & $6$ & $99.2^{+9.0}_{-8.0}$ & $88.945$ \nl
\tablevspace{.6ex}
\ion{C}{4}~EW~(\AA) & $13$ & $22.6^{+3.4}_{-3.4}$ & $13$ & $20.0^{+2.2}_{-2.5}$ & $53.288$ \nl
\tablevspace{.6ex}
\ion{C}{3}$]$~EW~(\AA) & $10$ & $17.9^{+1.6}_{-1.6}$ & $13$ & $17.9^{+0.7}_{-0.7}$ & $26.370$ \nl
\tablevspace{.6ex}
\ion{Mg}{2}~EW~(\AA) & $7$ & $23.1^{+2.3}_{-2.3}$ & $6$ & $29.1^{+4.1}_{-4.0}$ & $9.339$ \nl
\enddata
\tablenotetext{a}{Two-sided K-S probability that the RLQ and RQQ distributions of a given property were not drawn from the same parent sample.}
\end{deluxetable}
\end{scriptsize}
\clearpage

\subsection {$[$\ion{O}{3}$]$ Strength Difference}
We find that all measures of \ox3\ emission are stronger in the RLQ subsample
than in the RQQ subsample at $\geq97.2\%$ confidence.  The
$[$\ion{O}{3}$]/$H$\beta$ distributions are plotted for comparison in 
Figure 12b, notice
that the RLQ and RQQ sample means do not intersect within the 
$1\sigma$ errors.
BG92 found a similar, yet much weaker, result.
A manifestation of this result was already illustrated by the strong positive
correlation between \oiii\ strength and \logr\ presented in $\S4.1$.  
We propose that narrow $[$\ion{O}{3}$]$,
in RLQs at least, is in fact emitted from a non-spherically symmetric region 
and that its strength
is physically tied to the strong radio emission, perhaps
the NLR ionization is anisotropic due to preferential continuum emission
aligned with the radio axis (\cite{bower95}; \cite{simpson96}; and \cite{wilson97}).
The correlation between \oiii\ strength and the ionizing continuum slope 
$\alpha_{ox}$ (also in $\S4.1$) may be evidence for this idea.

\subsection {Broad H$\beta$ Width Difference}
We find that the \hb\ broad component FWHM is
narrower in the RLQ subsample, at the $97.8\%$ confidence level, compared to 
the RQQ subsample (see Figure 12c).  Notice again that
the means do not overlap within the $1\sigma$ errors.  Many higher redshift
samples have shown that RLQs have narrower BELR high ionization species, such
as \civ\ and Ly$\alpha$ ({\it e.g.}~\cite{corbin91}; and FHI93), than RQQs.
However, one low redshift study found RLQs to have wider, rather than
narrower, \hb\ profiles than RQQs (\cite{corbin96}).  It has been shown that
the \hb\ line width is anti-correlated with viewing angle for QSOs in general
(\cite{wills95}; and \cite{baker95}), and we corroborate this with the 
H$\beta_{{\rm broad}}$ FWHM -- $\alpha_{ox}$ and H$\beta_{{\rm broad}}$ FWHM --
$\log(L_{\nu}(V))$ correlations of $\S4.2$.  Yet if one assumes that
\oiii\ strength is a measure related to radio loudness, then the strong \oiii\
strength -- broad \hb\ width anti-correlation ($\S4.1$) suggests an additional
factor in the broad 
Balmer width difference we observe.  It would seem that the width of this line
is dependent upon orientation for both RLQs and RQQs, but that it is also
governed, to a lesser degree,
by the mechanism that controls radio loudness.

\subsection {Other Significant Differences}
We find that the line width of the \ion{C}{3}$]\lambda1909$ emission is narrower
in RLQs than RQQs.  Though this result is $<95\%$ confident,
the difference between the means of the two sets is quite large
($\sim2000$~km~s$^{-1}$) and statistically compelling --- no overlap of the $1\sigma$ error
bars.  This result has been strongly confirmed in
many other studies (\cite{corbin91}; \cite{wills93}; \cite{brotherton94a}; 
\cite{corbin94}; and 
\cite{vestergaard97}; but not found by FHI93).  

Finally, we find that soft X-ray luminosity (see Figure~12d) 
is significantly ($98.7\%$) greater, and that
the mean optical-to-X-ray slope is somewhat flatter and harder
in RLQs compared to RQQs.
This result agrees well with the findings of a large sample study of 
Green~({\it et al.} 1995).

\subsection {Important Null Result}
The optical \fe2\ EW distributions are not statistically different
between radio types, but the means of 
the two distributions are different at the $1\sigma$ level, such that RLQs 
have less \fe2\ emission than RQQs (see Figure 12e).  This has been found in
lower redshift ($z<0.8$) 
samples (BG92 and \cite{corbin96}).  As mentioned in $\S4.1$, BG92 suggest
that the strength of \fe2\ is tied to some unknown property governing 
radio loudness
and thus the two types of QSOs are physically different.

\section {BALQSO vs. nonBALQSOs}
To determine whether BALQSOs differed significantly from other radio quiet
nonBALQSOs, we
performed essentially the same analysis as described in $\S6$ on two subsets
drawn from the RQQ subsample.  BALQSO membership (see Table~2) 
was based upon classifications
by Turnshek (1984), WMFH91, and Thompson~{\it et al.} (1989).  Due to
insufficient data points, we dropped the Ly$\alpha$ and \ion{Mg}{2}$\lambda2798$
EW distributions.  The results of our comparative analysis are presented, as
before, in Table~11.  Again, we tested the validity of the K-S test under these
conditions of even smaller numbers (typically $7$ vs. $10$) and found that a
difference probability of $91.6\%$ is repeatable to within $\pm1.5\%$.  We find
no statistically significant ($\geq 95\%$) differences between any BALQSO
and nonBALQSO optical emission property subsets measured from our sample.  One property, the
EW of the \ion{C}{4}$\lambda1549$ is less in the BALQSO subset at $>99.6\%$ 
significance, even though all but one of 
the BALQSOs making up this subset were classified
as having ``detached'' blueward \ion{C}{4} absorption, there is still an issue
of its contamination of the \ion{C}{4} emission.

\clearpage
\begin{scriptsize}
\begin{deluxetable}{lccccc}
\tablewidth{0pt}
\tablenum{11}
\tablecaption{NonBALQSO~vs.~BALQSO~Emission~Property~Comparison}
\tablehead{\colhead{$$} & \multicolumn{2}{c}{NonBALQSO~Distribution} & \multicolumn{2}{c}{BALQSO~Distribution} & \colhead{$
$}\\
\colhead{Emission~Property} & \colhead{$N$} & \colhead{Mean} & \colhead{$N$} & \colhead{Mean} & \colhead{$(\%)_{\rm~diff}$\tablenotemark{a}
}}
\startdata
$[$\ion{O}{3}$]$~EW~(\AA) & $10$ & $11.6^{+2.1}_{-2.2}$ & $7$ & $8.1^{+2.6}_{-2.1}$ & $32.896$ \nl
\tablevspace{.6ex}
H$\beta$~EW~(\AA) & $10$ & $74.5^{+5.8}_{-5.5}$ & $7$ & $71.5^{+7.5}_{-7.3}$ & $48.550$ \nl
\tablevspace{.6ex}
\ion{Fe}{2}~EW~(\AA) & $10$ & $27.8^{+5.3}_{-5.6}$ & $7$ & $43.9^{+6.5}_{-7.1}$ & $84.672$ \nl
\tablevspace{.6ex}
$[$\ion{O}{3}$]$~FWHM~(km~${\rm~s}^{-1}$) & $10$ & $1130^{+180}_{-170}$ & $5$ & $1200^{+140}_{-150}$ & $13.811$ \nl
\tablevspace{.6ex}
H$\beta_{\rm~total}$~FWHM~(km~${\rm~s}^{-1}$) & $10$ & $4560^{+610}_{-460}$ & $7$ & $5870^{+540}_{-670}$ & $73.771$ \nl
\tablevspace{.6ex}
H$\beta_{\rm~broad}$~FWHM~(km~${\rm~s}^{-1}$) & $10$ & $11490^{+730}_{-760}$ & $7$ & $12460^{+660}_{-1020}$ & $27.632$ \nl
\tablevspace{.6ex}
$[$\ion{O}{3}$]$/H$\beta$ & $10$ & $0.19^{+0.07}_{-0.07}$ & $7$ & $0.12^{+0.04}_{-0.05}$ & $48.550$ \nl
\tablevspace{.6ex}
\ion{Fe}{2}/H$\beta$ & $10$ & $0.38^{+0.06}_{-0.06}$ & $7$ & $0.64^{+0.11}_{-0.09}$ & $91.605$ \nl
\tablevspace{.6ex}
Peak$\lambda5007$ & $10$ & $0.94^{+0.40}_{-0.40}$ & $7$ & $0.58^{+0.21}_{-0.23}$ & $66.549$ \nl
\tablevspace{.6ex}
$z_{\rm~sys}$ & $10$ & $2.221^{+0.048}_{-0.046}$ & $7$ & $2.225^{+0.022}_{-0.020}$ & $82.361$ \nl
\tablevspace{.6ex}
$\log(R^{\prime})$ & $10$ & $0.267^{+0.084}_{-0.091}$ & $7$ & $0.100^{+0.128}_{-0.143}$ & $48.550$ \nl
\tablevspace{.6ex}
$\log(L_{\nu}(V))\;$~(ergs~${\rm~s}^{-1}~{\rm~Hz}^{-1}$) & $10$ & $32.32^{+0.07}_{-0.08}$ & $7$ & $32.27^{+0.06}_{-0.06}$ & $22.525$ \nl
\tablevspace{.6ex}
$V$~magn & $10$ & $16.88^{+0.24}_{-0.27}$ & $7$ & $17.32^{+0.18}_{-0.17}$ & $22.525$ \nl
\tablevspace{.6ex}
$H$~magn & $10$ & $15.11^{+0.18}_{-0.15}$ & $7$ & $15.26^{+0.12}_{-0.11}$ & $22.525$ \nl
\tablevspace{.6ex}
$\log(L_{\nu}$(2keV))$\;$~(ergs~${\rm~s}^{-1}~{\rm~Hz}^{-1}$) & $7$ & $28.16^{+0.20}_{-0.17}$ & $5$ & $27.65^{+0.21}_{-0.21}$ & $90.923$ \nl
\tablevspace{.6ex}
$\alpha_{ox}$ & $7$ & $1.43^{+0.06}_{-0.06}$ & $5$ & $1.57^{+0.09}_{-0.09}$ & $55.721$ \nl
\tablevspace{.6ex}
\ion{C}{4}~FWHM~(km~${\rm~s}^{-1}$) & $8$ & $6330^{+800}_{-660}$ & $7$ & $5650^{+970}_{-740}$ & $13.918$ \nl
\tablevspace{.6ex}
\ion{C}{3}$]$~FWHM~(km~${\rm~s}^{-1}$) & $9$ & $8100^{+580}_{-660}$ & $5$ & $8990^{+1310}_{-910}$ & $63.599$ \nl
\tablevspace{.6ex}
\ion{C}{4}~EW~(\AA) & $8$ & $25.3^{+2.6}_{-1.8}$ & $6$ & $13.1^{+1.3}_{-1.5}$ & $99.639$ \nl
\tablevspace{.6ex}
\ion{C}{3}$]$~EW~(\AA) & $8$ & $18.1^{+0.6}_{-0.7}$ & $5$ & $17.6^{+1.9}_{-1.9}$ & $41.425$ \nl
\enddata
\tablenotetext{a}{Two-sided K-S probability that the RLQ and RQQ distributions of a given property were not drawn from the same parent sample.}
\end{deluxetable}
\end{scriptsize}
\clearpage

\subsection {Possible Differences}
We find that the BALQSO subset has stronger optical \fe2\
emission than the nonBALQSO subset (see Figure 12e).  This difference is obvious
in both the \ion{Fe}{2}/H$\beta$ and \fe2\ EW pairs of means.
It has been found that loBALs, in particular, are stronger in \ion{Fe}{2}, as
well as weaker in $[$\ion{O}{3}$]$, compared
with nonBALQSOs (\cite{bm92}; \cite{turnshek94}; and \cite{wills96b}); however, our BALQSO subset
contains only one known loBAL (Q1011+091 -- WMFH91).  A visual inspection of
this object's rest-frame optical spectrum shows that it is indeed \fe2\ rich,
as well as \oiii\ weak.  Yet, the same can be said for known hiBALs Q1246-057
and Q1309-056, as well as known radio quiet nonBALQSOs Q0049+007 (UM287) and
Q1346-036, while other BALQSOs like Q0043+008 (UM275 -- known hiBAL) and 
Q2212-179 (unknown BAL type) show the inverse trend of weak \fe2\ and strong
\oiii\ emission, common among most of the nonBALQSO subsample.  Thus it appears
that the \oiii\ -- \fe2\ anti-correlation of $\S4.1$ is valid for
{\it all} QSO classifications, but no distinction between loBALs and hiBALs
is apparent in our data.  Whether these statistically weak results
point towards any differences, physical or aspect dependent, between BALQSOs and
nonBALQSOs is not clear.  A larger sample with
better S/N might shed some light on this.  We can only say that
at the sensitivity of our observations, the BALQSO and radio quiet nonBALQSO
subsamples appear similar in their optical rest-frame spectral
properties.  

We also find that BALQSOs are somewhat weaker in soft X-ray luminosity 
(see Figure 12d)
as confirmed by Turnshek (1984), and Green \& Mathur (1996a).  It has been
suggested that strong absorption from the dense BAL clouds is responsible for the
weaker X-ray flux (\cite{green95}; and \cite{murray95}).

\section {Summary}
We have presented a detailed study of the emission line properties of 32 high 
luminosity QSOs at high redshift, drawing on new $H$-band spectra.  Of the
sample members, $15$ are classified as RLQ and $17$ are RQQ, but both are
similar in redshift and luminosity distributions.  The spectral coverage for the
entire sample included the forbidden \odoublet\ narrow lines, the allowed
Balmer \hb\ broad line, and the blended optical \fe2\ broad emission in the
neighborhood of H$\beta$.  We caution the reader that even though our data
are allowing an unprecedented view of the rest-frame optical spectrum in
luminous, high redshift QSOs, the
data are still of low S/N and only moderate spectral resolution.  The 
summary of our findings follows:

\begin{enumerate}
\item We find that, aside from their radio properties, the RLQ and RQQ subsets 
are significantly ($>97.2\%$)
different in several observables, implying
that the two types come from intrinsically separate populations of QSOs.

\item At the sensitivity of our observations, we find no statistically
significant ($>95\%$) differences between any rest-frame optical 
emission line properties, of the BALQSO and nonBALQSO subsets drawn from the
radio quiet subsample.  This is consistent with the hypothesis that any observed
differences on an object-to-object basis can be explained by a combination
of BAL cloud covering factor and orientation.

\item We report a previously unknown possible luminosity 
dependency of the forbidden \ox3\ NLR emission
velocity width over the range $0<z<2.5$.  In addition, we confirm a 
similar dependency for the \hb\ broad line width.  We propose that these
findings might be
evidence for a physical connection between the continuum and line emitting
regions at similar energies.  Furthermore, we report a 
$[$\ion{O}{3}$]\lambda5007$ ``Baldwin Effect'' for
the RQQ-only sample over this same redshift range.
\end{enumerate}

\section {Discussion}
The most significant correlation we found is the RLQ-RQQ dichotomy.  To put our 
observations into context we present a model, largely drawing on existing work,
illustrating the physical differences supported by the majority of the
significant correlations found in this statistical study.
For the RLQ population we assume the ``standard model'' for the central power source --- a super-massive
black hole and associated accretion disk that produces the ionizing continuum,
from thermal soft X-rays to UV photons (\cite{laor97}).  A powerful,
relativistic, and tightly collimated jet of synchrotron radio flux
(\cite{stocke92}) is emitted roughly perpendicular to the accretion disk.
The existence of this jet explains why RLQs are $\sim 1000$ times more radio
luminous than RQQs.  An extra component of harder X-rays is produced by inverse
Compton scattering of some of the beamed radio emission (\cite{elvis94};
and \cite{green95}).
The existence of this additional X-ray component explains why RLQs are
somewhat more X-ray bright, and why their ionizing continuum is flatter than
in RQQs.  We assume that this
component strengthens and hardens the ionizing continuum preferentially near
the radio jet axis.  Surrounding the central engine are the
BELR clouds in a semi-flattened, 
gravitationally bound system (\cite{jackson91}). 
For RLQs we adopt the VBLR+ILR model proposed by
Wills and Brotherton (see \cite{wills93}; \cite{brotherton94a}; 
etc.), such that emission from
the VBLR is equivalent to the broad ($v_{FWHM} \sim 10^4$~km~s$^{-1}$)
component of emission from the BELR, and that the VBLR is
stratified with the highest ionization species closest to the nucleus
(\cite{baldwin97}).  Extending away from the VBLR, we assume the existence
of bipolar ionization cones
roughly aligned with the radio axis as seen in some low redshift Seyfert
galaxies ({\it e.g.}~\cite{simpson96}).
At the apex of each cone is the low velocity ($v_{FWHM} \sim {\rm few} 10^3$~km~s$^{-1}$)
ILR that we assume is also stratified.  Finally, further outward, at
the wider end of each cone, is the non-spherically symmetric NLR.

The statistical differences between RLQs and RQQs, as well as many significant
correlations between emission properties, found in this study can be explained
in the context of our proposed RLQ model as follows:

\begin{itemize}
\item[(i)] The more strongly collimated continuum emission photo-ionizing a greater
volume of NLR gas, that is concentrated in a non-spherically symmetric region,
could produce the stronger \oiii\ emission observed in RLQs compared to RQQs.
 
\item[(ii)] As with the NLR, we argue that an increased flux from the ILR is expected,
due to more completely ionized gas by the collimated continuum focused on a
tight region at the apex of the bipolar ionization cones.  Permitted \ciii\ 
and \civ\ emission originating in sufficient amounts from the ILR could
produce the overall narrower line widths seen in RLQs relative to RQQs, as
well as explain the two inverse correlations between \oiii\ strength and
the widths of \ciii\ and \ion{C}{4}, and the positive relation between 
\oiii\ EW and \civ\ EW.

\item[(iii)] The association between the existence of the radio jet and the production
of the more powerful and harder ionizing continuum can physically explain why
\logr\ and $\alpha_{ox}$ are anti-correlated.  Coupling the preceding idea
with (i) should produce the observed positive \oiii\ strength -- \logr\ 
and negative \oiii\ strength -- $\alpha_{ox}$ relations.

\item[(iv)] The broad component of the H$\beta$, \ciii\ and \ion{C}{4} emission lines
is presumed to come from the VBLR, while a significant flux from the ILR
makes up the remaining component of the \civ\ and \ciii\ broad-emission-lines.
Thus, one should expect the strong \civ\ FWHM -- \ciii\ FWHM correlation
observed, in addition to weaker correlations between the {\it total} line
widths of the carbon ions and H$\beta_{{\rm broad}}$ FWHM.

\item[(v)] Assuming $\alpha_{ox}$ is a measure of viewing angle with respect to the
radio jet axis, then the H$\beta_{{\rm broad}}$ FWHM -- $\alpha_{ox}$ 
correlation may be evidence of a gravitationally bound VBLR.  Furthermore,
assuming the luminosity is directly related to the mass of the central
massive black hole, an increase in luminosity should correlate, on average,
with an increase in line width as observed in the \hb\ FWHM -- 
$\log(L_{\nu}(V))$ relation.
\end{itemize}

The above discussion was aimed towards an explanation of the RLQ population
properties.  The RQQ version of our model differs in its lack of a
strong, collimated radio jet.  Thus, we argue that though RQQs have a
similar nuclear source in mass and accretion rate as the RLQ population 
(\cite{hooper96}), but
instead of a collimated jet, the RQQs produce
a sub-relativistic, uncollimated wind (\cite{stocke92}), or a weak, non-boosted
jet (\cite{falcke96}).  This proposed model has several implications:

\begin{itemize}
\item[(i)] Assuming there is a connection between the increased
coverage of optically thick, iron emitting VBLR clouds, and the lack of a
strong, collimated radio jet, the VBLR covering factor in RQQs should be larger,
increasing the shielding of ionizing photons to the NLR (\cite{wills96a}),
as well as the ILR.

\item[(ii)] Combining the increased shielding of ionizing radiation to the ILR and NLR
with the absence of an additional hard, collimated component of thermal soft
X-rays in RQQs, may enhance emission from low ionization species like
Fe$^+$ in the VBLR, while diminishing emission from higher
ionization species in the ILR and NLR.
\end{itemize}

These two effects could be responsible for the observed negative correlation
between \oiii\ strength and \fe2\ strength, plus the \hb\ EW and \fe2\ EW
correlations with $\alpha_{ox}$, and the \civ\ and \ciii\ line width to \fe2\
strength relations.

\acknowledgments{}
We would like to thank George Rieke for useful discussion and help in obtaining
these data.  Fred Chaffee is thanked for a generous allotment of Director's
time on the MMT.  Todd Boroson kindly made his spectra of the low redshift
sample available in digital form, as well as the optical \fe2\ template from
I Zw 1.
We thank Bev Wills for faxing us a UV rest spectrum of Q2310+385.
We acknowledge helpful discussions and correspondence with Gary Schmidt,
Jim Condon, Jill Bechtold, Eric Hooper, and Pat Hall.  We are grateful to
Chad Engelbracht and Kevin Luhman for answering endless data reduction
questions.  And we thank Margaret Hanson for help with atmospheric corrections.
This research has made use of the NASA/IPAC Extragalactic 
Database (NED) which is operated by the Jet Propulsion Laboratory,
California Institute of Technology, under contract with the National Aeronautics
and Space Administration.  We also made use of NASA's Astrophysical Data
System Abstract Service (ADS).
We acknowledge support from NSF grant (AST93-20715).  Acquisition of the NICMOS3
array was made possible by NASA.

\clearpage


\newpage


\begin{figure}[p]
\centering
\plotone{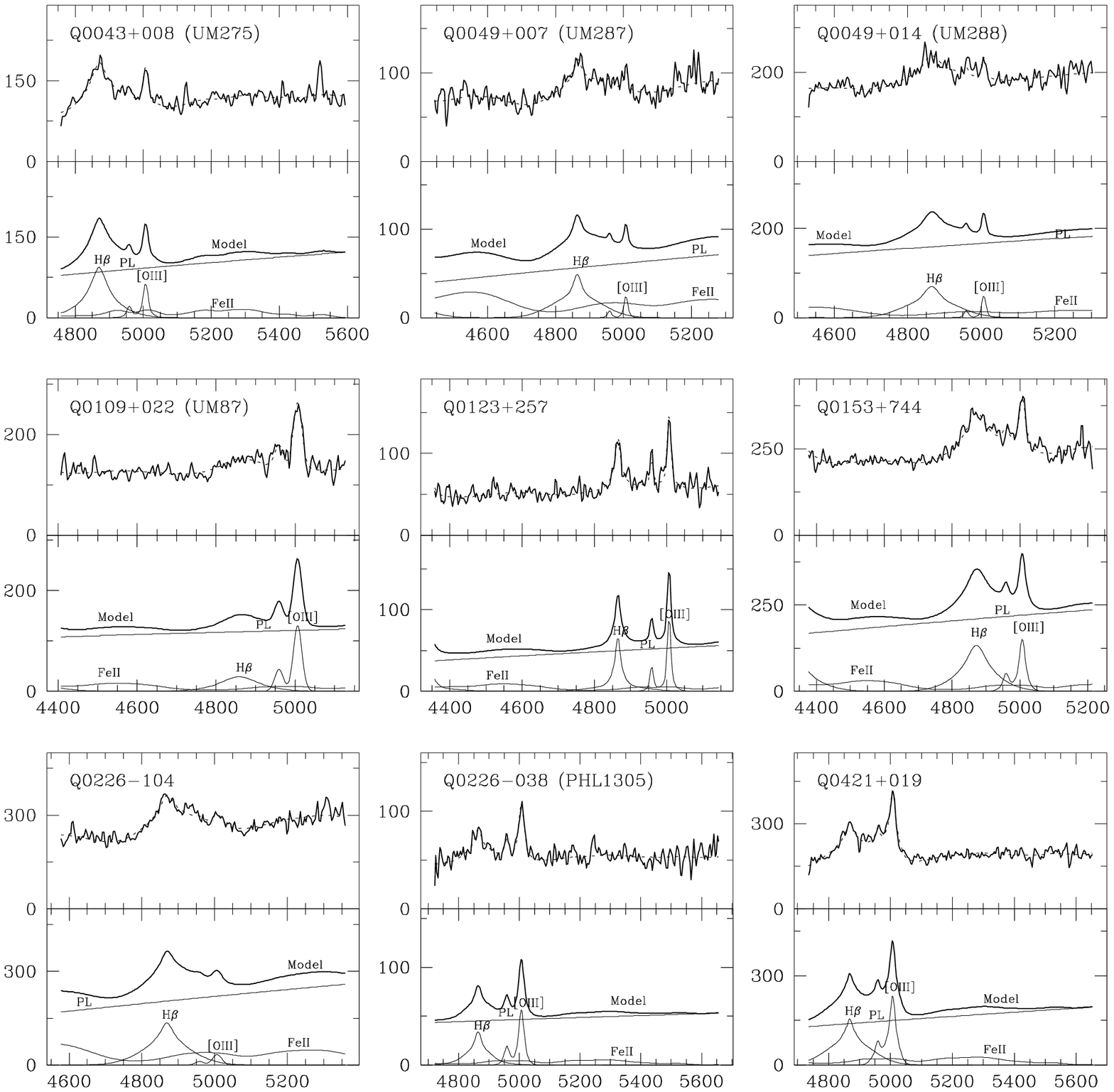}
\figcaption{Rest-frame spectra of
all 32 QSOs with their minimized $\chi^2$ best fit
model (dashed line) in top panel.  The lower panel shows the model (bold line)
with its H$\beta$, $[$\protect\ion{O}{3}$]$, \protect\ion{Fe}{2} and power-law components.  
The horizontal axis is wavelength in \AA, while the verticle axis shows flux
levels in arbitrary units.}
\end{figure}

\begin{figure}[p]
\centering
\plotone{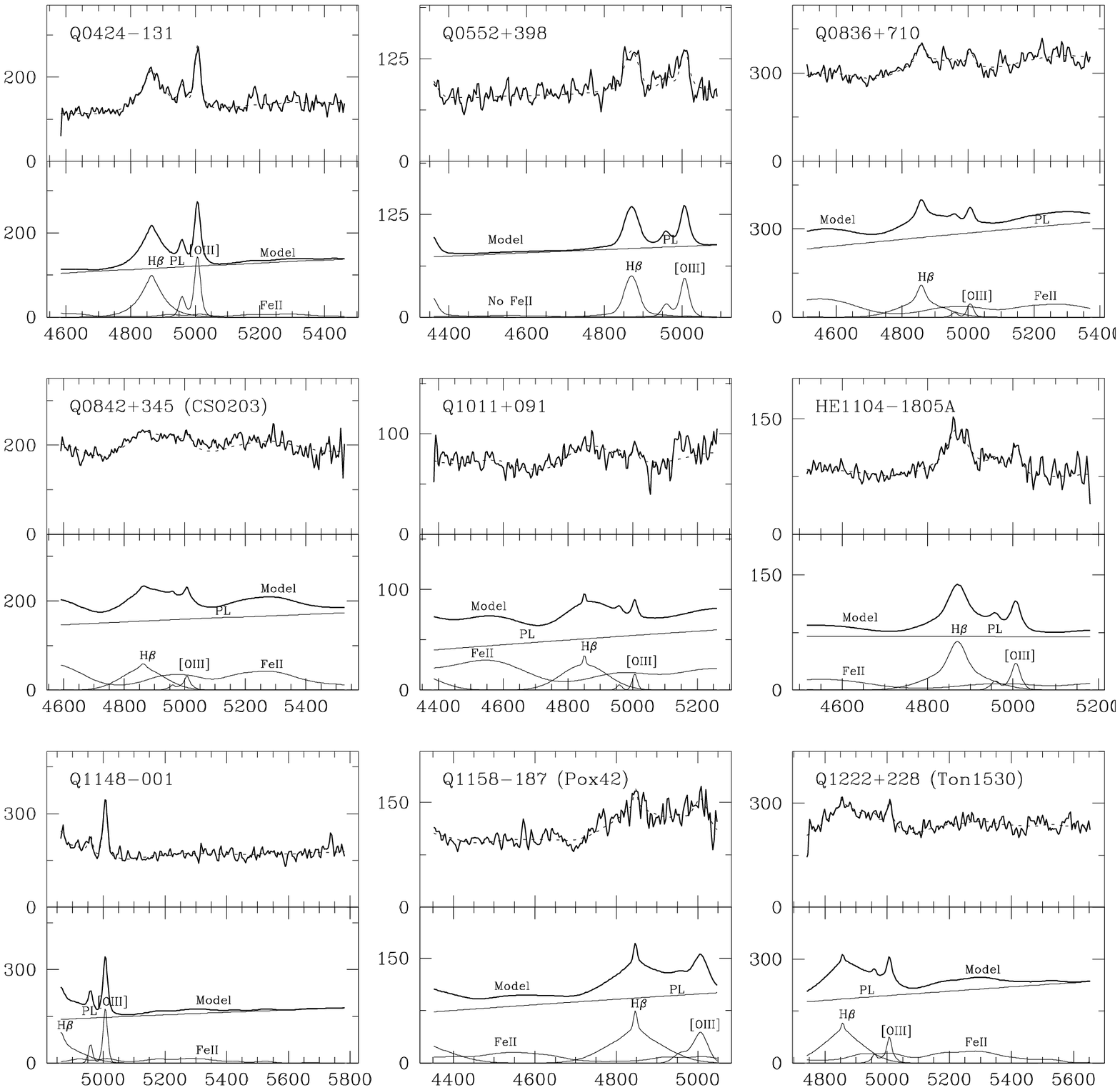}
\end{figure}

\begin{figure}[p]
\centering
\plotone{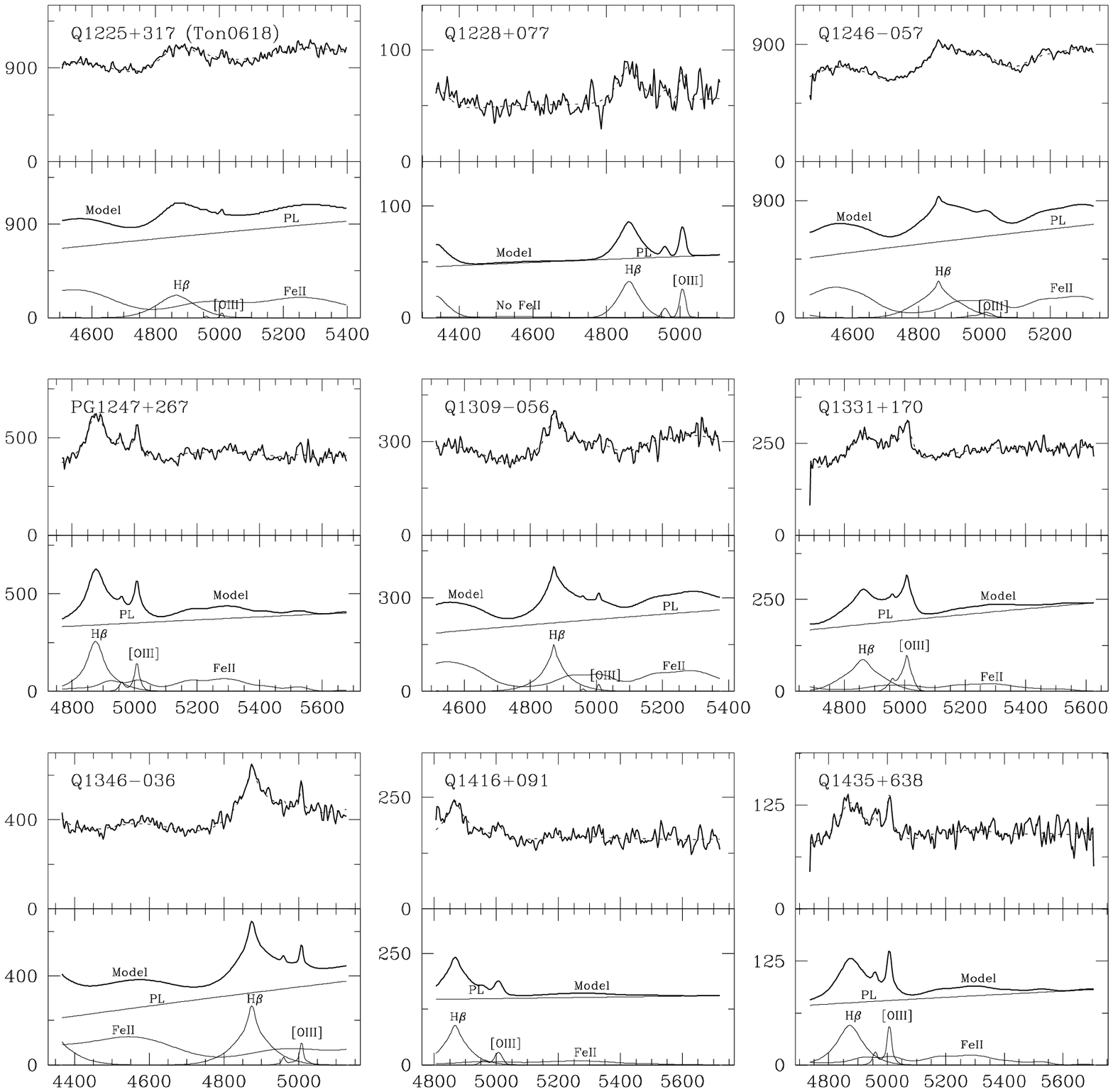}
\end{figure}

\begin{figure}[p]
\centering
\plotone{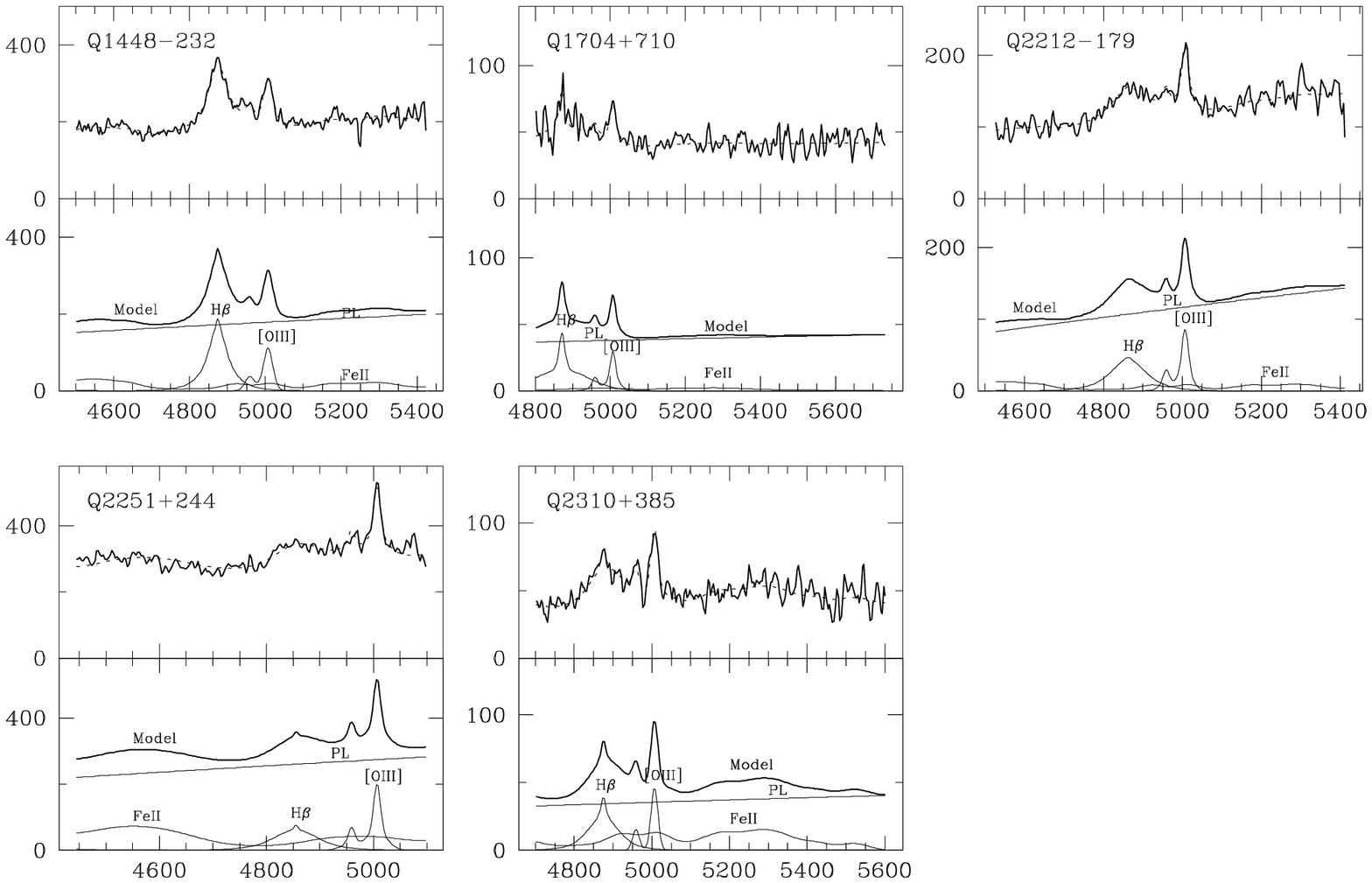}
\end{figure}

\begin{figure}[p]
\centering
\plotone{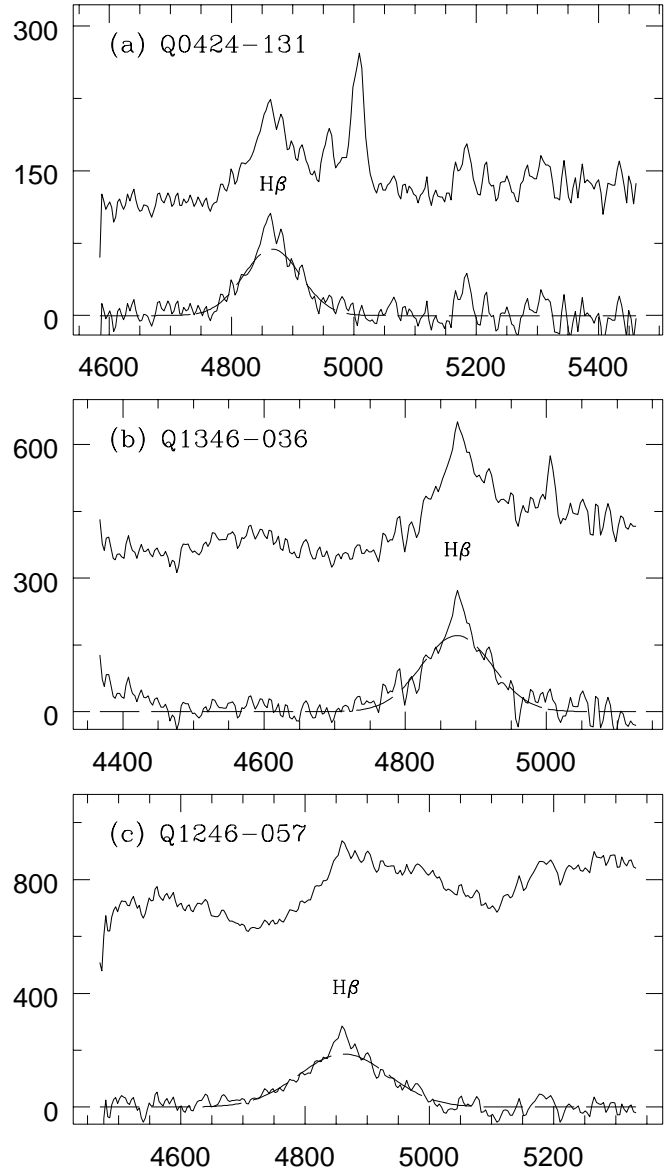}
\figcaption{Representative examples of a rest-frame spectrum (upper),
with extracted H$\beta$ line (lower), and single Gaussian {\it broad} component 
fit (dashed).  For: (a) a RLQ; (b) a RQQ; and (c) a BALQSO.  The horizontal
axis is wavelength in \AA, while the verticle axis shows flux levels in
arbitrary units.}
\end{figure}

\begin{figure}[p]
\centering
\plotone{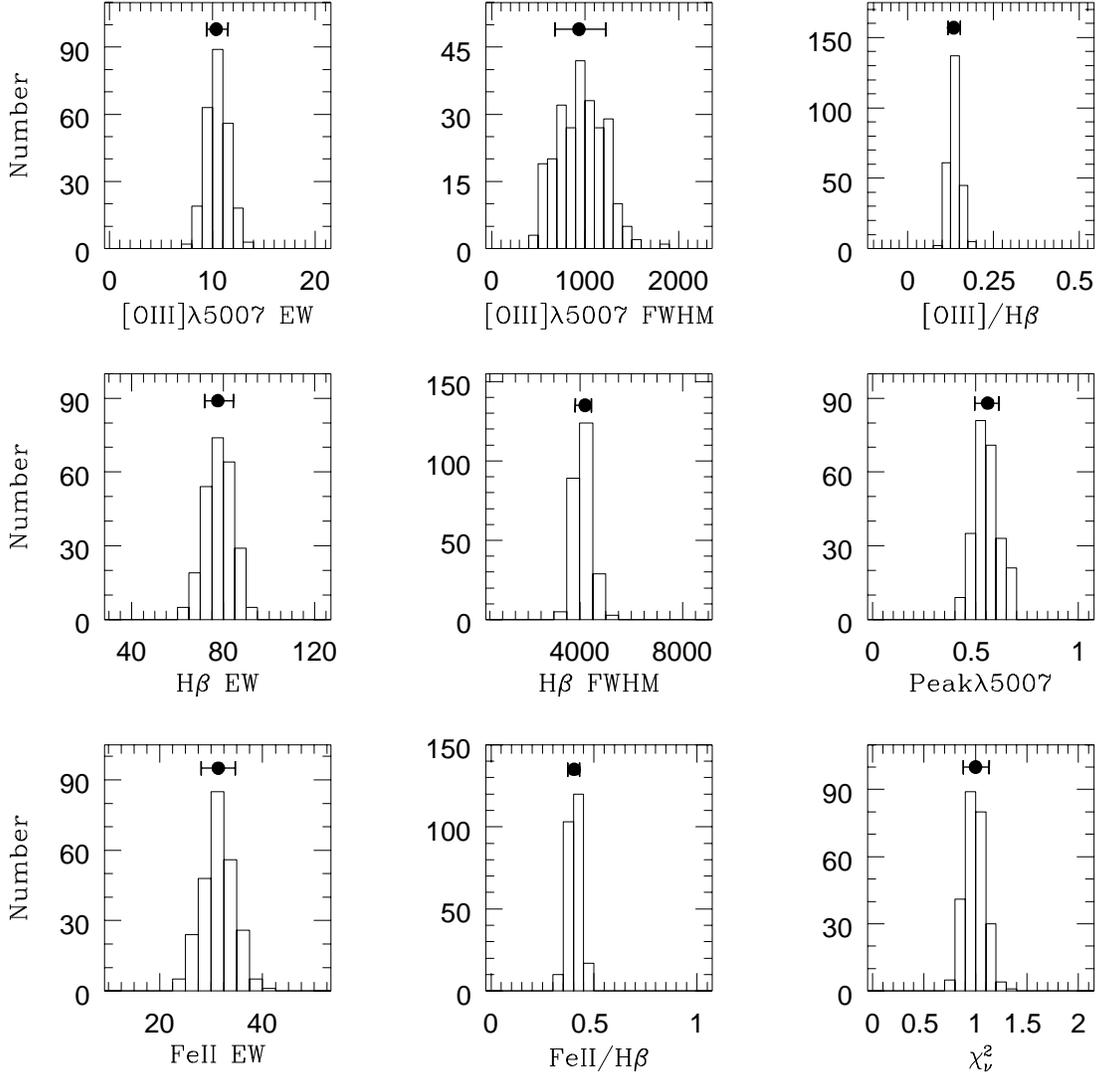}
\figcaption{The Monte Carlo error distributions for minimized $\chi^2$
best fits to 250 random
synthetic representations of PG1247+267.  The nine boxes represent eight
of the parameters that we measured for each QSO, plus the reduced chi-squared
$\chi_{\nu}^2$ best fit parameter.  The solid circle above each parameter
distribution represents the value measured from the best fit to the real
data, while the right and left $1\sigma$ error bars are calculated from
the error distributions.}
\end{figure}

\begin{figure}[p]
\centering
\plotone{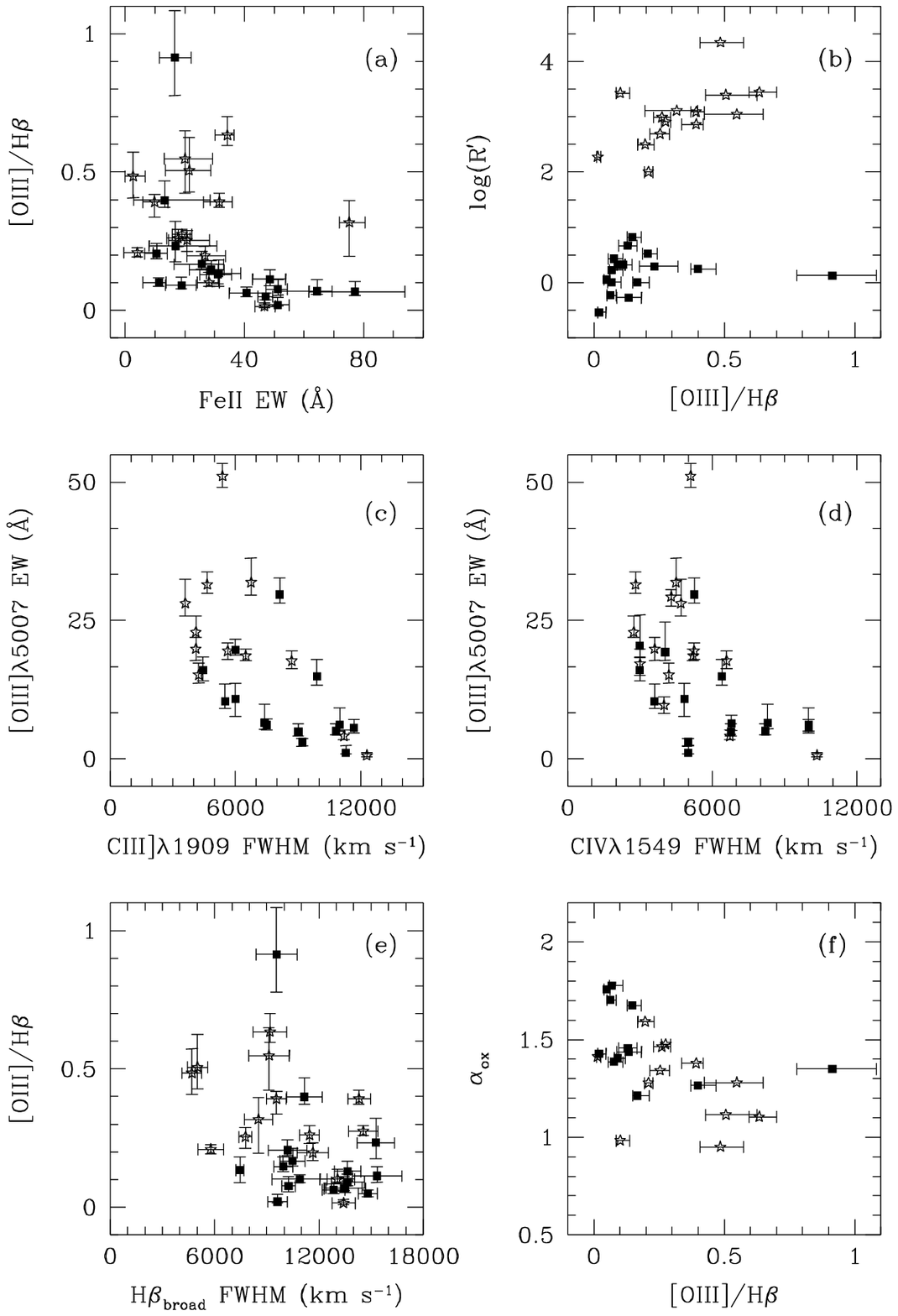}
\figcaption{The significant correlations with the strength of 
$[$\protect\ion{O}{3}$]\lambda5007$ emission: (a) negative $[$\protect\ion{O}{3}$]$/H$\beta$
to \protect\ion{Fe}{2} EW; (b) positive $[$\protect\ion{O}{3}$]$/H$\beta$ to 
${\log(R^{\prime})}$; (c) negative $[$\protect\ion{O}{3}$]$ EW to 
\protect\ion{C}{3}$]\lambda1909$ FWHM; (d) negative $[$\protect\ion{O}{3}$]$ EW to
\protect\ion{C}{4}$\lambda1549$ FWHM; (e) negative $[$\protect\ion{O}{3}$]$/H$\beta$ to 
broad H$\beta$ FWHM; and (f) negative $[$\protect\ion{O}{3}$]$/H$\beta$ to 
$\alpha_{ox}$.  Solid squares are RQQs, while open stars are RLQs.}
\end{figure}

\begin{figure}[p]
\centering
\plotone{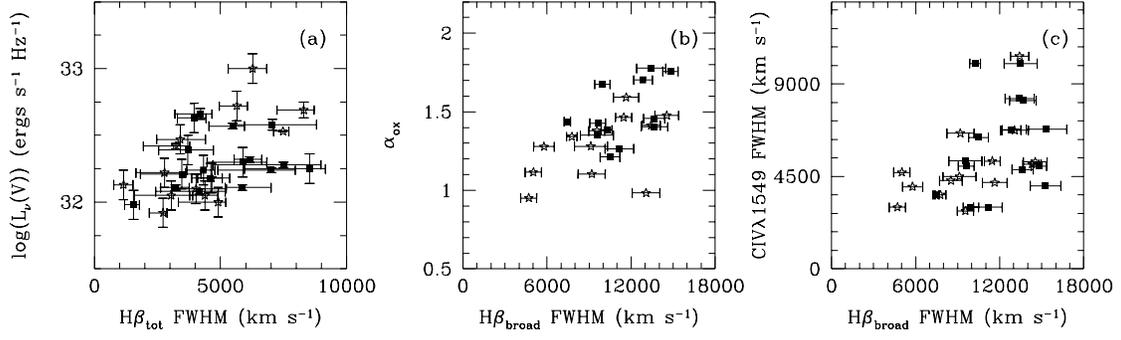}
\figcaption{The significant positive correlations with the width of 
H$\beta$ emission: (a) total H$\beta$ FWHM to $L_{\nu}(V)$; (b) broad
H$\beta$ FWHM to $\alpha_{ox}$; and (c) H$\beta$ FWHM to 
\protect\ion{C}{4}$\lambda1549$ FWHM.  Solid squares are RQQs, while open 
stars are RLQs.}
\end{figure}

\begin{figure}[p]
\centering
\plotone{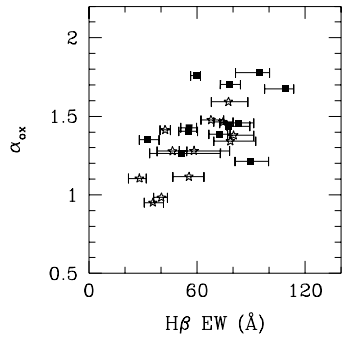}
\figcaption{The significant positive correlation between 
H$\beta$ EW and $\alpha_{ox}$.  Solid squares are RQQs, while open stars
are RLQs.}
\end{figure}

\begin{figure}[p]
\centering
\plotone{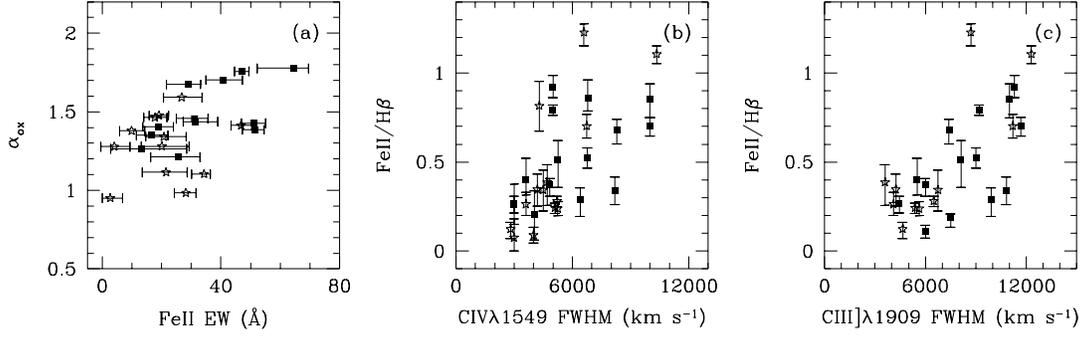}
\figcaption{The significant positive correlations with the strength 
of optical \protect\ion{Fe}{2} emission: (a) \protect\ion{Fe}{2} EW to $\alpha_{ox}$; (b)
\protect\ion{Fe}{2}/H$\beta$ to \protect\ion{C}{4}$\lambda1549$ FWHM; and (c) 
\protect\ion{Fe}{2}/H$\beta$ to \protect\ion{C}{3}$]\lambda1909$ FWHM.  Solid squares are RQQs,
while open stars are RLQs.}
\end{figure}

\begin{figure}[p]
\centering
\plotone{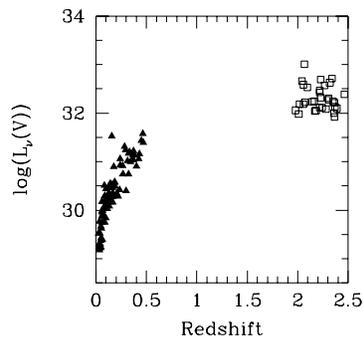}
\figcaption{The Hubble Diagram for the combined sample of 87 low 
redshift QSOs from BG92 (solid triangles)
and our 32 high redshift QSOs (open squares).  The luminosity densities
(in units of ergs~s$^{-1}$~Hz$^{-1}$) were calculated adopting 
H$_{0} = 50$~km~s$^{-1}$~Mpc$^{-1}$ and $q_{0}=0.1$.}
\end{figure}

\begin{figure}[p]
\centering
\plotone{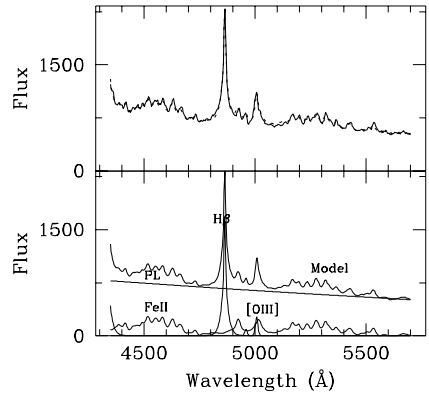}
\figcaption{Rest-frame spectrum of a representative QSO (PG1404+226) from
the low redshift sample of BG92, with its minimized $\chi^2$ best fit
model (dashed line) in top panel.  The lower panel shows the model (bold line)
with its H$\beta$, $[$\protect\ion{O}{3}$]$, \protect\ion{Fe}{2} and 
power-law components.}
\end{figure}

\begin{figure}[p]
\centering
\plotone{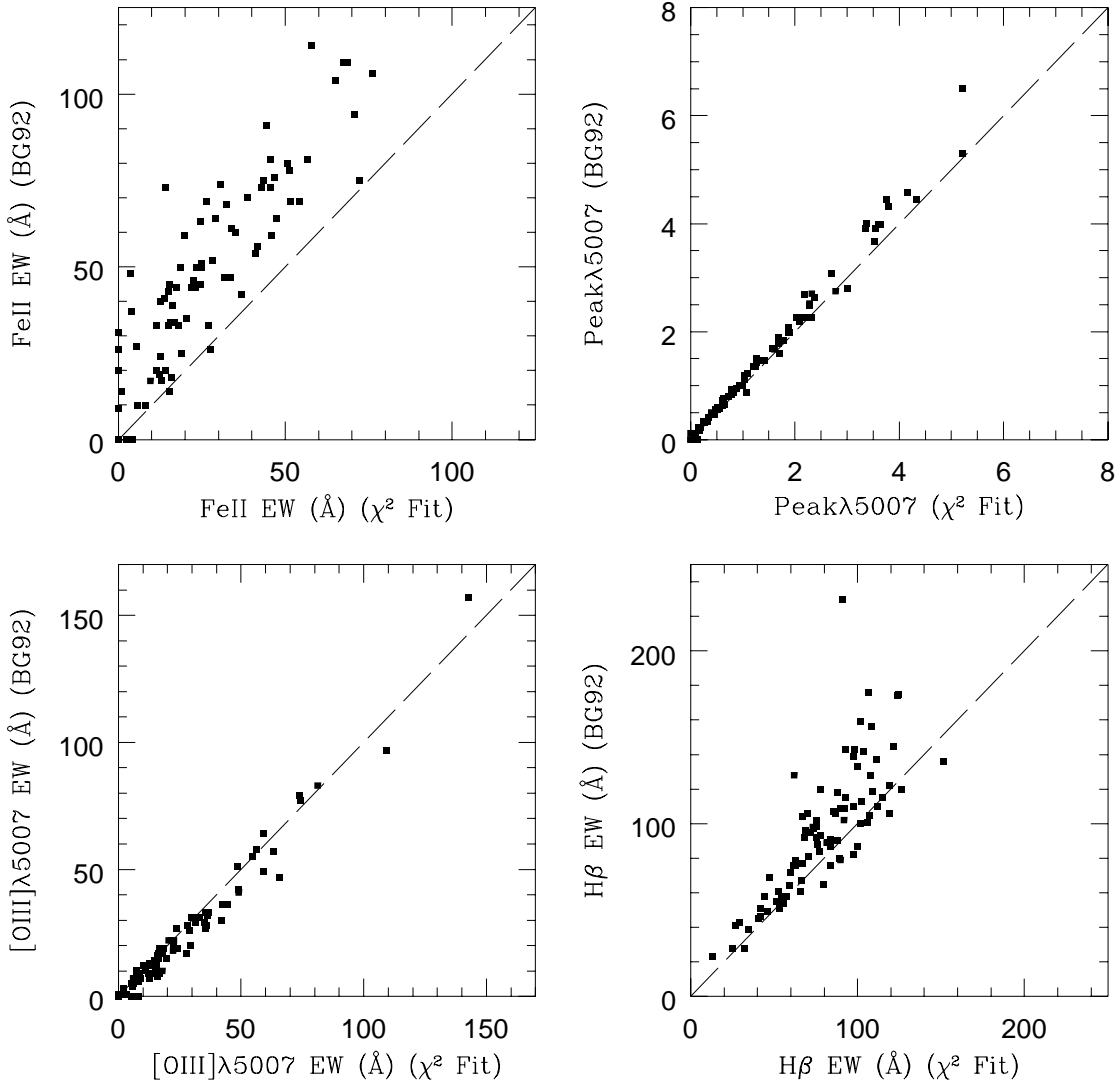}
\figcaption{Comparison between our minimized $\chi^2$ best fit derived
values versus the published BG92 results for the EWs of $[$\protect\ion{O}{3}$]$,
and H$\beta$, the {\it relative} EW of optical \protect\ion{Fe}{2}, and the
Peak$\lambda5007$ parameter.}
\end{figure}

\begin{figure}[p]
\centering
\plotone{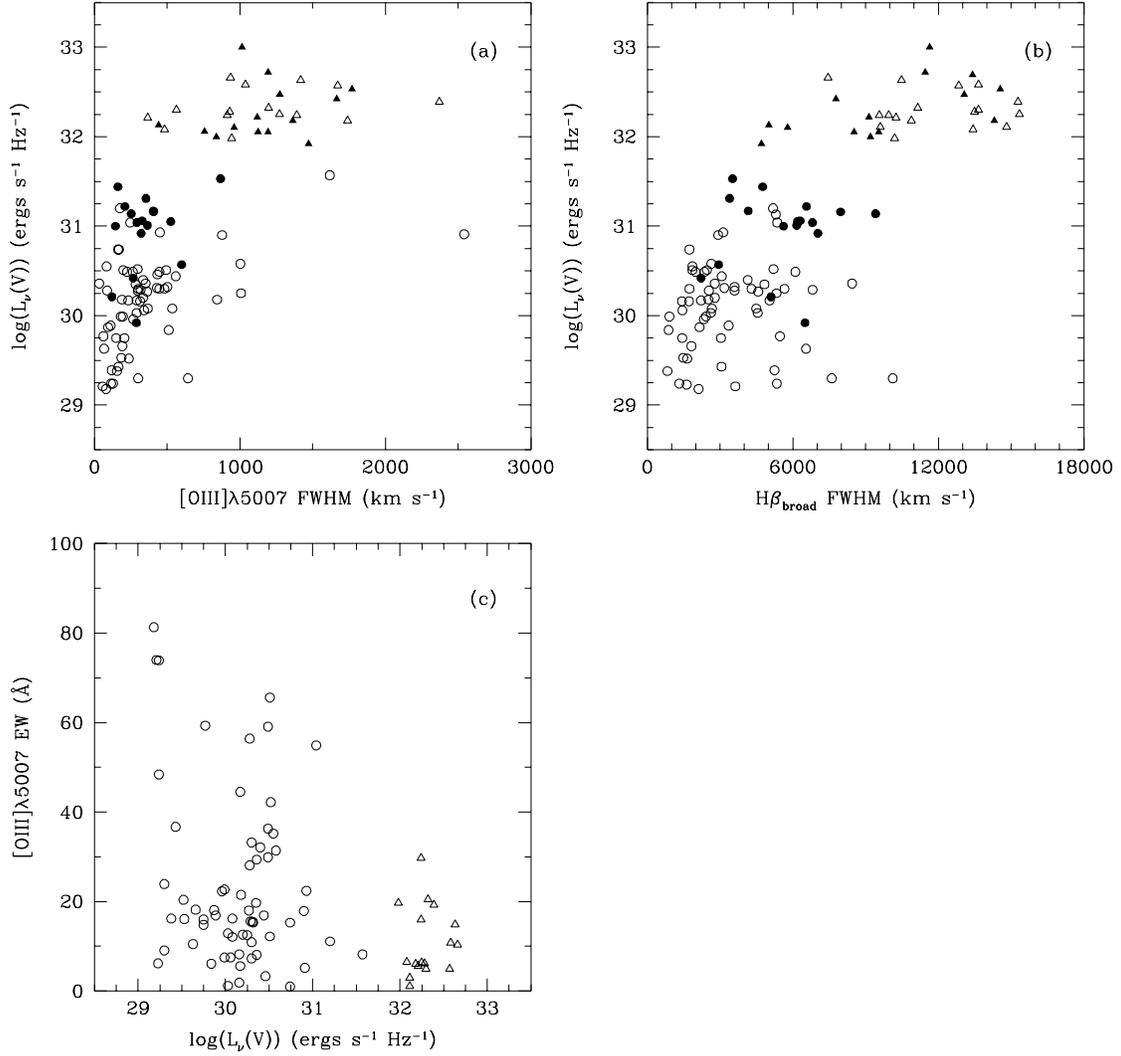}
\figcaption{The strong luminosity dependencies of the combined 
low redshift (circles) and high redshift (triangles) $0<z<2.5$
sample for (a) rest-frame $[$\protect\ion{O}{3}$]\lambda5007$ FWHM, and (b) 
rest-frame broad H$\beta$ FWHM.
Solid symbols are RLQs, while open symbols are RQQs.  The Baldwin Effect for
the $[$\ion{O}{3}$]\lambda5007$ line in the (c) RQQ only subsample.}
\end{figure}

\begin{figure}[p]
\centering
\plotone{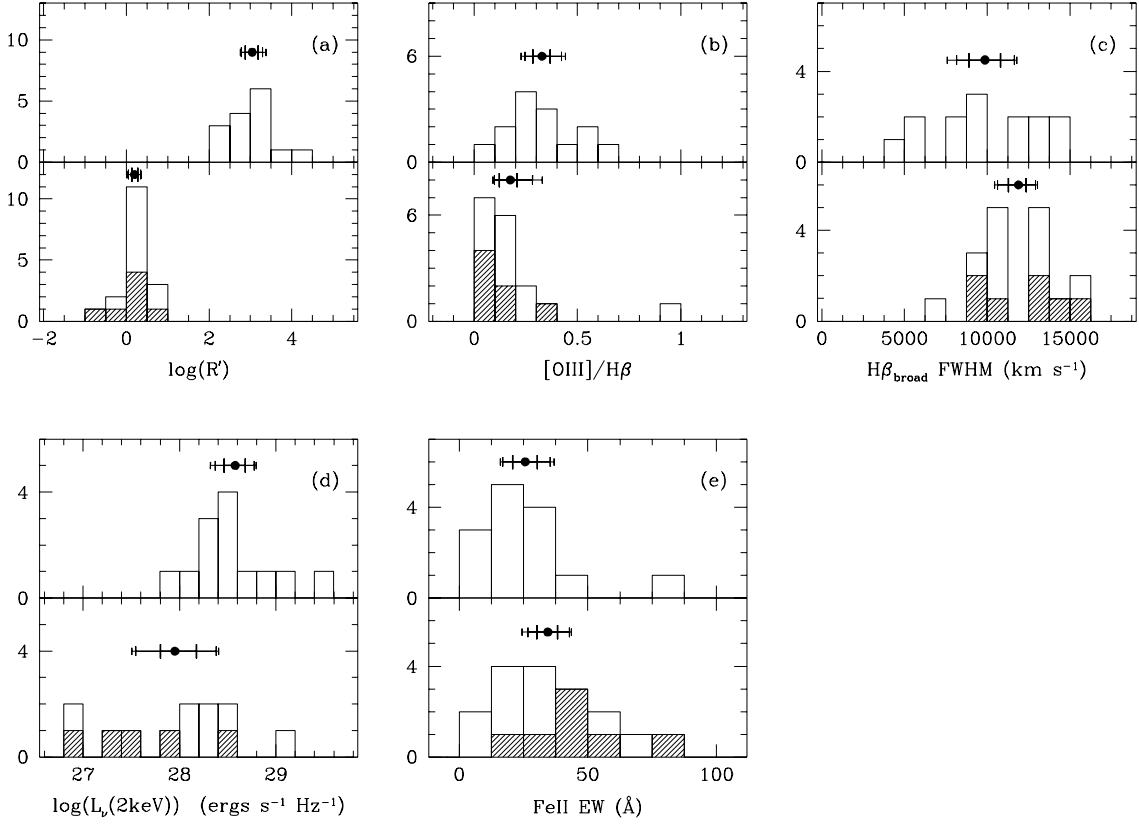}
\figcaption{The RLQ (upper panel) and RQQ (lower panel -- with 
shaded BALQSO subset
over-plotted) subsample distributions: 
(a) ${\log(R^{\prime})}$; (b) $[$\protect\ion{O}{3}$]/$H$\beta$; (c) broad H$\beta$
FWHM; (d) $\log(L_{\nu}(\rm 2keV))$; and (e) \protect\ion{Fe}{2} EW.  The verticle 
axis gives the number per bin.  The bin sizes are approximately 
$2\langle \sigma \rangle$ from the Monte Carlo error analysis.  The mean
of each panel's distribution is plotted (solid circle) along with the 
corresponding right and left $1-3 \sigma$ uncertainties.}
\end{figure}

\end{document}